%% file: ms.tex
\documentclass[12pt,preprint]{aastex}
\usepackage{lscape}
\usepackage{multirow}

\newcommand{\degree}{\mbox{$^{\circ}$}}

\newcommand{\as}{\mbox{\arcsec}}

\newcommand\cmv{\mbox{cm$^{-3}$}}

\def\lsim {$\rlap{\raise.4ex\hbox{$<$}}\lower.55ex\hbox{$\sim$}\,$}

\newcommand{\neff}{n$_{\mathrm{eff}}$}

\newcommand{\lsun}{\mbox{L$_\odot$}}
\newcommand{\msun}{\mbox{M$_\odot$}}
\newcommand{\ta}{{$T_A^*$}}

\newcommand{\lbol}{\mbox{$L_{bol}$}} 
\newcommand{\tiso}{\mbox{$T_{iso}$}} 

\newcommand{\mvir}{\mbox{$M_{vir}$}} 
\newcommand{\miso}{\mbox{$M_{iso}$}} 
\newcommand{\mean}[1]{\mbox{$\langle#1\rangle$}} 

\newcommand{\water}{H$_2$O}

\newcommand{\methanol}{CH$_3$OH}

\newcommand{\soo}{SO$_2$}

\newcommand{\hcop}{HCO$^+$}
\newcommand{\hcsp}{HCS$^+$}
\newcommand{\hcopi}{H$^{13}$CO$^+$}

\newcommand{\nthp}{N$_2$H$^+$}
\newcommand{\nthpo}{N$_2$H$^+$ $J = 1 \rightarrow 0$}

\input{epsf}

\begin{document}

               
\title {\bf The Physical Properties of High-Mass Star-forming Clumps: A Systematic Comparison of Molecular Tracers}
\author {Megan Reiter, Yancy L. Shirley\altaffilmark{1}}
\affil{Astronomy Department, The University of Arizona,
       933 N. Cherry Ave., Tucson, AZ 85721}
\affil{mreiter@as.arizona.edu, yshirley@as.arizona.edu}

\author {Jingwen Wu}
\affil{Jet Propulsion Laboratory, 
       M/S 264-723, 4800 Oakgrove Drive, Pasadena CA 91109}
\affil{jingwen.wu@jpl.nasa.gov}

\author{Crystal Brogan, Alwyn Wootten}
\affil{National Radio Astronomy Observatory, 520 Edgemont Road,
       Charlottesville, VA 22903}
\affil{cbrogan@nrao.edu, awootten@nrao.edu}
\and
\author{Ken'ichi Tatematsu}
\affil{National Astronomical Observatory of Japan, 2-21-1 Osawa, Mitaka
       Tokyo 181-8588, Japan}
\affil{k.tatematsu@nao.ac.jp}
\altaffiltext{1}{Adjunct Astronomer at the National Radio Astronomy 
Observatory. The National Radio Astronomy Observatory is a facility of the 
National Science Foundation operated under a cooperative agreement by 
Associated Universities, Inc.}
 
\begin{abstract}

We present observations of \hcop\ and \hcopi, \nthp, 
\hcsp, HNC and HN$^{13}$C, 
SO and $^{34}$SO, 
CCH, \soo, 
and CH$_3$OH-E 
towards a sample of 27 high-mass clumps coincident with water maser emission.
All transitions are observed with or convolved to nearly identical resolution ($30$\arcsec), allowing for inter-comparison of the clump properties derived from the mapped transitions. 
We find \nthp\ emission is spatially differentiated compared to the dust and the other molecules towards a few very luminous cores (10 of 27) and the \nthp\ integrated intensity does not
correlate well with dust continuum flux.  
We calculate the effective excitation density, \neff, the density required to excite a 1 K line in T$_{kin}$=20 K gas for each molecular tracer.  
The intensity of molecular tracers with larger effective excitation densities 
($\mathrm{n_{eff}} \ge 10^5$ cm$^{-3}$) appear to correlate more strongly 
with the submillimeter dust continuum intensity. 
The median sizes of the clumps are anti-correlated with the \neff\ of the tracers (which span more than three orders of magnitude). 
Virial mass is not correlated with \neff, especially where the lines are optically thick as the linewidths may be broadened significantly by non-virial motions. 
The median mass surface density and median volume density of the clumps is correlated with \neff\ indicating the importance of understanding the excitation conditions of the molecular tracer when deriving the average properties of an ensemble of cores.
\end{abstract}



\section{Introduction}

Throughout their short lives, massive stars make significant contributions to their host galaxy's energy budget. From strong winds and intense radiation during their brief main sequence existence to their spectacular end in powerful supernovae, massive stars provide substantial energetic feedback (Motte \& Hennebelle 2008). On galactic scales, massive stars enrich the Interstellar Medium (ISM) with heavier elements, and their powerful winds may trigger successive episodes of star formation. Understanding the role of massive stars in galaxy evolution is essential for improving galaxy evolution models and properly treating feedback from high-mass stellar evolution. Despite their importance, many fundamental questions remain to be answered about the formation of massive stars. 

While the theory of low-mass star formation is relatively mature (e.g. Shu, Adams \& Lizano 1987), that theory does not simply extend to the formation of stars above $\sim 8$ \msun; indeed, the basic formation mechanisms for high-mass stars are still debated. 
Three competing theories attempt to describe the origin of massive stars (Zinnecker \& Yorke 2007). 
The core accretion model essentially scales the theory of low mass star formation to higher masses by assuming that massive stars form from massive cores (Zinnecker \& Yorke 2007). 
Core accretion models, such as those studied by Krumholz et al. (2009a) require non-spherical accretion to overcome the intense radiation pressure that massive protostars create (and that causes theories for low mass star formation to fail for larger masses). 
%
%
Core accretion models must explain why a massive cloud would collapse to form a single, massive core, rather than fragment in to several less-massive cores. Recent work by Krumholz (2006) and Krumholz et al. (2007a,b) suggests that radiation feedback may solve this problem by changing the equation of state of the gas, which determines whether the clump will fragment, and thus allows a more massive core to collapse. 
 
Competitive accretion models, on the other hand, assume that fragmentation produces many low-mass stellar cores while high mass stars form subsequently due to continued accretion (e.g. Bonnell \& Bate 2006). This model may also explain why massive stars are found preferentially in the center of clusters --- residing deep in the cluster potential allows cores to grow more massive, providing greater gravity to attract the gas, thus leading to runaway accretion (Bonnell et al. 2001). 
It is unclear whether this mass segregation is a result of how massive stars form (Bonnell \& Davies 1998; Huff \& Stahler 2006) or if it is a consequence of cluster dynamics (Tan, Krumholz \& McKee 2006; McMillian, Vesperini \& Portegies Zwart 2007).
%
%
For the densest star clusters, massive stars may form via stellar collisions, although this process requires densely packed objects of relatively high mass to proceed (Zinnecker \& Yorke 2007).

Current observational constraints are not sufficiently quantitative to discriminate between models of high mass star formation (Krumholz \& Bonnell 2007; Motte \& Hennebelle 2008). 
Theoretical models are fundamentally limited by an incomplete understanding of the properties of the dense gas in high-mass (M $>$ few M$_{\sun}$) star-forming clumps.
While the high spatial resolution available with ALMA is required to observe massive star formation on the scale of individual stars (Krumholz \& Bonnell 2007), on scale of individual star clusters, progress can be made by improving our understanding of the properties of massive clumps. 
Recent mapping surveys of dust continuum emission (Mueller et al. 2002, Beuther et al. 2002, Williams et al. 2005) as well as molecular emission from
multiple transitions of CS and HCN (Plume et al. 1992, Plume et al. 1997, Shirley et al. 2003, Wu et al. 2010) have attempted to characterize the large-scale 
density and temperature structure of the massive progenitors of clusters.
Blind surveys of continuum emission in the galactic plane (e.g. GLIMPSE, Benjamin et al. 2005;  
BGPS, Aguirre et al. 2011; 
ATLASGAL, Schuller et al. 2009;  
HIGAL, Molinari et al. 2010; etc.) 
reveal new sites of potential star formation. 
Molecular line observations are essential for determining the basic properties of these newly discovered sources.
From molecular line observations, we can calculate kinematic distances and derive physical properties, such as size, mass and density, of these sources.  
Proper interpretation of those results requires that we understand how the derived clump properties depend on the tracer and the excitation conditions required to excite that transition.

In this paper, we analyze observations of a sample of 27 high-mass clumps associated with \water\ masers in the dense gas tracers \nthp , \hcop , HCS$^+$, HNC, CCH, SO, \soo , and CH$_3$OH in a total of 12 transitions among the isotopologues.  We study spatial information in emission maps of transitions of 
\nthp , \hcop , HNC, CCH, SO, and for previously published CS transitions.
This particular set of transitions covers more than three orders of magnitude in
effective excitation density with nearly identical resolution ($30^{\prime\prime}$) 
for all twelve
transitions.  Matching the resolution between tracers is of critical
importance for systematic statistical comparisons of the properties of the sources.
The transitions studied in this work represent diverse 
probes of the chemistry and physical structure in these clumps.  
They include molecular ions, sulfur-bearing species, the simplest carbon chain
molecule, and the isomer of a commonly observed dense gas tracer.
%
The rich chemistry in high mass star forming regions may provide clues to the evolutionary state of the clump and there is a rich literature of line surveys studying the chemistry in high mass star forming regions in detail (e.g. Sakai et al. 2008; Beuther et al. 2009; Vasyunina et al. (2010). Though many of our molecular species have been observed in other studies, we do not study the chemistry of these clouds in detail here. 
Instead, we use the results of this mapping survey
to study in detail how fundamental properties of the clumps (e.g. size,
mass, surface density, etc.) vary with the effective excitation density of the tracer.

In our analysis we use the effective density \neff, rather than critical density $n_{crit}$, to characterize the environment where a transition is excited.  
As discussed in Evans (1999), the ability to detect a transition is usually taken to mean that $n > n_{crit}$, but in truth lines are easily excited in subthermally populated gas with densities more than an order of magnitude lower than $n_{crit}$.
Molecular filling fraction, line frequency, optical depth, multilevel excitation effects and trapping can all affect the observed line strength. To account for these effects, we use the Evans (1999) definition of effective density as the density needed to excite a $T_R = 1$ K line for a given kinetic temperature as calculated with a non-LTE radiative transfer code. In this paper we update and expand the table of effective densities given in Evans (1999). 
We use the online version of RADEX (van der Tak et al. 2007), 
a non-LTE radiative transfer code,
assuming log($N / \Delta v$) = $13.5$ cm$^{-2}$ / (km s$^{-1}$) and a kinetic temperature of 20 K for all species (see Table 1; Figure 1).

The sample of sources is drawn from the Plume et al. (1992, 1997) survey of multiple CS transitions (see Table 2). 
Most of these sources have been mapped and modeled in 350 \micron\ dust continuum emission (Mueller et al. 2002). Maps of CS J = $5 \rightarrow 4$ emission offer additional constrains on size and virial mass of these dense clumps (Shirley et al. 2003). Wu et al. (2010) extended the mapped dataset to include two CS transitions (J = $7 \rightarrow 6$ and J = $2 \rightarrow 1$) as well as adding two transitions of HCN (J = $1 \rightarrow 0$ and J = $3 \rightarrow 2$). 
Our sample of 27 sources represents a range of bolometric luminosities from 
$1.1 \times 10^3 L_{\sun}$ to $1.6 \times 10^6 L_{\sun}$ (median $4.4 \times 10^4 L_{\sun}$). CS clump sizes are $\sim 0.25$ pc on average. 
The median virial mass, calculated from CS J = $5 \rightarrow 4$ emission in (Shirley et al. 2003) is $600 M_{\sun}$. 
Sources are at distances ranging from $0.7$ kpc to $8.5$ kpc (median $2.8$ kpc) so that with a $30^{\prime \prime}$ single dish beam we have a corresponding sensitivity to spatial scales of $0.05$ pc to $0.62$ pc respectively (median $0.21$ pc). All of our maps have nearly identical resolution ($30\arcsec$) allow for intercomparison of the source properties derived from those maps. 
The diversity of this sample reflects the very heterogeneous populations uncovered in Galactic plane surveys (i.e. BGPS, although this sample is skewed toward more massive / more luminous sources than what is found in BGPS on average) that will be prime candidates for this kind of spectroscopic follow-up.

We describe the observations in \S2.  
Integrated intensity maps and correlation between dust emission and molecule emission are explored in \S3. 
The physical properties derived from spatial intensity distributions (e.g. size, mass, surface density) are analyzed in \S4.
We summarize our results in \S5.

\begin{figure}[h!]
\epsscale{0.9}
\includegraphics[angle=90, scale=0.45]{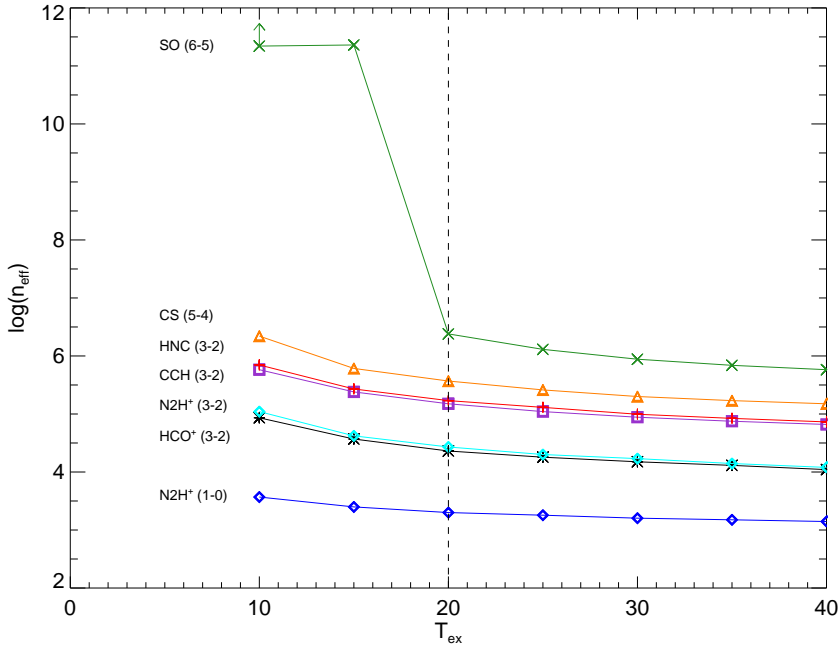}
\figcaption{The effective densities as a function of temperature for transitions of CS, HCN, \hcop , \nthp , SO, HNC, and CCH. Collision rates for CCH are assumed to be those of H$^{13}$CN. \neff\ for SO at T=10K is an upper limit. }\label{f:neff}
\end{figure}

\section{Observations}

\subsection{Nobeyama Observations}

Twenty-seven sources were mapped with BEARS on April 22 through 26, 2004
using the Nobeyama Radio Observatory 45-meter telescope.  
BEARS is a 25 element focal plane
array of SIS mixers (Sunada et al. 2000).  
The beams are spaced at $41\farcs 1$
on the sky with a beamsize of approximately $17\farcs 8$ (see
Tatematsu et al. 2004). To map a source at nearly full 
beam resolution, the array must be dithered in $4 \times 4$ grid 
with a spacing of $20\farcs 55$.
Also, to minimize systematic calibration difference between the 
individual beams, the source was mapped with the array rotated 
by $+90$\degree\ and $-90$\degree.  Each source
was observed in position-switched mode with OFF positions that were 
determined to be clean of emission in CS $J = 5 \rightarrow 4$ 
(Shirley et al. 2003).  The final map covers approximately 
$3$\arcmin\ $\times$ $3$\arcmin\ with 100 spectra.

The BEARS backend is a digital autocorrelator with 1024 channels.  The
spectral resolution is 37.8 kHz resulting in $0.12$ km/s resolution at $93.7$
GHz.  Each spectrum was centered at $v_{\rm{LSR}}$ for the strongest
hyperfine component of \nthpo\ 
($F F_1 = 2 3 \rightarrow 1 2$ at 93.1737767 GHz).
During data reduction, the spectra were smoothed by $4$ channels to improve
the signal-to-noise; the final spectral resolution is $0.49$ km/s.

The spectra were placed onto the \ta\ scale by comparing observations of each
beam in the array with the Nobeyama S100 receiver.  
T$_{mb}$ is calculated by using a main-beam efficiency of the telescope with S100 of $0.515$.
The S100 receiver uses the
standard chopper-wheel calibration method (Penzias \& Burrus 1973).  
The bright source, G40.50, was observed
with each beam in the array on 2 separate days.  
The peak of the \nthpo\ emission
was used to compute the calibration corrections for each beam.  We estimate
that the calibration is consistent with \ta\ determined by S100 to within 
$15$\% .  Weather conditions were good to excellent during the observations.  The
typical wind speed was below $10$ m/s.

The spectra were reduced using the NEWSTAR software package of the
Nobeyama Radio Observatory. 
A linear baseline was removed from each spectrum and the spectra at
the same offset position were averaged.  Since the actual beam
centers on the sky do not form a perfectly regular grid, the spectra were
interpolated onto a regular grid using a Gaussian smoothing function.
The final spectra were written out as a SPFITS file (Greisen \& Harten 1981)
and were read into \textit{AIPS++} where the individual spectral
arrays could be manipulated.  Gaussian fits were performed using
the profile fitter in \textit{AIPS++}.

\subsection{CSO Observations}

Observations of \nthp\ $J = 3 \rightarrow 2$, \hcop\ $J = 3 \rightarrow
2$, and \hcsp\ $J = 6 \rightarrow 5$ of the 27 sources were performed at the 
Caltech Submillimeter Observatory 10.4 m telescope on 10 nights
between June 2002 and July 2004.  The 230 GHz single pixel receiver
(Kooi et al. 1992, 1998) with a 50 MHz AOS backend was used for all observations.
The average spectral resolution was 0.15 km/s; however,
the spectra were smoothed to a velocity resolution of 0.21 km/s (\nthp , \hcop ) 
and 0.14 km/s (\hcsp ).
We made oversampled, boustrophedonic (alternating scan direction between left-to-right and right-to-left lines) On-The-Fly maps of \nthp\ and \hcop\
on a grid with $10$\as\ resolution (Mangum 2000) while \hcsp\
observations were single position-switched spectra toward the 350 \micron\ 
dust continuum peak position.

The standard chopper calibration-wheel method was used to calibrate the
spectra on the \ta\ scale (Penzias \& Burrus 1973).  Calibration on
planets (Mars, Jupiter, and Venus) was used to convert to
the T$_{mb}$ scale assuming that the sidelobes do not contribute
significantly (Kutner \& Ulrich 1981).  
Pointing was performed every hour on planets or
bright high-mass star-forming cores (W28A2(1)) when planets were
not available.  We estimate our pointing is better than $10$\as\
when no planets were available and is better than $5$\as\ with
planets.

\subsection{HHT Observations}





All $27$ sources were mapped using the $1$mm ALMA prototype sideband-separating receiver on the 10-meter Heinrich Hertz Telescope (HHT) on Mt. Graham, Arizona over six nights between October 2008 and June 2009.
We observed in dual polarization 4 IF mode ($5.0$ GHz IF), with HNC J $= 3 \rightarrow 2$ ($271.9811$ GHz) centered in the upper sideband and SO N$_J = 6_7 \rightarrow 5_6$, C$_2$H N$_J = 3_J \rightarrow 2_J$ and CH$_3$OH-E J$_K = 2_1 \rightarrow 1_0$ in the lower sideband. We also observed the isotopologues HN$^{13}$C J $= 3 \rightarrow 2$ ($261.263$ GHz) centered in the upper sideband and $^{34}$SO N $= 6_7 \rightarrow 5_6$ ($256.877$ GHz) also centered in the upper sideband between November 2009 and April 2010. 
The backend filter banks have $256$ channels with 1 MHz width and separation. 
At these frequencies, we have velocity resolution of $\sim 1$ km s$^{-1}$ and $\sim 30^{\prime \prime} $ resolution. 

For each source, we made two arcminute square OTF maps with $10$\as\ spacing. All sources were scanned in RA and DEC and most were also scanned again in RA and DEC with the observational scheme rotated by an arbitrary position angle or offset by $5$\as\ from the original map center. 
Pointing was performed using Jupiter, Saturn or Mars and the main beam efficiency was calibrated with Jupiter or bootstrapped from the intensity of S140 if Jupiter was not available. Pointing accuracy for the HHT is typically $5$\as\ RMS. 

Spectra were reduced using GILDAS CLASS reduction software. 
The ALMA prototype receiver is a dual polarization receiver. We scale the horizontal polarization ($\mathrm{H}_{\mathrm{pol}}$) data to the vertical polarization ($\mathrm{V}_{\mathrm{pol}}$) data using the main beam efficiency determined for each feed. 
After a linear baseline was removed, the spectra were convolved onto a $11 \times 11$ 10\as\ grid using a Gaussian-tapered Bessel function of the form
$ \frac{J_1 ({\frac{r}{a}})} {({\frac{r}{a}})} \mathrm{exp}- (\frac{r}{b})^2 $ where
$a = 1.55 (\frac{\mathrm{\theta_{mb}}}{3})$ and $b = 2.52 (\frac{\mathrm{\theta_{mb}}}{3})$ (Mangum, Emerson \& Greisen 2007). 
A Bessel function is a more accurate representation of the spatial frequency response of a single dish. Tapering with a Gaussian is a compromise to improve computation time, although the spatial frequencies lost in doing so are not significantly different from those probably already tapered by the response of the receiver feed horns (Mangum, Emerson \& Greisen 2007).

\subsection{Dust Continuum Observations}

We also downloaded SCUBA 850 \micron\ archived data (Di Francesco et al. 2008) for the 23 of our sources observed with SCUBA. We display a portion of the image centered on the continuum peak for each available source (see Appendix).

\section{Analysis of Molecular Emission}

\begin{figure}[t!]
\epsscale{0.9}
\includegraphics[angle=90,scale=0.60]{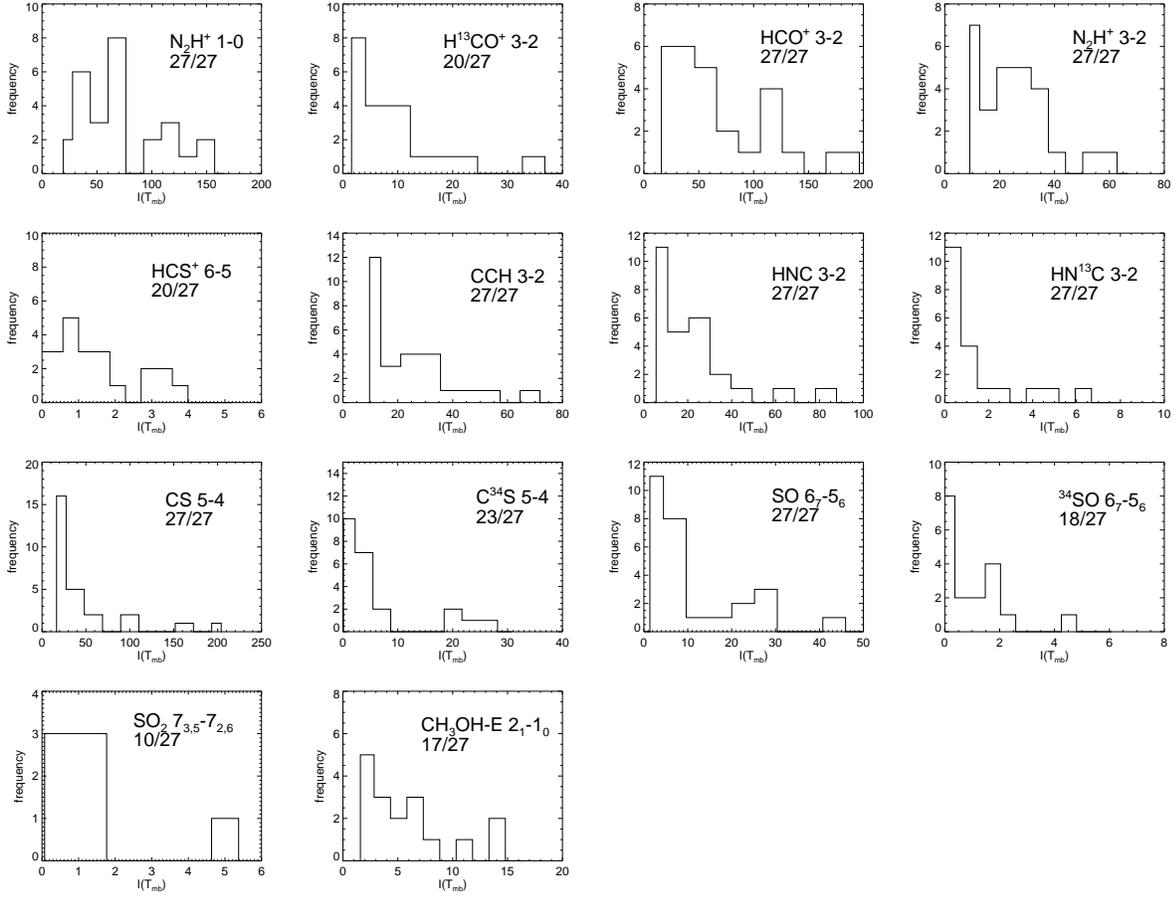}
\figcaption{Integrated intensity histograms and detection statistics for every transition observed. Fractions in upper right-hand corner indicate fraction of sources detected in that transition. 
Integrated intensities are in units of K km s$^{-1}$ on the main beam scale.}\label{f:detect_hist}
\end{figure}

\subsection{Integrated Intensity Maps}\label{ss:maps}
We present maps of the integrated intensity of two wavelengths of dust continuum emission and seven molecule transitions. 
We calculate the integrated intensity from
\begin{equation}
I(T_{mb}) = \int T_{mb} dv \;\; \pm \sqrt{\delta v_{line} \delta v_{chan}} \sigma_{T_{mb}}
\end{equation}
where $\delta v_{line}$ is the full velocity width of the line, the width of the line above $\sigma_{T_{mb}}$, 
and $\delta v_{chan}$ is the channel width of the spectrometer.
Emission maps of integrated intensity, $I(T_{mb})$, are
shown in the Appendix. 
All 27 sources are detected in the mapped transitions of
\hcop, \nthp, HNC, CCH, and SO but not every one is detected in 
the rarer isotopologues (\hcopi, H$^{13}$NC, $^{34}$SO) nor the transitions of \hcsp, CH$_3$OH and SO$_2$. 
The \hcop\ J = $3\rightarrow2$ transition typically has the
highest integrated intensity with a median intensity, $I(T_{mb}) = 61.5$ K km s$^{-1}$. 
Figure 2 shows histograms of the integrated intensity for every transition we observed.

Most transitions trace a single emission peak centered near the
dust continuum peak position. The emission maps of \nthp\ are significant exceptions. 
For 10 of our sources (W3(OH), RCW142, M8E, G8.67, W43S, W44, ON2S, W75(OH), S140, and CepA)
we find that \nthp\ emission shows significant spatial differentiation that is not reflected in the dust emission or emission from other molecules. 
These 10 sources not only have multiple intensity peaks in the \nthp\ emission, but these peaks are also spatially separated from the peak of the dust and other molecule emssion. In general, the dust and other molecules peak near the water maser position while the \nthp\ emission in the differentiated sources peaks several arcseconds ($\sim 15$\arcsec) away from the water maser position --- more than can be accounted for with pointing errors alone. 
Both \nthp\ J = $3 \rightarrow 2$ and J = $1 \rightarrow 0$ show this differentiation, indicating that this is a spatial variation in the abundance of \nthp\ 
and not a radiative transfer effect.

The spatial variation in the \nthp\ emission may be understood from the dependence of \nthp\ chemistry on temperature. 
\nthp\ chemistry and \hcop\ chemistry are directly related by the gas phase abundance of CO. 
\hcop\ is created by CO while \nthp\ is more abundant when CO is frozen out onto dust grains (see J{\o}rgensen et al. 2004). 
Each source with spatial variations in the \nthp\ emission also has centrally peaked \hcop\ emission, supporting the conclusion that these sources are chemically differentiated. 
Warmer temperatures ($T > 20$ K) are required to return CO to the gas phase, suggesting that \nthp\ is preferentially differentiated toward warmer, more evolved sources.

To test this hypothesis, we looked at several physical properties that might indicate the presence of at least one evolved protostar embedded in the clump. 
The isothermal dust temperature, \tiso, defined as the temperature required to match the mass of the best-fit dust continuum radiative transfer model (Mueller et al. 2002),
\begin{equation}\label{e:Tiso}
T_{iso} = 41 (K) \left[ ln \left( \frac{M_{iso}}{5.09 \times 10^{-8} M_{\sun} S_{\nu} (Jy) D^2 (pc)} + 1 \right) \right] ^{-1} \;\;\;,
\end{equation} 
should be higher for warmer, more evolved clumps. The spread in \tiso\ is within a few K of the median \tiso\ ($22$ K), reflecting the unresolved temperature gradients found for these sources by Mueller et al. (2002; see Table 2). 
Warmer clumps will also be more luminous, although more massive clumps will have larger luminosities even at lower temperatures. We make a simple correction for clump mass by taking the ratio of clump luminosity to the dust-determined isothermal mass, \miso, defined as 
\begin{equation}
M_{iso} = \frac{S_{\nu} D^2}{B_{\nu} \kappa_{\nu}} = 5.09 \times 10^{-8} M_{\sun} S_{\nu}(Jy) D^2(pc)(e^{41K/T_{iso}} - 1)
\end{equation}
where $S_{\nu}$ is the source dust flux density in a $120$\arcsec\ aperture (Mueller et al. 2002). The median \lbol/\miso\ value for the differentiated sources is somewhat larger than for the remaining sample ($110 \pm 58 \frac{L_{\sun}}{M_{\sun}}$ compared with $50 \pm 23 \frac{L_{\sun}}{M_{\sun}}$). The difference suggests that the differentiated clumps are warmer on average, although this conclusion is not robust.

If the hypothesis that \nthp\ is differentiated with respect to the dust in more evolved clumps is correct, then the ionizing radiation we expect from more evolved protostars should lead to radio continuum (HII) emission from the clump, provided there is a protostar of early enough type ($>B0$). 
Using the Shirley et al. (2003) clump classifications (see Table 2), we find that eight of the strongly differentiated sources are associated with HII regions, indicating the presence of a more evolved massive protostar (see table 2). 
Care must be taken in interpreting the evolutionary state of the clump as a whole as recent interferometric work has shown that many of these regions have multiple cores in different evolutionary states that are unresolved in the single dish beam (e.g. Hunter et al. 2008; Brogan et al. 2009).

Strong outflows in sources with \nthp\ differentiation may shock heat gas in the clump, leading to increased \nthp\ destruction. Most observations of massive star forming regions, including those studied here, are observed with resolution insufficient to determine the outflow structure in the clump. It is possible that a single protostellar object powers some outflows, though others may reflect the combined effect of outflows from multiple protostars in the clump (Bally 2008). Interferometric follow-up is required to directly test the effect of outflows on \nthp\ chemistry.

The majority of our sources show \nthp\ peak emission that is spatially coincident with the dust peak at $30$\arcsec\ resolution, indicating that emission within the central beam is dominated by cold ($T < 25$ K), dense ($n > 10^4$ cm$^{-3}$) gas. 
However, this does not mean that \nthp\ is not chemically differentiated in these sources. Unresolved chemical differentiation may be reflected in the poor correlation between \nthp\ emission and emission from the dust and other molecules. We explore these correlations in \S3.2 and \S3.3.

Spatial differentiation in \nthp\ emission has been
observed toward other high-mass clumps. 
Pirogov et al. (2003, 2007) studied 35 dense molecular cloud cores known to be forming massive stars or clusters of stars and found \nthp\ differentiation in $\sim \frac{1}{3}$ of their sources. 
This differentiation is not what we would expect based on studies of \nthp\ emission in less evolved, low mass star forming regions, where \nthp\ emission is well correlated with the dust emission 
(Lee, Myers \& Tafalla 2001, Crapsi et al. 2005, Emprechtinger et al. 2009). 
For a single core, \nthp\ differentiation indicates a warmer source that is likely in a more advanced evolutionary state. 
However, as Pirogov et al. point out, 
in high mass star forming regions, emission from cores in several different evolutionary states overlap in the single dish beam, 
leading to a more complicated emission profile. 
In addition, more massive stars evolve faster, so emission from the hot region surrounding a massive YSO may drive the observed differentiation while we simply cannot spatially resolve the colder, less evolved regions where \nthp\ is still well coupled to the dust. 
Dynamical activity in high mass clumps may generate additional pressure on an individual collapsing core causing the collapse to proceed faster than free fall, thus producing high densities before the molecules responsible for \nthp\ destruction (e.g. CO) are frozen out (Lintott et al. 2005). Lintott et al. (2005) suggest that preventing freeze out facilitates \nthp\ destruction near massive young stellar objects and 
enhances emission from other species that freeze out in more quiescent low mass cores (Myers \& Benson 1983; Evans et al. 2001).

Most of our sources do not show large-scale spatial differentiation in any transitions other than the two \nthp\ transitions, with the notable exception of ON2S. 
The 850 \micron\ continuum emission for ON2S shows two peaks in the larger, more southern source. 
The northeast peak is brighter than the southwest peak in both 350 \micron\ and 850 \micron\ emission indicating that the northwest source is warmer.
\hcop, CS, and HNC emission appears to trace the more northern of the two peaks in the 850 \micron\ continuum, while the position of the \nthp, SO and CCH emission is more consistent with the position of the more southern peak (see Figure in the appendix). 
A colder southern source with strong \nthp\ emission is consistent with the predictions for \nthp\ chemical differentiation, however the chemical differentiation in SO and CCH is not easily explained. 
Higher spatial resolution interferometric observations are required to resolve the individual northeastern and sourthern sources. In a future work, we will use the SMA to study the temperature and the chemistry of these sources individually in order to better understand what may be causing the unusual SO spatial differentiation in this source (Reiter et al., in prep). 

Two of the mapped molecules in this survey, CCH and SO,
are molecules with unpaired electrons and therefore sensitive to the Zeeman effect and potential tracers of the magnetic fields (e.g. Bel \& Leroy 1989, 1998).
Zeeman splitting is too small to be observed with our spectral resolution for the transitions considered. 
The similarity of the CCH and SO emission to the dust continuum emission in this survey indicates that observations of these Zeeman tracers at higher resolution should typically be centered within $15$\as\ of the dust continuum position.
Ultimately, observing magnetic fields in these clumps via Zeeman splitting in lower frequency transitions of these molecules will be best be done with the sensitivity and high spatial resolution available with ALMA (e.g. 3mm band for CCH, 9mm band for SO).

\subsubsection{Molecular Tracers with Non-Detections}

\hcsp, \methanol\ and \soo, as well as the isotopologues, are detected in some, but not all, of the the sources in our sample. In this section we examine the detection statistics and chemistry of \hcsp, \methanol\ and \soo.

In the lower sideband of the HNC setup, we have serendipitously detected the \methanol-E J$_K = 2_1 \rightarrow 1_0$ transition toward 17 clumps (63\%). 
The typical gas phase abundance of \methanol\ observed toward high-mass cores cannot be explained by purely gas phase reactions (Leurini et al. 2007). 
\methanol\ is formed in the solid state via ultraviolet irradiation of ices and is released into the gas phase in regions of the core envelope where temperatures are above the gas sublimation temperature of H$_2$O (T $> 90$ K; Charnley et al. 2004). 
As a result, \methanol\ is an excellent tracer of the hot core phase, and the warm emitting regions that are very likely unresolved within our $30^{\prime\prime}$ beam. 
Sources with \methanol-E J$_K = 2_1 \rightarrow 1_0$ detections have a slightly higher median \lbol\ ($5.7 \pm 4.3 \times 10^4$ \lsun) than the median \lbol\ for the
remaining sources ($3.6 \pm 3.0 \times 10^4$ \lsun), suggesting that a hot core is embedded in the clump. 
With higher spatial resolution, we might expect a better correlation between the intensity of \methanol\ and \lbol. 
Eleven of the 17 sources with a \methanol\ detection are associated with HII emission, although the fraction of the detections with an associated HII region is comparable to the fraction of the total sample with associated HII emission (roughly two thirds in both cases). 
This is not surprising given that a protostar with spectral type $\ge$ B0 is required to produce enough ionizing radiation to create an observable HII region while a wider range of protostar masses will pass through the hot core phase.

We detect \hcsp\ J = $6\rightarrow5$ (in the lower sideband for the H$^{13}$CN J = $3\rightarrow2$ setup at the CSO; Wu \& Evans 2003) towards 20 of our 27 sources (74\%). 
The abundance of \hcsp\ relative to CS is found to be high in cold, dense dark clouds, suggesting that the dissociative recombination of \hcsp\ to form CS is a slow process, leading to high ratios even in hot ($T \ge 300$ K) environments (Irvine, Good \& Schloerb 1983). 
The 7 sources with non-detections also have smaller CS integrated intensities than those with a detection (median $I(T_{mb})=20.2 \pm 2.0$ K km s$^{-1}$ for the non-detections compared to a median $I(T_{mb})=44.3 \pm 15.2$ K km s$^{-1}$ for sources with an \hcsp\ detection). 
The two other sulfur-bearing molecules we observe also have higher intensities for the sources with an \hcsp\ detection (SO median $I(T_{mb})=9.8 \pm 4.7 $ K km s$^{-1}$ for \hcsp\ detections, compared with $I(T_{mb})=6.6 \pm 0.7$ K km s$^{-1}$ for \hcsp\ non-detections; \soo\ median $I(T_{mb})=1.4 \pm 0.3$ K km s$^{-1}$ for \hcsp\ detections, compared to $I(T_{mb})=1.1$ K km s$^{-1}$ for CepA, the only \soo\ detection without an \hcsp\ detection).
The median \lbol\ for sources with an \hcsp\ detection is $9.5 \pm 7.6 \times 10^4$ while it is $2.2 \pm 0.94 \times 10^4$ for the remaining sources. 
Sources where we do not detect \hcsp\ are associated with weak or no radio continuum (2 of 7 non-detections) or UCHII emission (4 of 7 non-detections; the remaining source is associated with CHII emission). 
\hcsp\ non-detections are preferentially observed towards the earlier evolutionary states.

Our spectral configuration for observing $^{34}$SO included SO$_2$ in the bandpass. We detect SO$_2$ towards 10 of our sources (37\%). 
SO$_2$ is expected in more chemically evolved clumps, as it is created from multiple gas phase reactions of OH with sulfur-bearing molecules (e.g. CS) that are released from grains as protostars heat the envelope (see \S3.2 for a discussion of SO and SO$_2$ formation). 
Sources with a SO$_2$ detection tend to have higher \lbol\ --- half of the sources with a detection have an L$_{bol}$ that is nearly a factor of three larger than the median \lbol\ for sources with no SO$_2$ detection (median \lbol\ for SO$_2$ detections is $1.0 \pm 0.78 \times 10^5$ \lsun\ compared to $0.36 \pm 0.23 \times 10^5$ \lsun\ for the remaining sources). 
Six of the 10 sources with an SO$_2$ detection are associated with HII emission, again a fraction that is consistent with that observed in the entire sample.

High resolution observations of high-mass star forming regions reveal many cores in different evolutionary states within a single clump (e.g. Hunter et al. 2008; Brogan et al. 2009), so it may be that a few young cores are responsible for all hot core molecule emission. 
Many of the sources with a \methanol\ or a SO$_2$ detection are also among the nearer sources and more distant sources with a detection are among the more luminous in our sample. This suggests that these transitions are excited in all of the clumps in our sample, but in many cases beam dilution renders the transition unobservable in at 30\arcsec beam.

\subsection{Integrated Intensity Correlations}\label{ss:int_corr}

\begin{figure}[h!]
\epsscale{0.9}
\includegraphics[angle=0,scale=0.90]{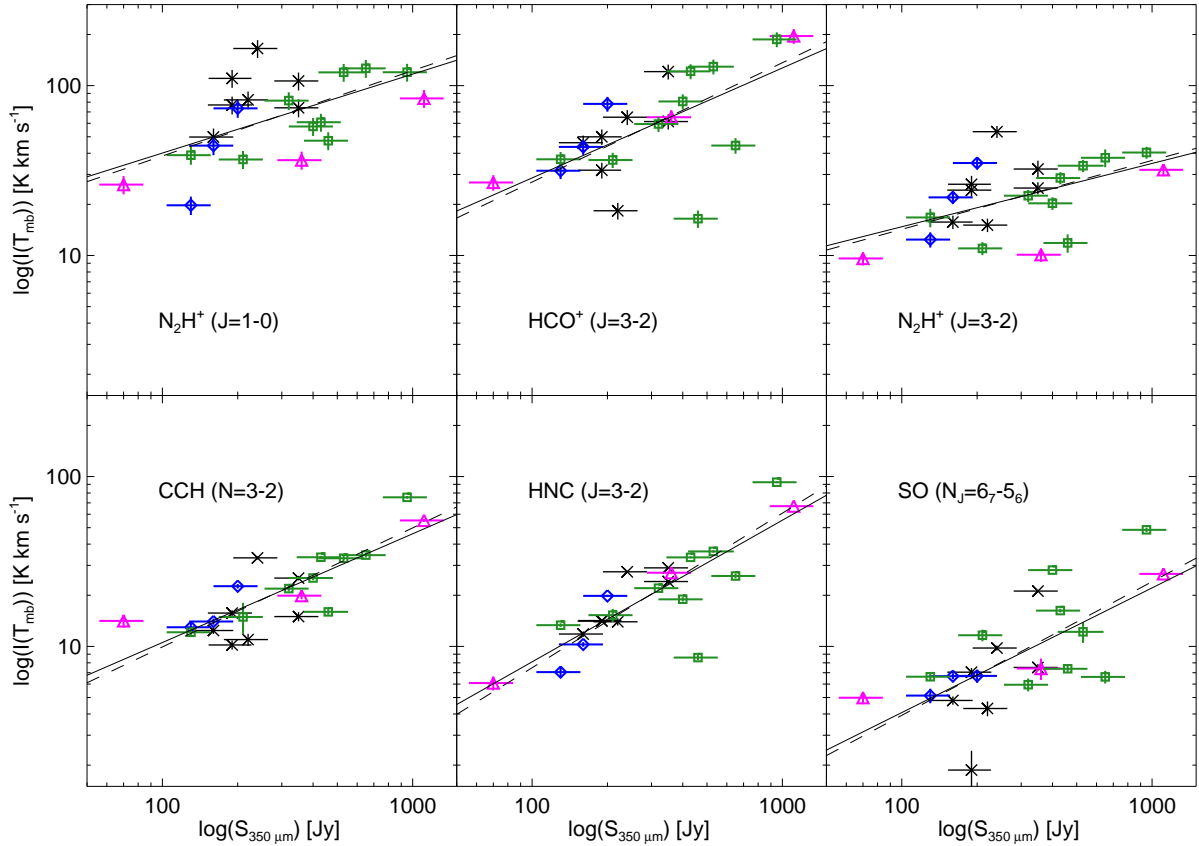}
\figcaption{Correlation between the beam-averaged $350 \micron$ dust intensity and the integrated intensity of the molecules. The Spearman rank correlation coefficients are listed in table 4. The blue diamonds represent HII regions, magenta triangles represent compact HII regions, green squares represent UCHII regions. Sources plotted with black crosses have weak or no radio continuum. In some cases, error bars are smaller than the plotting symbol.}\label{f:pp_corr}
\end{figure}

We quantify how well the molecules trace the dust (and each other) by calculating the correlation between peak molecular emission and the dust emission at both $350$ \micron\ and $850$ \micron\ (correlations with $350$ \micron\ dust emission are plotted in Figure 3). 
We use the Bayesian IDL routine LINMIX\_ERR (Kelly et al. 2008) to perform a linear regression to find the slope of the best fit line. 
The relationship between dust emission and molecule emission may not be strictly linear, however, given the scatter in the correlation plots (see Figure 3), we are not justified in fitting a more complex function. 
We also calculate the Spearman Rank coefficient for every combination of dust and molecule intensity. The Spearman rank analysis determines the degree of correlation between two distributions nonparametrically, and is therefore less sensitive to outliers in the data. Larger values of the correlation coefficient ($r_{corr} \geq 0.7$) indicate a more robust correlation.

Correlations of each molecule with the 350 \micron\ and the 850 \micron\ dust emission are in excellent agreement (see columns 1 and 2 respectively of Table 4), which is encouraging given that the dust observations have different resolutions ($15^{\prime\prime}$ versus $10^{\prime\prime}$), and were taken with different telescopes using different observing techniques (jiggle mapping versus scanning).
High \neff\ tracers CCH, HNC, CS and SO are strongly correlated ($r_{corr} \geq 0.7$) with both $350$ \micron\ and $850$ \micron\ dust emission. 
In general, transitions with higher \neff\ do a good job of tracing the dust continuum flux density. 

In contrast, both transitions of \nthp\ show the poorest correlation with the dust emission ($r_{corr} \le 0.5$), in agreement with the Zinchenko, Caselli \& Pirogov (2009) result that \nthp\ does not trace the dust emission in high-mass star forming regions. As discussed in the previous section, this is likely due to the destruction of \nthp\ in the warm, CO-rich gas in the center of these sources, unresolved within the $30\arcsec$ beam. This divergent \nthp\ chemical behavior is more apparent when compared to the other molecules.

\begin{figure}[h!]
\epsscale{0.9}
\includegraphics[angle=90,scale=0.60]{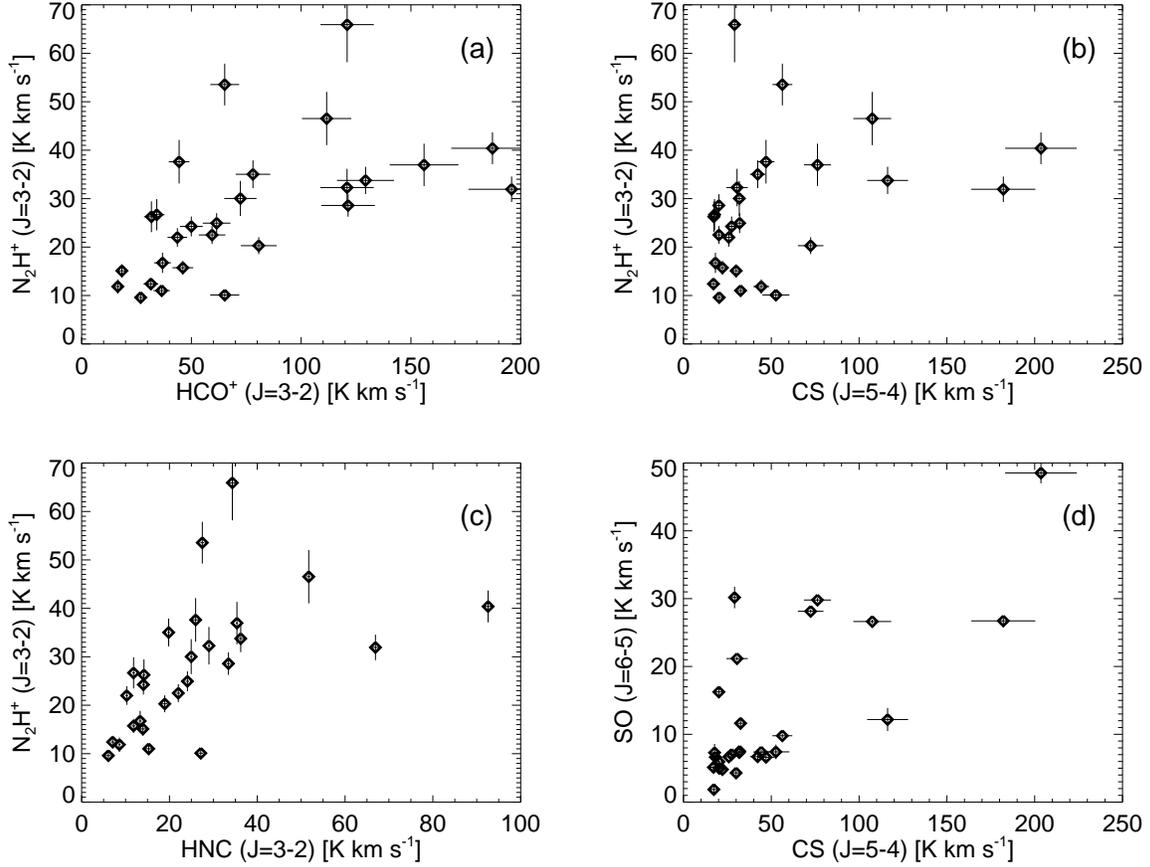} 
\figcaption{Integrated intensity correlations 
(a) Chemically opposite species \hcop\ and \nthp\ are well correlated ($r_{corr}=0.68$) inspite of observed spatial differentiation in some sources. 
(b) Molecules found to trace the dust for low mass clumps (\nthp) and high mass clumps (CS). 
(c) Nitrogen-bearing molecules that Zinchenko, Caselli \& Pirogov (2009) find to be differentiated. We observe \nthp\ differentiation, but our HNC observations do not trace the same peaks as \nthp.
(d) The two sulfur-bearing molecules mapped in this study ($r_{corr}=0.67$). In some cases, the error bars are smaller than the plotting symbol.}
\end{figure}

We also intercompare the integrated intensities for each molecule transition (columns 4 through 10 in Table 4). 
While a detailed understanding of these molecule-molecule correlations requires multi-dimensional, clumpy radiative transfer modeling with realistic (and probably complicated) molecular abundance profiles, several striking trends are apparent among the correlations.
Molecular tracers with \neff\ $> 10^5$ cm$^{-3}$ (CCH, HNC, CS, and SO) have integrated intensities that are generally well correlated ($r_{corr} \ge 0.7$) with each other. 
CCH and HNC (\neff$=1.5\times10^5$ cm$^{-3}$ and \neff$=1.7\times10^5$ cm$^{-3}$ respectively) are strongly correlated with every transition. 
The two densest gas tracers, CS and SO, are poorly correlated with \nthp, again supporting the hypothesis that \nthp\ is destroyed in the warm gas in the center of these clumps (see \S3.1).

CS and SO have somewhat weaker correlation coefficients overall compared to CCH and HNC even though CS and SO are the highest \neff\ tracers in this survey. This may be due to sulfur's complicated chemistry, which, in part, remains poorly understood due to the difficulty observing S$^+$ or SII. 
Beuther et al. (2009) studied the chemistry of sulfur in molecules in their study of high mass protostellar chemical evolution. As discussed in that work, CS is released from grains early (CS desorbs from grains at temperatures of $\sim30$ K), then reacts with OH to form SO and eventually SO$_2$. 
It is possible that shock heating from strong outflows may also drive up gas temperatures, facilitating this chemical pathway. 
With higher spatial resolution, such as that used in Beuther et al. (2009), we may be able to use the relative abundances of CS, SO and SO$_2$ to determine the chemical evolutionary state of the individual cores in our sources. Without resolving individual cores, we can say that this kind of chemical evolution likely contributes to the weaker correlations with S-bearing molecules. 
CS and SO integrated intensities are only modestly correlated ($r_{corr}=0.67$) as we would expect for unresolved chemical differentiation due to the destruction of CS in the production of SO.

Tracers with \neff\ $\approx 10^4$ cm$^{-3}$ (\hcop\ and \nthp\ J = $3 \rightarrow 2$) also show relatively weak correlations with CS and SO (except for \hcop\ with SO), though they are well correlated with the other transitions. 
Given these poor correlations and the fact that \nthp\ is the only molecule to show substantial differentiation at $30$\arcsec\ resolution, we conclude that the \nthp\ abundance structure
within the central telescope beam is significantly different than for the other molecule species. 
These poor correlations suggest that \nthp\ is chemically differentiated with respect to the other molecules in many of our sources, but this differentiation is spatially unresolved for the majority of the sources in this survey.

\subsection{Column Density Correlations}\label{ss:ncol_corr}

\begin{figure}[h!]
\epsscale{0.9}
\includegraphics[angle=90,scale=0.40]{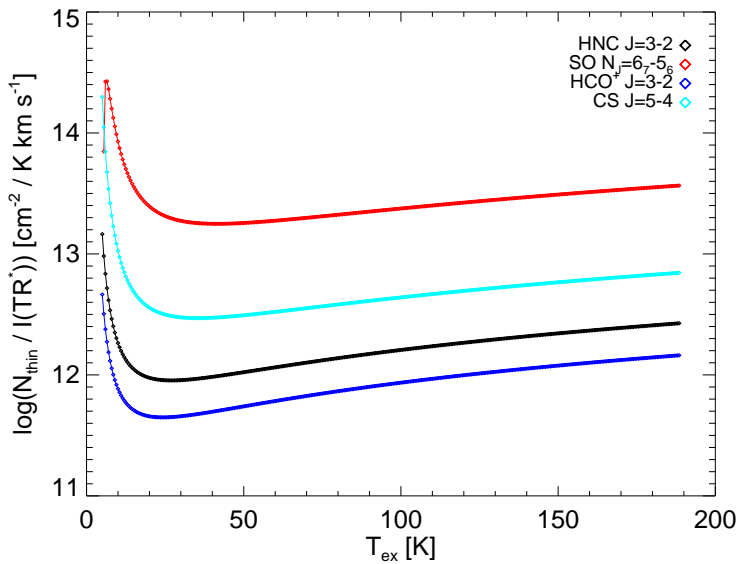}
\figcaption{Temperature sensitivity of the optically thin column density, assuming local thermodynamic equilibrium.}\label{fig:n_tex}
\end{figure}

To compare the behavior of the molecules, we calculate the column density. 
Molecules with similar \neff\ will be excited in the same gas, so we expect that as one transition gets brighter, so should other transitions with similar \neff, in the absence of strong abundance gradients. Poor correlations may indicate chemical differentiation or the development of substructure (cores within the clump), even if the scales on which these things are occurring are too small to be resolved with the single dish beam.

\begin{figure}[t!]
\epsscale{0.9}
\includegraphics[angle=0,scale=0.90]{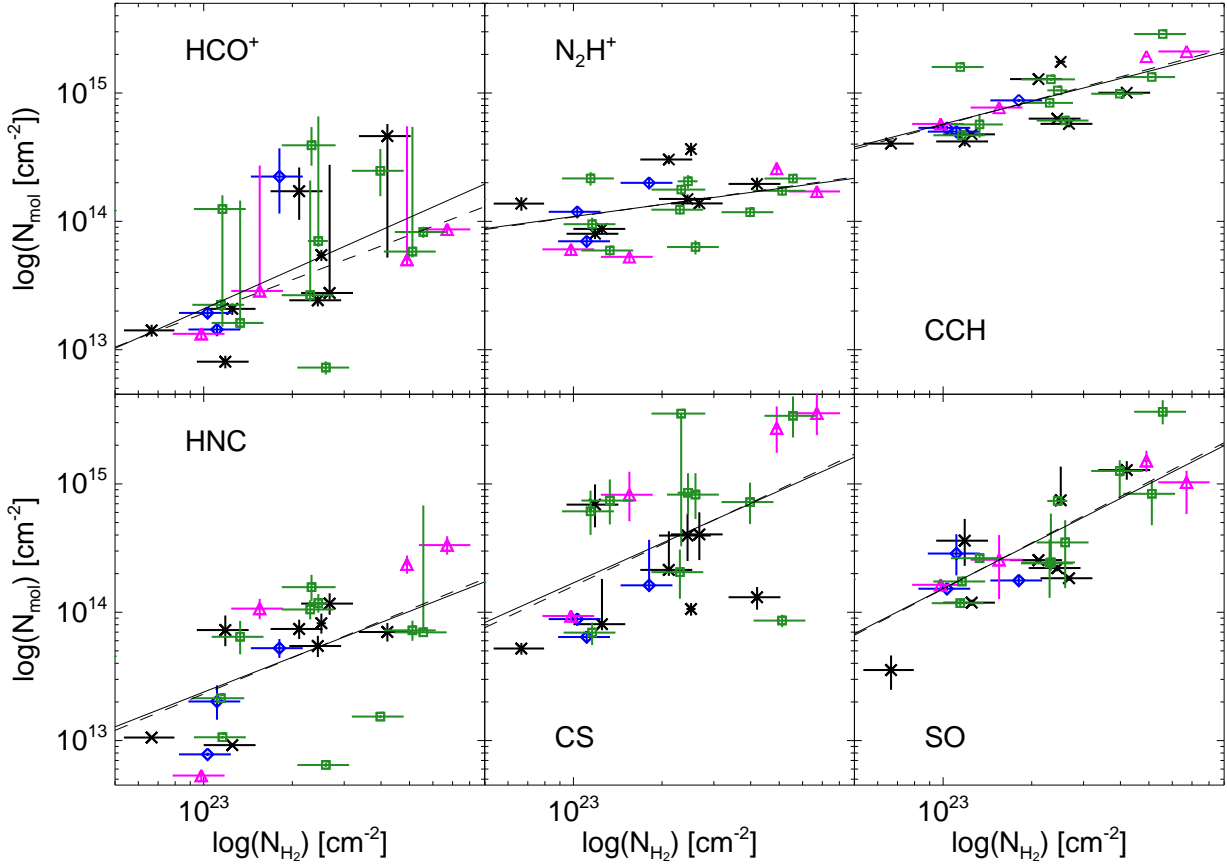}
\figcaption{Molecule column densities plotted versus H$_2$ column density.
The Spearman rank correlation coefficients are listed in Table 6. The blue diamonds represent HII regions, magenta triangles represent compact HII regions, green squares represent UCHII regions. Sources plotted with black crosses have weak or no radio continuum.}\label{fig:ncol_350_corr}
\end{figure}

We calculate the column density from
\begin{equation}
N_{ul} = \frac{4\pi Q(T_{ex}) e^{E_u/kT_{ex}} B_{\nu}(T_{ex})}{hcg_u A_{ul} [J_{\nu}(T_{ex}) - J_{\nu}(T_{cmb})] } I(T_{mb}) \frac{\tau}{1-e^{-\tau}}  \;\; \rm{cm}^{-2}
\end{equation}
where $g_u$ is the statistical weight for an energy level $E_u$, 
$A_{ul}$ is the transition Einstein A, 
$B_{\nu}$ is the Planck function, $J_{\nu}(T) = (h\nu/k) / [\exp(h\nu/kT) - 1]$ is
the Planck function in temperature units (K), 
$Q$ is the partition function for the molecule in LTE with excitation temperature $T_{ex}$, and 
$\tau / (1 - e^{-\tau})$ is the correction for non-zero optical depth. 
We approximate the average excitation temperature along the line-of-sight
with the isothermal dust temperature \tiso\ (see \S3.1). 
\tiso\ averages over significant temperature gradients in the clump, and may not represent the temperature of the gas where a specific transition is excited. 
However, for the range of \tiso\ for these clumps, the column density is not very sensitive to the differences in temperature (see Figure 4), changing by at most a factor of two.

\begin{figure}[t!]
\epsscale{0.9}
\includegraphics[angle=90,scale=0.60]{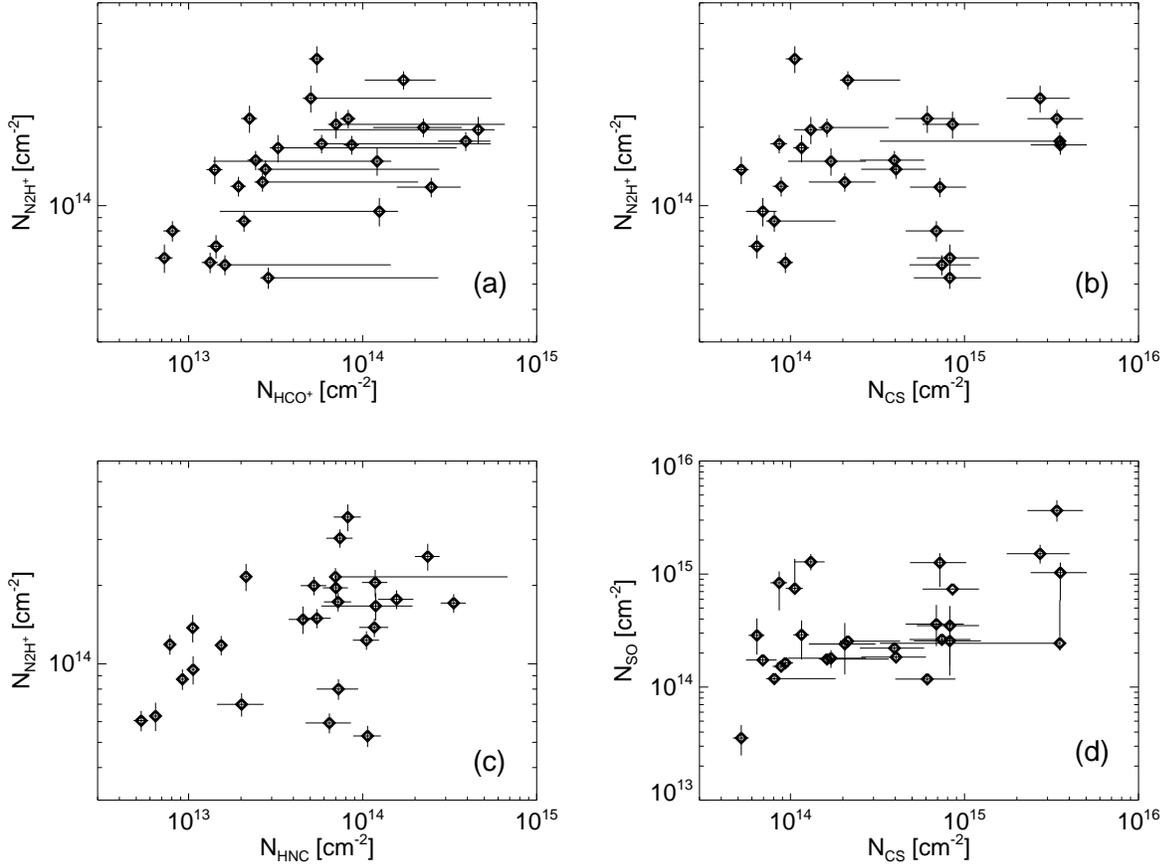} 
\figcaption{Column density correlations for the same molecules examined in Figure 4. Note the worsening of the correlation especially for molecules where an optical depth correction has been applied.  
(a) Chemically opposite species \hcop\ and \nthp. 
(b) Good dust tracers for low mass (\nthp) and high mass clumps (CS). 
(c) Nitrogen-bearing molecules \nthp\ and HNC. 
(d) The two sulfur-bearing molecules mapped in this study. The modest correlation of the column densities ($r_{corr}=0.61$) supports the Beuther et al. (2009) suggestion that SO is formed from CS.}
\end{figure}

We attempt to correct for optical depth, $\tau$, for those molecules for which we have observed a rarer isotopologue. 
With a value for $\tau$, we can correct the column density for optical depth 
by multiplying the optically thin column density with a correction factor, $\tau / (1 - e^{-\tau})$ (Goldsmith \& Langer 1999).  
If we assume that the isotopologue (\hcopi\ for example) is optically thin, then we can use the ratio of the peak temperature of the rarer isotopologue (\hcopi) to the peak temperature of the main isotopologue (\hcop) to solve for the optical depth. 
To do so, we assume that the molecular abundance of \hcopi\ relative to \hcop\ is equal to the Interstellar Medium (ISM) abundance of atomic $^{13}$C relative to $^{12}$C ([$^{13}$C]/[$^{12}$C]$= \frac{1}{77}$; Wilson \& Rood 1994). 
With this assumption and ignoring background radiation, we can calculate the optical depth $\tau$ as 
\begin{equation}
\frac{[T_{\mathrm{H^{13}CO^+}}]}{[T_{\mathrm{HCO^+}}]} 
= \frac{1-e^{-\tau_{\mathrm{H^{13}CO^+}}}}{1 - e^{-\tau_{\mathrm{HCO^+}}}}
= \frac{1 - e^{-X \tau}}{1 - e^{-\tau}}
\end{equation}
where $X$ is the ratio of [$^{13}$C]/[$^{12}$C], $\tau$ is the optical depth of the \hcop\ line, and \hcopi\ is assumed to be thin. 
For HNC, where both isotopologues show linewidths that are approximately Gaussian, we find a median $\tau=3.26 \pm 0.72$.
We know that \hcop\ J = $3\rightarrow2$ must be optically thick in at least some of our sources because we see self-absorption in the line profiles. Measuring the temperature of a self-absorbed line is difficult, and leads to considerable uncertainty in the temperature of the main isotopologue. 
ISM abundance ratios for [$^{13}$C]/[$^{12}$C] and [$^{34}$S]/[$^{32}$S] are also uncertain to at least 10\%, further increasing our $\tau$ uncertainty. 
Not surprisingly, our highly uncertain optical depth correction dramatically increases the scatter in our column densities, and thus decreases the correlation in many cases. We have calculated correlation coefficients for both the optically thin and the optically thick column density and report both in Tables 5 and 6 respectively.

We calculate total H$_2$ column density from the $350$ \micron\ dust emission 
or the $850$ \micron\ dust emission for sources without $350$ \micron\ data 
from 
\begin{equation}
N_{\mathrm{H}_2} = \frac{S_{\nu}^{obs}}{B_{\nu}(T_d) \kappa_{\nu} \mu m_H \Omega_{beam}}
\end{equation}
where $S_{\nu}^{obs}$ is the observed flux in a $30$\arcsec\ aperture and $\kappa_{\nu}=0.10$ cm$^2$ g$^{-1}$ is the opacity of gas and dust at $350$ \micron, and $\kappa_{\nu}=0.0197$ cm$^2$ g$^{-1}$ at $850$ \micron\ (Ossenkopf \& Henning 1994) 
We use \tiso\ as the dust temperature and assume a mean molecular weight of $\mu=2.34$ as is appropriate for dense molecular gas. 
We plot the correlation between the molecular column densities and H$_2$ column density in Figure 5.

Correlations between the H$_2$ column density and the optically thick molecule column densities follow the same general trends as correlations of the molecule integrated intensities with the dust emission, although this is expected since $N_{ul} \sim I(T_{mb})$ in the optically thin limit. 
Indeed, the most pronounced differences are for correlations of optical-depth corrected column densities with the H$_2$ column density. These are made notably worse by the optical depth correction, dropping from strongly correlated ($r_{corr} = 0.78$ for the integrated intensity correlation) to relatively poorly correlated ($r_{corr} = 0.54$ for the column densities) in the case of HNC.

\section{Discussion}
With a wealth of galactic plane surveys targeting hitherto unexplored regions of the electromagnetic spectrum coming out in the next few years (e.g. BGPS, ATLASGAL, HIGAL, etc.), it is important to understand how best to follow up the newly discovered sources in these surveys. 
Molecular line surveys of sources discovered in millimeter and submillimeter dust emission allow us to study the properties of the dense gas in these sources. 
As discussed in \S3.1, dense gas tracers such as \nthp\ that are well correlated with the dust for low mass clumps are not necessarily good tracers of the dust for high mass sources. 
Proper assessment of the physical properties of those sources requires an understanding of how the choice of molecular tracer affects the derived source properties. 
With 7 mapped transitions probing more than 3 orders of magnitude in \neff, we may begin to study how \neff\ of the tracer affects the derived physical properties of the clump.

\subsection{Size, Mass and Densities}\label{ss:size}
Our suite of transitions represent
a range of \neff, allowing us to examine how the derived clump properties change as a function of the gas a specific transition is tracing. 
In this section, we examine how size $R$, linewidth $\Delta v$, virial mass \mvir, mass surface density $\Sigma$ and average density $\mean{n}$ vary with \neff. 

Our sample of high mass star forming regions represents a range of masses, chemistries and evolutionary states. These source-to-source variations are evident in the scatter in the relationships between molecule emission and dust emission (see \S3.2 and \S3.3 and Figures $3-7$). These variations also lead to a broad range of values for the physical properties derived from this sample (though they tend to be positively skewed). 
We therefore take the median value of each physical property and examine its behavior as a function of \neff\ and characterize the spread of values around that median with the median absolute deviation. Because we analyze trends observed in a population of 27 sources, we emphasize that these trends are primarily qualitative as the uncertainties on the median values are large. 
However, we note that trends in the median value of a given physical property that persist despite the substantial scatter in the values for individual sources are remarkable given the diversity of our sources and relatively small sample size.

\begin{figure}[t!]
\epsscale{0.9}
\includegraphics[angle=90,scale=0.65]{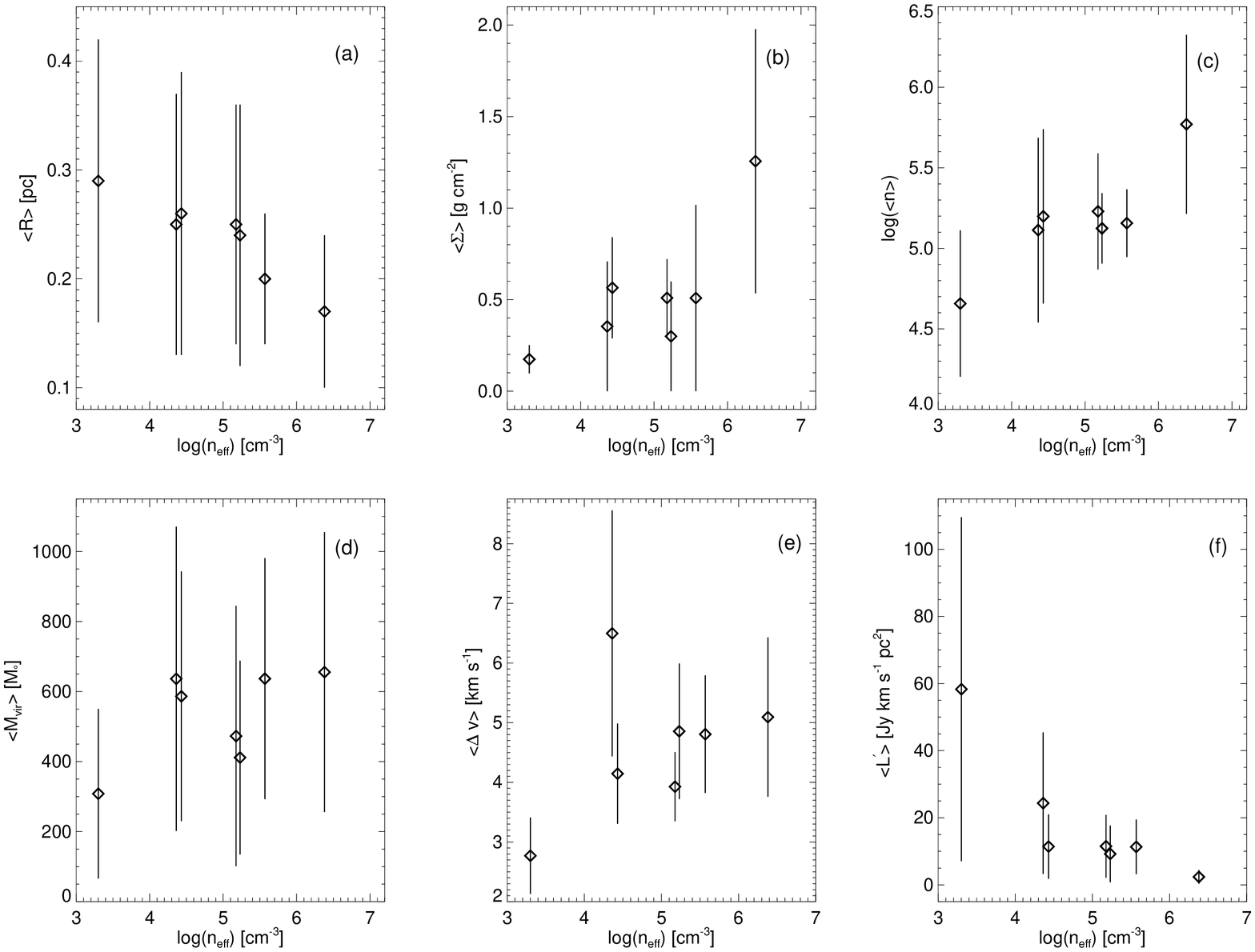}
\figcaption{As a function of \neff\ (at 20 K), we show 
(a) median clump size 
(b) median $\Sigma$ 
(c) median $\mean{n}$ 
(d) median \mvir\ 
(e) median $\Delta v$ and 
(f) median L$^{\prime}$ for each tracer.
Median clump size and L$^{\prime}$ appear to be anti-correlated with \neff\ while $\Sigma$ and $\mean{n}$ show a positive correlation. \mvir\ and $\Delta v$ do not appear to be correlated with \neff. 
\nthp\ J$= 1 \rightarrow 0$ observations have been convolved to $30\arcsec$ resolution to match the resolution of the other transitions.}\label{fig:RSMn_neff}
\end{figure}

\begin{figure}[h!]
\epsscale{0.9}
\includegraphics[angle=90,scale=0.60]{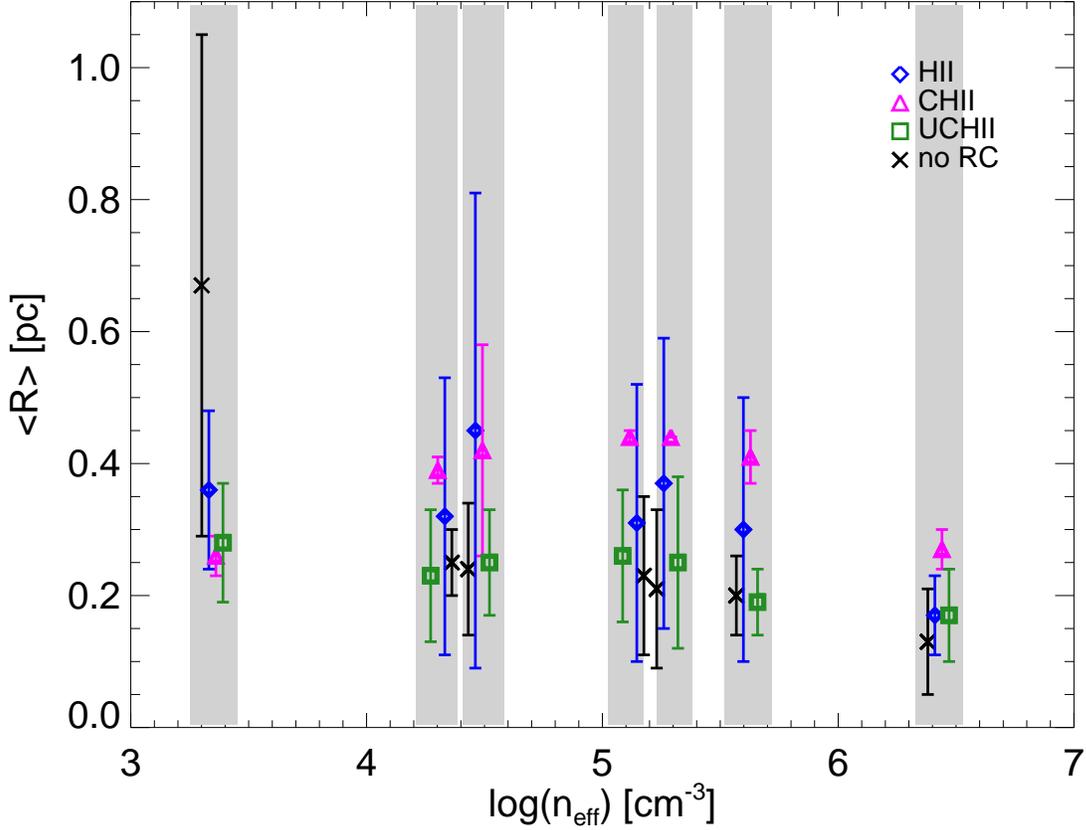}
\figcaption{Median clump size for each evolutionary state as a function of tracer \neff\ (at 20 K). Points for each evolutionary state are plotted near the tracer \neff\ and gray boxes indicate points belonging to a single tracer. Younger evolutionary states (weak or no radio continuum --- ``no RC'', UCHII) are denoted with black crosses and green squares, respectively. More evolved stages (CHII and HII) are denoted with magenta triangles and blue diamonds respectively. More evolved clumps appear to be $\sim 0.1$ pc larger on average.}\label{fig:RSMn_neff_by_type}
\end{figure}

\subsubsection{Size}\label{sss:size}
Clump size, as determined from a molecular intensity map, depends on the integrated intensity threshold chosen. In this paper, we use the half-peak intensity contour though other definitions that depend on a fixed intensity level are possible.
Comparisons between the different definitions generally give the same qualitative conclusions (see Shirley et al. 2003).
We find the deconvolved angular radius of a circle with the same area as is enclosed within the half-peak intensity contour,  
\begin{equation}
\theta = \sqrt{ \frac{A}{\pi} - \frac{\theta^2_{\mathrm{beam}}} {4}}
\end{equation}
where A is the area enclosed in the contour in arcsec$^2$ and $\theta_{\mathrm{beam}}$ is the beam size in arcsec. The angular radius can be converted to a linear radius using the distance, $R = \theta D$.

The observed clump size also depends on the underlying density structure of the clump. 
The large-scale structure of the clumps
can be described, on average, with a power law density distribution
($n \propto r^{-p}$; Mueller et al. 2002, see also Shirley et al. 2008). 
The convolution of a Gaussian telescope beam pattern with a power-law intensity  distribution results in a FWHM core size slightly larger than the beamsize with the observed FWHM larger for flatter power-law indices, $p$ (see Shirley et al. 2003).
To account for this, we convolved the \nthp\ J $= 1\rightarrow0$ data to $30\arcsec$ resolution (originally $18\arcsec$) in order to study all the mapped transitions at the same resolution. 
With these caveats in mind, we analyze the distribution of sizes for each core in the mapped transitions.

An anti-correlation is observed between the median clump size and the effective
density of the tracer (Figure 8).  This is the expected trend for a power-law
density distribution; the higher density tracers are preferentially excited in a smaller emitting
region within the clumps.  A similar trend was observed in the HCN and CS multi-transition
study of Wu et al. (2010); however, their study contained observations with 
very different spatial resolutions, from $60^{\prime\prime}$ to $30^{\prime\prime}$, while the observations
in this paper are all similar ($30^{\prime\prime}$).

We also see a trend when we look at clump sizes as a function of \neff\ for each evolutionary state. 
If we separate sources into two categories, one for less evolved sources (weak/no radio continuum and UCHII) and one for more evolved sources (CHII and HII), 
we find that more evolved sources are $\sim 0.1$ pc larger than less evolved sources, although the difference in size as a function of evolutionary state becomes less pronounced for higher \neff\ where source sizes get closer to the beam size. We note that this is a qualitative trend as this analysis requires separating 27 sources in to 4 categories. With so few sources entering in to the medians, the median absolute deviation is dominated by small number statistics.

\subsubsection{Masses}\label{sss:masses}
Assuming these clumps are self-gravitating, we calculate the virial mass from
\begin{equation}
M_{vir} = \frac{5 R \Delta v^2}{8 a_1 a_2 G ln(2)} \approx 209 \rm{M}_{\odot} \;\; \frac{(R / 1 \mathrm{pc}) (\Delta v / 1 \mathrm{km s^{-1}})^2 }{ a_1 a_2 }
\end{equation}
\begin{equation}
\begin{array}{cc}
a_1 = \frac{1 - p/3}{1 - 2p/5} , & p < 2.5 \\ 
\end{array}
\end{equation}
where $a_1$ is the correction for power-law density distribution and $a_2$ is the correction for non-spherical shape (Bertoldi \& McKee 1992). For aspect ratios less than 2, $a_2 \approx 1$, so we do not include its effects here. 
We use power-law indices $p$ found in Mueller et al. (2002) to calculate $a_1$. We assume a value of $a_1=1.39$ (corresponding to the median power law index, $p=1.75$) for sources without $350$ \micron\ modeling.

In the \mvir\ calculation, the linewidth, $\Delta v^2$, reflects the thermal energy and turbulent pressure supporting the cloud against gravity. 
Linewidths may be broadened by optical depth and systematic collapse motions (for example), leading to an artificially large estimate of the mass. 
Observing a rarer isotopologue of the molecule mitigates the effects of optical depth in some cases, as the optically thin linewidth from the rarer isotopologue may be substituted for the optically thick linewidth of the main isotopologue. 
Other systematic motions can still broaden the linewidth and, in many cases, 
we cannot correct for those effects. Thus, any mass estimate dependent on the linewidth is only an upper bound.

\begin{figure}[t!]
\epsscale{0.9}
\includegraphics[angle=90,scale=0.60]{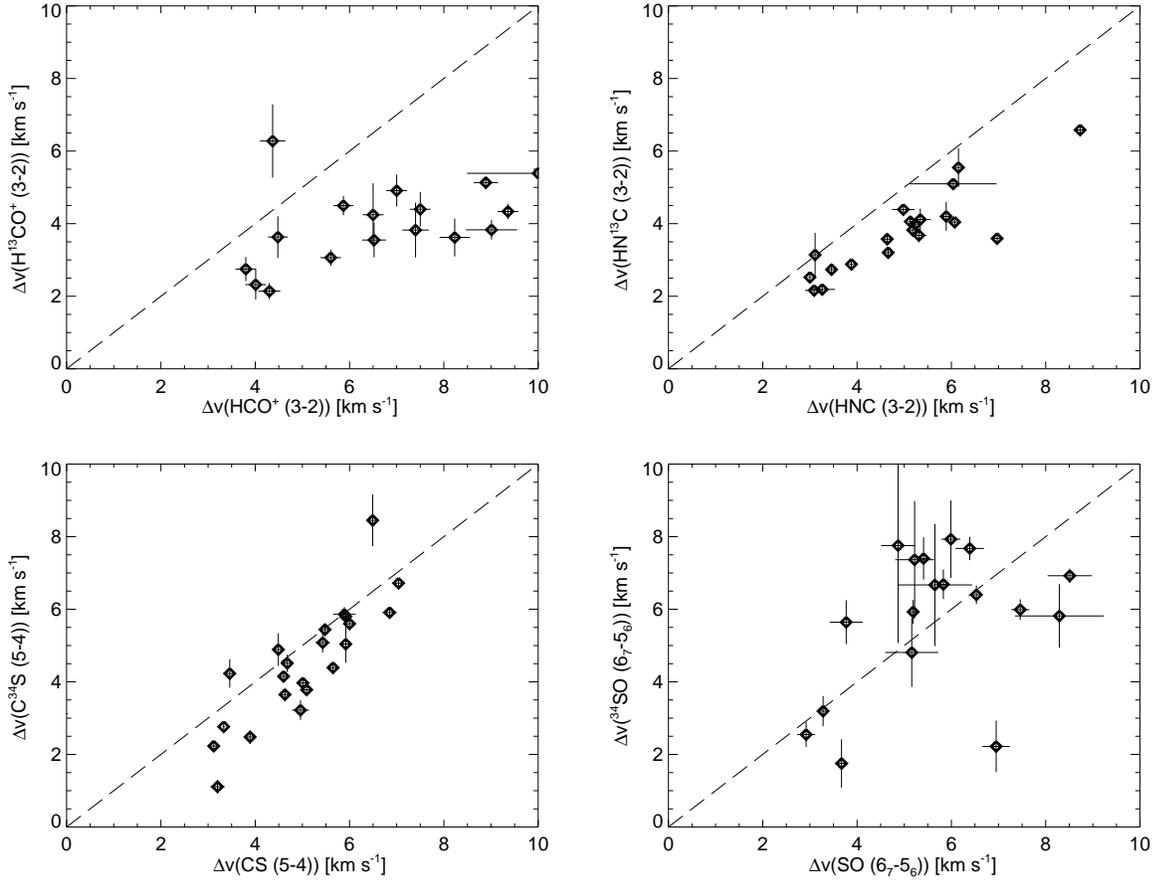}
\figcaption{Isotopologue linewidths plotted against the linewidth for the main species. Dashed line is the one-to-one line. }\label{f:Dv_main_iso}
\end{figure}

We have observed isotopologues of four species in this survey, \hcopi, HN$^{13}$C, C$^{34}$S and $^{34}$SO, and have derived reliable linewidths for the first three. 
Comparing the median \mvir\ calculated with the main and the rarer isotopologue linewidths, we find that the broader linewidths of the main isotopologues leads to a systematic over-estimation of the mass by a factor of up to 1.83 (see Figure 9). 

$\Delta v$ is also not correlated with \neff\ (see Figure 8). 
For every source with a reliable fit for the linewidth, the optically thin isotopologue has a smaller linewidth than the main isotopologue, although this difference is most pronounced for \hcop, which we expect to be optically thick (see Figure 11). No correlation emerges when we compare the optically thin isotopologues with \neff, we find only that the scatter in the lack of correlation is somewhat decreased.

\subsubsection{Mass Surface Density and Volume Density}\label{sss:sigma}
McKee \& Tan (2003) first noted that almost all regions forming massive stars have mass surface densities $\Sigma \sim 1$ g cm$^{-2}$ (equivalent to a column density of $> 2 \times 10^{23}$ cm$^{-2}$). 
Recent numerical simulations by Krumholz, Klein \& McKee (2007) and Krumholz \& McKee (2008) have shown that $\Sigma \ge 1$ g cm$^{-2}$ is a minimum threshold to prevent fragmentation of a massive clump in to several smaller objects. $\Sigma$ of order 1 corresponds to an opacity high enough for gas to be heated substantially (hundreds of K) and to substantial distances ($\ge 1000$ AU) from the protostar (Krumholz, Klein \& McKee 2007). Warmer temperatures are crucial to prevent the clump from fragmenting into smaller objects, as the Jeans Mass decreases for higher densities for a fixed temperature ($M_J \sim T^{3/2}$). 

We calculate the mass surface density of our sources using
\begin{equation}
\Sigma = \frac{M_{vir}}{\pi R^2} \approx 0.665 \frac{(M_{vir} / 1.0 \times 10^4 M_{\sun})}{(R / 1 \mathrm{ pc})^2} \mathrm{ g cm^{-2}}.
\end{equation}
The median surface density for each transition is positively correlated with \neff. This is as expected; higher \neff\ transitions trace denser regions of the clump, corresponding to larger surface densities. 
Surface densities are $>1$ for $\sim \frac{1}{3}$ of our sources when calculated from transitions with \neff$\approx 10^4$ cm$^{-3}$. 
When measured with a tracers with \neff\
above a few $10^6$ cm$^{-3}$, $\sim \frac{2}{3}$ of the clumps have $\Sigma \ge 1.0$ g cm$^{-2}$.
Unlike median size, median 
surface density shows no apparent trend with evolutionary state.

Krumholz \& McKee (2008) point out that even though suppression of fragmentation allows massive stars to form from massive cores, some low mass stars will form before radiation halts fragmentation.
Therefore, massive stars are expected to be found in clusters. 
Volume-averaged density, $\mean{n}$, for a clump with clustered star formation embedded within will be dominated by the diffuse gas between the high density cores, leading to smaller values of $\mean{n}$ than would be expected for an individual star-forming core (Dunham et al. 2010). 
$\mean{n}$ is then a valuable tool for 
determining whether objects discovered in new and upcoming surveys (e.g. BGPS) are individual star forming cores or clumps with cores embedded within. 
For this survey, all of our sources are clumps rather than individual cores, so we focus on characterizing the variation in $\mean{n}$ with \neff. 

Assuming these clumps are spherical, we can calculate the volume-averaged density, 
\begin{equation}
\mean{n} = \frac{M_{vir}} {\frac{4}{3} \pi R^3 \mu m_H}
\end{equation}
 where $\mu$ is the mean molecular weight of molecular gas ($\mu=2.34$), $m_H$ is the mass of hydrogen and $R$ is the radius of the half-peak intensity contour.
We find that $\mean{n}$
increases with \neff, exactly as we expect for transitions with higher \neff\ tracing denser gas (see Figure 7). As a function of evolutionary state, we find that the $\mean{n}$ values for sources associated with no HII emission or (U)CHII emission are essentially the same, while $\mean{n}$ for sources associated with HII regions tends to be higher. The difference is small and may simply reflect small number statistics.

\subsection{Comparison of Molecular Column Density with Physical Variables}\label{ss:phys_var}
We find poor correlations ($r_{corr} < 0.7$) between column density and the physical variables \lbol, \miso, $\mean{n}$, and $\Sigma$. Several factors contribute to these poor correlations. 
A specific molecular transition will trace out the parts of a clump with conditions sufficient to excite that transition. 
Clumps likely contain unresolved cores that, if they are forming stars, have strong gradients with temperature and density increasing towards the center of the core. 
Chemical variability from source to source as well as within our beam increases the scatter in comparisons of the column densities. 

Qualitatively, the column density of higher \neff\ tracers tend to be larger for larger \lbol\ and \miso, though the correlation never becomes robust 
($r_{corr} > 0.7$).
Both volume-averaged density, $\mean{n}$, and mass surface density, $\Sigma$, 
are completely uncorrelated with molecular column densities regardless of \neff. 
Ratios of column densities also fail to correlate with any of the physical variables. 
As discussed in the previous section, the various tracers in this study are sensitive to different excitation conditions and we believe that analyzing the median value of the clump properties is the best way to understand how derived clump properties depend on the tracer chosen.

\subsection{L$^{\prime}_{mol}$ Comparisons}\label{ss:Lprime}

\begin{figure}[t!]
\epsscale{0.9}
\includegraphics[angle=0,scale=0.70]{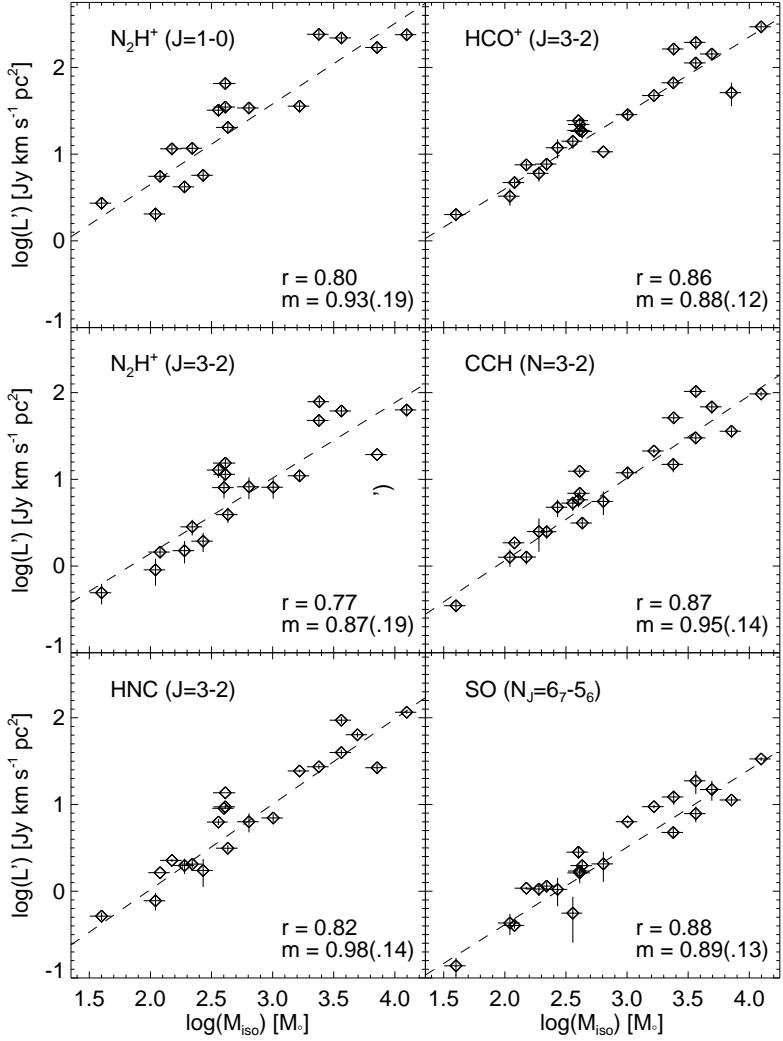}
\figcaption{L$^{\prime}$ versus \miso. Slopes (m) are all slightly sublinear. Spearman Rank coefficient $r_{corr}>0.7$ in every case, indicating that L$^{\prime}$ does reliably trace the mass of dense gas. In some cases, error bars are smaller than the plot symbol.}\label{f:lpmiso}
\end{figure}

Understanding clustered, massive star formation
in our own galaxy is an essential step to understanding star formation in other galaxies. Except for the nearest galaxies, star formation can only be assessed via the unresolved star formation indicators across a galaxy. 
Beginning with Schmidt (1959), 
star formation prescriptions in other galaxies have relied on the 
correlation between the galaxy's gas content and its star formation rate (SFR). Kennicutt (1998) refined this relationship using HI and CO emission to measure the amount of atomic and molecular gas available for star formation, yielding a power law index of $1.4 - 1.5$. Molecular gas is essential to star formation, 
but much of the molecular gas traced by CO (n$_{crit} = 2.8 \times 10^3$ cm$^{-3}$, \neff=$30-50$ cm$^{-3}$) is probably diffuse, and therefore may not be directly involved in the star formation process (Gao \& Solomon 2004). 

More recent work has focused on higher density tracers such as HCN J = $1\rightarrow0$ (n$_{crit} = 1.2 \times 10^6$ cm$^{-3}$, \neff$=1.4\times10^4$ cm$^{-3}$) and HCN J = $3\rightarrow2$ (n$_{crit} = 6.8 \times 10^7$ cm$^{-3}$, \neff$=1.5\times10^5$ cm$^{-3}$) to study the dense molecular gas that is much more likely to be associated with star formation (Gao \& Solomon 2004; Bussmann et al. 2008). 
The star formation law derived from studying dense molecular gas yields a power law index much closer to unity. 
Wu et al. (2009) extended the HCN J = $1\rightarrow0$ sample to galactic cloud cores and found that the data were still well fit with a power law index of 1. 
This led to the interpretation that the linear correlation between dense molecular gas and the star formation rate 
may be understood as the cumulative contribution of individual star forming ``units.''

Extragalactic star formation studies have used the molecular line luminosity, L$^{\prime}$, to estimate the mass of dense gas available for star formation (see Juneau et al. 2009 for a summary of dense gas tracers). 
L$^{\prime}$ is defined as 
\begin{equation}
L^{\prime}  = 23.504 \pi \Omega_{s \ast b} (1+z)^{-3} D_L^2 I(T_{mb})
\end{equation} 
where $D_L$ is the luminosity distance in Mpc, $z$ is the redshift, and $\Omega_{s \ast b}$ is the solid angle of the source convolved with the beam in arcsec$^2$. As described in Mangum et al. (2007), two lines with the same brightness temperature and spatial extent will have the same L$^{\prime}$, such that ratios of L$^{\prime}$ indicate the physical conditions in the gas. 

Empirically, L$^{\prime}$ is found to trace the mass of molecular gas, 
\begin{equation}
M_{dense} = \alpha L^{\prime}_{mol}
\end{equation}
where $\alpha$ is a scaling factor that has traditionally been based on the H$_2$-mass-to-line-luminosity found in the Milky Way (Downes, Solomon \& Radford 1993). 
A direct test of how well L$^{\prime}$ actually traces the mass of dense gas is not possible without an independent determination of the mass. 
For our galactic clumps, we use \miso\ (see \S3.1; Mueller et al. 2002), which is determined from the optically thin $350$ \micron\ dust flux in a $120$\arcsec\ aperture. \miso\ is independent of the integrated intensity of line and independent of the source size determined from the map of that line, making this method of determining mass completely independent from L$^{\prime}$.

Comparing L$^{\prime}$ with \miso, we find that the two are highly correlated ($r > 0.7$; see Figure 11) for every transition.  
Using LINMIX\_ERR to do a linear regression, we find 
power law indices that are slightly less than 1 in every case. 
There is no trend in power law index with \neff. 
Even \nthp, which shows 
strong chemical differentiation in the integrated intensity maps and 
the poorest correlations with the dust flux, 
has an L$^{\prime}$ that is well correlated with \miso\ (with somewhat more scatter than the other transitions).
All of the dense gas tracers we have observed, despite chemical and kinematical behavior that affects the other derived properties of these clumps, are linearly related to the mass of dense gas available for star formation. 
Power law indices slightly smaller than 1 indicate that L$^{\prime}$ will tend to underestimate the mass, and we show that using higher \neff\ tracers will not resolve this issue. 

Looking at L$^{\prime}$ versus \mvir\ for multiple transitions of HCN and CS, 
Wu et al. (2010) also found that L$^{\prime}$ is an excellent tracer of the dense gas. As those authors discuss, questions have been raised concerning how reliably HCN, CS and other high \neff\ molecules trace the mass, given the high densities required for excitation. 
The power indices near unity we find 
supports the Wu et al. (2010) conclusion that 
L$^{\prime}$ is linearly related to the mass 
regardless of \neff\ of the tracers in these two studies because they trace only the dense gas most relevant to star formation.

\section{Conclusions}

We have observed 27 sources in 12 transitions of 8 molecules spanning three orders of magnitude in \neff. These transitions are observed with nearly identical resolution ($30$\arcsec), allowing for inter-comparison of the clump properties derived from the mapped transitions. 
Understanding how derived clump properties depend on \neff\ of the tracer is essential for designing and interpreting molecular line follow-up observations of sources discovered in continuum surveys of the Galactic plane.

We confirm the Zinchenko, Caselli \& Pirogov (2009) result that \nthp\ does a poor job of tracing the dust emission in our massive star forming clumps. This is evident in the spatial differentiation of \nthp\ emission in both the J = $1\rightarrow0$ and J = $3\rightarrow2$ transitions, and the weak correlations between \nthp\ and dust emission. 
\nthp\ is a better tracer of the dust in early stages of evolution, but for our more evolved clumps we find that clump luminosity and associated HII emission support the conclusion that \nthp\ is differentiated in these sources because they are warmer and thus CO, which destroys \nthp, is in the gas phase. 

Unlike \nthp, high \neff\ molecules are well correlated with the dust emission and each other. The correlation between CS and SO column densities is somewhat lower ($r_{corr} = 0.61$), as we would expect if CS is destroyed in the production of SO, as suggested by Beuther et al. (2009).

Derived clump properties vary with \neff\ as we would expect. The highest \neff\ tracers are sensitive to dense gas in the center of these clumps, corresponding to a smaller emitting region (size) but a higher surface density. Virial masses should not depend on \neff\ of the tracer because the calculation is dominated by $\Delta v^2$, which represents thermal and turbulent support in virialized gas. 
We note that virial mass is particularly sensitive to optical depth effects in the line, leading to an over-estimate of the mass by almost a factor of 2 if an optically thin isotopologue is not used to determine $\Delta v$.

Comparing median clump properties --- size $R$, mass \mvir, surface density $\Sigma$, and volume density $\mean{n}$ --- to \neff, we find 
larger median sizes for lower \neff\ transitions. More evolved sources appear to be $\sim 0.1$ pc larger than younger sources. 
Virial masses derived from optically thick lines may not depend
directly on \neff\ since these linewidths may be significantly broadened 
by non-virial motions. Such effects could lead to possible over (or
under) estimates of \mvir. Virial masses derived from more
optically thin lines trace greater pathlengths through the clumps and
may therefore be more reliable.
Mass surface density is correlated with \neff, as we expect from virial masses that are not correlated with \neff\ and sizes that decrease with increasing \neff. 
Approximately $\frac{1}{3}$ of our sources have $\Sigma \ge 1$ g cm$^{-2}$ (as measured by tracers with \neff$>10^4$) which 
Krumholz \& McKee (2008) identify as a minimum surface density required to suppress fragmentation. 
Volume density is correlated with \neff\ and may be useful for distinguishing between higher density cores (dominated by the high column density of the forming protostar) and lower density clumps (dominated by the diffuse gas between cores embedded in the clump).
Molecule column densities are not well correlated with any of the physical properties we derive for these clumps.

We find a nearly linear relationship between L$^{\prime}$ and \miso, independent of \neff\ of the tracer, confirming that L$^{\prime}$ does reliably trace the mass of dense molecular gas.

\section*{Acknowledgments}

We wish to thank the staffs of the HHT, CSO, NRO, and JCMT for their excellent
assistance.  We thank Ruisheng Peng, Hiroshige Yoshida, Katelyn Allers,
and Jingwen Wu for their assistance with CSO observations.  
Special thanks to  Joe McMullian for help with glish scripting in AIPS++.  
We also thank the HHT operators Patrick Fimbres, Sean Keel, Bob Moulton and John Downey for their assistance with HHT observations. 
Guest User, Canadian Astronomy Data Centre, which is operated by the 
Dominion Astrophysical Observatory for the National Research Council of 
Canada's Herzberg Institute of Astrophysics.
Yancy Shirley is partially supported by NSF grant AST-1008577.

\appendix

\section{Appendix material}

Contour maps are presented for the 27 high-mass star forming regions. 
Beamsizes are indicated in the lower left hand corner of each panel. 
The lowest contour as a percentage of the peak intensity and the corresponding $\sigma$ value are listed at the bottom of each panel. Contours are in steps of percentages of the peak. In cases where a percentage less than that of the lowest contour were used, step size is indicated in percent of the peak after the $\sigma$ value. 

Source characteristics are listed in Table 2.



\newpage

\begin{deluxetable}{l|ccccccc}
\tablecolumns{7}
\tablecaption{Properties of Observed Transitions \label{t:tran_properties}}
\tablewidth{0pt} 
\tablehead{
\colhead{Transition}                   &
\colhead{$\nu$}      &
\colhead{$\theta_{mb}$}       &
\colhead{$g_u$}               & 
\colhead{$E_u/k$}               & 
\colhead{$A_{ul}$}      &
\colhead{\neff (20K) }    &
\colhead{n$_{\mathrm{crit}}(20$K)}             \\
\colhead{} &
\colhead{(GHz)} &
\colhead{(\as )} & 
\colhead{} &
\colhead{(K)} & 
\colhead{(s$^{-1}$)} &
\colhead{\cmv } &
\colhead{\cmv }               
}
\startdata 
\nthp\ $1-0$	& 93.1737767	& 18  &  3 & 4.47 & $3.63 \times 10^{-5}$ & $2.0 \times 10^3$  &  $1.6 \times 10^5$ \\
H$^{13}$CO$^+$ $3-2$ & 260.2553390  &  27.9  &  7 &  24.98  &  $1.34 \times 10^{-3}$  &  \nodata  &  $3.2 \times 10^6$ \\
\hcop\ $3-2$ & 267.5576190  &  27.1  &  7  &  25.68  &  $1.48 \times 10^{-3}$  &  $2.3 \times 10^4$  &  $3.5 \times 10^6$ \\
\nthp\ $3-2$ & 279.5117010  &  26.0  &  7  &  26.83  & $1.26 \times 10^{-3}$  &  $2.7 \times 10^4$  &  $3.0 \times 10^6$ \\
HCS$^+$ $6-5$ & 256.0271000  &  28.3  &  13  &  43.01  & $3.46 \times 10^{-4}$  &  \nodata  &  $9.1 \times 10^5$ \\
CCH $3_{J} - 2_{J}$ & 262.07167\tablenotemark{c}  &  28.8  &  7  &  21.32  &  $5.75 \times 10^{-5}$  &  $1.5 \times 10^5$\tablenotemark{b}  &  $5.15 \times 10^7$\tablenotemark{b} \\
HNC $3-2$ & 271.9811420  &  27.7  &  7  &  26.11  &  $9.34 \times 10^{-4}$  &  $1.7 \times 10^5$  &  $6.2 \times 10^7$ \\
HN$^{13}$C $3-2$ & 261.2633101  &  28.9  &  7  &  25.08  &  $6.48 \times 10^{-4}$  &  \nodata  &  $4.3 \times 10^7$ \\
CS $5-4$\tablenotemark{a} & 244.9355565  &  29.6  &  11  &  35.27  &  $2.98 \times 10^{-4}$  &  $3.7 \times 10^5$  &  $7.3 \times 10^6$ \\
C$^{34}$S $5-4$\tablenotemark{a} & 241.0161940  &  30.1  &  11  &  34.70  &  $2.84 \times 10^{-4}$  &  $$  &  $6.9 \times 10^6$ \\
SO $6_7 - 5_6$ & 261.84296479  &  28.8  &  11  &  47.551  &  $2.24 \times 10^{-4}$  &  $2.4 \times 10^6$  &  $1.2 \times 10^8$ \\
$^{34}$SO $6_7 - 5_6$ & 256.8774563  &  29.4  &  15  &  36.23  &  $2.19 \times 10^{-4}$  &  \nodata  &  $1.2 \times 10^8$ \\
SO$_2$ $7_{3,5} - 7_{2,6}$ & 257.0999658  &  29.3  &  29  &  47.84  &  $1.13 \times 10^{-4}$  &  \nodata  &  $3.42 \times 10^7$ \\
CH$_3$OH-E $2_1 - 1_0$ & 261.805736  &  28.8  &  5  &  23.30  &  $6.71 \times 10^{-5}$  &  \nodata  &  $4.0 \times 10^6$ \\
\enddata
\tablenotetext{a}{These transitions were observed in Shirley et al. 2003, but are presented
here for completeness.}
\tablenotetext{b}{Assuming H$^{13}$CN J = $3\rightarrow2$ collision rates since no data are available for CCH.}
\tablenotetext{c}{Average frequency for the blended hyperfine components.}
\end{deluxetable}

\begin{deluxetable}{lccccccc}
\tablecolumns{8}
\footnotesize
\tablecaption{Observed Sources \label{t:sources}}
\tablewidth{0pt} 
\tablehead{
\colhead{Source}                  &
\colhead{$\alpha$ (J2000.0)}      &
\colhead{$\delta$ (J2000.0)}      &
\colhead{D}               &
\colhead{\lbol\ }                       &
\colhead{\tiso\ }                  &
\colhead{$p$\tablenotemark{a}}                      &
\colhead{Evolutionary\tablenotemark{b}}          \\                      
\colhead{}                         &
\colhead{($^h$~~$^m$~~$^s$~)~}            &  
\colhead{($\degree$ ~\arcmin\ ~\arcsec )}  &  
\colhead{(kpc)}                          &
\colhead{(\lsun )}                           &
\colhead{(K)}                             &
\colhead{}                          &
\colhead{State}                                
}
\startdata 
121.30+0.66	&	00 36 47.5	&	+63 29 01	&	0.9	&	1.1E+03	&	22\tablenotemark{d}	&	1.25	&	NRC	\\
123.07-6.31	&	00 52 25.1	&	+56 33 54	&	2.2	&	6.0E+03	&	21.8	&	1.75	&	HII	\\
W3(OH)	&	02 27 04.7	&	+61 52 26	&	2.4	&	9.5E+04	&	18.1	&	1.50	&	UCHII	\\
S231	&	05 39 12.9	&	+35 45 54	&	2.0	&	1.3E+04	&	16.4	&	1.50	&	NRC	\\
S252A	&	06 08 35.4	&	+20 39 03	&	1.5	&	6.4E+03	&	19.2	&	1.75	&	HII	\\
RCW142	&	17 50 14.6	&	-28 54 32	&	2.0	&	5.7E+04	&	26.1	&	2.25	&	UCHII	\\
W28A2	&	18 00 30.4	&	-24 03 58	&	2.6	&	2.0E+05	&	22.7	&	2.25	&	UCHII	\\
M8E	&	18 04 53.1	&	-24 26 42	&	1.8	&	1.7E+04	&	21.9	&	1.75	&	UCHII	\\
  8.67-0.36	&	18 06 18.8	&	-21 37 38	&	8.5	&	1.3E+05	&	44.2	&	2.00	&	UCHII	\\
10.60-0.40	&	18 10 28.6	&	-19 55 49	&	6.5	&	9.2E+05	&	22.4	&	2.50	&	CHII	\\
12.89+0.49	&	18 11 51.1	&	-17 31 30	&	3.5	&	3.9E+04	&	23.6	&	2.00	&	NRC	\\
W33A	&	18 14 39.2	&	-17 52 11	&	4.5	&	1.0E+05	&	20.2	&	1.50	&	NRC	\\
24.49-0.04	&	18 36 05.3	&	-07 31 22	&	3.5	&	4.9E+04	&	29.4	&	2.25	&	NRC	\\
W43S	&	18 46 03.5	&	-02 39 25	&	8.5	&	1.6E+06	&	26.4	&	2.50	&	CHII	\\
31.41+0.31	&	18 47 34.4	&	-01 12 46	&	7.9	&	2.3E+05	&	23.1	&	2.25	&	UCHII	\\
W44	&	18 53 18.4	&	+01 14 57	&	3.7	&	7.5E+05	&	22\tablenotemark{d}	&	...	&	CHII	\\
40.50+2.54	&	18 56 10.4	&	+07 53 14	&	2.1	&	2.8E+04	&	19.0	&	1.50	&		\\
35.20-0.74	&	18 58 12.7	&	+01 40 36	&	3.3	&	4.5E+04\tablenotemark{c}	&	22\tablenotemark{d}	&	...	&	HII	\\
59.78+0.06	&	19 43 11.5	&	+23 43 54	&	2.2	&	1.3E+04\tablenotemark{c}	&	22\tablenotemark{d}	&	...	&	UCHII	\\
ON1	&	20 10 09.1	&	+31 31 37	&	6.0	&	1.5E+05	&	20.7	&	1.75	&	UCHII	\\
ON2S	&	20 21 41.0	&	+37 25 29	&	5.5	&	3.7E+05	&	18.8	&	1.75	&	HII	\\
W75N	&	20 38 36.9	&	+42 37 38	&	3.0	&	1.4E+05	&	22\tablenotemark{d}	&	...	&	UCHII	\\
W75OH	&	20 39 00.9	&	+42 22 50	&	3.0	&	5.0E+04\tablenotemark{c}	&	22\tablenotemark{d}	&	...	&	NRC	\\
S140	&	22 19 18.2	&	+63 18 48	&	0.9	&	1.9E+04	&	16.9	&	1.25	&	NRC	\\
CepA	&	22 56 18.1	&	+62 01 47	&	0.7	&	2.2E+04	&	16.9	&	1.50	&	UCHII	\\
NGC7538-IRS9	&	23 14 01.6	&	+61 27 20	&	2.8	&	3.6E+04	&	18.9	&	1.75	&	UCHII	\\
S157	&	23 16 04.3	&	+60 01 41	&	2.5	&	2.9E+04	&	15.9	&	0.75	&	CHII	\\
\enddata
\tablenotetext{a}{power law index from Mueller et al (2002).}
\tablenotetext{b}{NRC = weak or No Radio Continuum; UCHII = Ultra-Compact HII; CHII = Compact HII; HII = free-free emission from ionized H.}
\tablenotetext{c}{Using L$_{IR}$ in place of L$_{bol}$.}
\tablenotetext{d}{\tiso\ is taken to be the median value ($=22 K$) for sources without 350 \micron\ modelling. } 
\end{deluxetable}

\include{medians_table}

\include{int_int_corrtables}

\include{ncol_corrtable}

\include{n2hp_obs_table}

\include{hcop32_obs_table}

\include{n2hp32_obs_table}

\include{cch32_obs_table}

\include{hnc32_obs_table}

\include{so65_obs_table}

\include{chemfigs1}

\include{chemfigs2}

\include{chemfigs3}

\end{document}

%% file: medians_table.tex
\begin{deluxetable}{l|cccccccc}
\tablecolumns{7}
\tablecaption{Median Source Properties \label{t:median_properties}}
\tablewidth{0pt} 
\tablehead{
\colhead{Transition}                   &
\colhead{log(\neff)}      &
\colhead{$R$}      &
\colhead{$\Sigma$}       &
\colhead{$\log{\mean{n}}$}               & 
\colhead{$M_{vir}$}               & 
\colhead{$\Delta v$}      &
\colhead{L$^{\prime}$}    & 
\colhead{$\frac{L_{bol}}{M_{vir}}$}             \\
\colhead{} &
\colhead{ \small (cm$^{-3}$) \small } &
\colhead{ \small (pc) \small } &
\colhead{ \small ($\frac{\mathrm{g}}{\mathrm{cm^2}}$) \small } &
\colhead{ \small ($\frac{\mathrm{g}}{\mathrm{cm^3}}$) \small } &
\colhead{ \small (\msun) \small } & 
\colhead{ \small ($\frac{\mathrm{km}}{\mathrm{s}}$) \small } &
\colhead{ \small ($\frac{\mathrm{Jy}}{\mathrm{pc}^{2}} \frac{\mathrm{km}}{\mathrm{s}}$) \small } & 
\colhead{ \small ($\frac{L_{\sun}}{M_{\sun}}$) \small } 
}
\startdata 
\nthp\ $1-0$  &  $3.30$  &  0.29  &  0.17  &  4.66  &  308  &  2.77  &  58.30  &  168.42  \\
\hcop\ $3-2$  &  $4.36$  &  0.25  &  0.36  &  5.12  &  640  &  6.50  &  24.35  &  74.57  \\
\nthp\ $3-2$  &  $4.43$  &  0.26  &  0.73  &  5.28  &  639  &  4.14  &  11.41  &  51.08  \\
CCH $3_{J} - 2_{J}$  &  $5.18$  &  0.25  &  0.55  &  5.28  &  524  &  3.93  &  11.51  &  96.06  \\
HNC $3-2$  &  $5.23$  &  0.24  &  0.35  &  5.19  &  446  &  4.86  &  9.23  &  61.06  \\
CS $5-4$  &  $5.57$  &  0.20  &  0.53  &  5.23  &  700  &  4.81  &  11.32  &  154.05  \\ 
SO $6_7 - 5_6$  &  $6.38$  &  0.17  &  1.43  &  5.82  &  689  &  5.09  &  2.37  &  102.09  \\
\enddata
\end{deluxetable}

%% file: int_int_corrtables.tex
\clearpage
\begin{landscape}

\begin{deluxetable}{cccccccccc}
\tabletypesize{\scriptsize}
\tablecaption{Correlation Coefficients for Integrated Intensities of all transitions \label{t:corr}}
\tablewidth{0pt} 
\tablehead{
\colhead{Transition}        &
\colhead{350 \micron}       &
\colhead{850 \micron}               & 
\colhead{\nthp}    &
\colhead{\hcop}             &
\colhead{\nthp}    &
\colhead{CCH}    &
\colhead{HNC}    &
\colhead{CS}    &
\colhead{SO}    \\
\colhead{} & 
\colhead{} & 
\colhead{} & 
\colhead{($1\rightarrow 0$)} &
\colhead{($3\rightarrow 2$)}   & 
\colhead{($3\rightarrow 2$)}   & 
\colhead{($3_J\rightarrow 2_J$)}   & 
\colhead{($3\rightarrow 2$)}   & 
\colhead{($5\rightarrow 4$)}   & 
\colhead{($6_7\rightarrow 5_6$)}   
}
\startdata
\nthp\ ($1 \rightarrow 0$)  &  
0.55  &  0.56  &    &  \textit{0.46}  &  \textbf{0.83}  &  0.54  &  0.66  &  \textit{0.44}  &  \textit{0.34}  \\
\hcop\ ($3 \rightarrow 2$)  &  
0.62  &  0.62  &  \textit{0.46}  &    &  0.68  &  \textbf{0.82}  &  \textbf{0.90}  &  0.61  &  \textbf{0.80}  \\
\nthp\ ($3 \rightarrow 2$)  &  
\textit{0.48}  &  0.55  &  \textbf{0.83}  &  0.68  &    &  0.68  &  \textbf{0.72}  &  \textit{0.43}  &  0.52  \\
CCH ($3 \rightarrow 2$)  &  
\textbf{0.81}  &  \textbf{0.74}  &  0.58  &  \textbf{0.82}  &  \textbf{0.73}  &    &  \textbf{0.86}  &  \textbf{0.72}  &  \textbf{0.79}  \\
HNC ($3 \rightarrow 2$)  &  
\textbf{0.78}  &  \textbf{0.73}  &  0.68  &  \textbf{0.90}  &  \textbf{0.76}  &  \textbf{0.86}  &    &  0.68  &  \textbf{0.76}  \\
CS ($5 \rightarrow 4$)  &  
\textbf{0.75}  &  \textbf{0.70}  &  \textit{0.44}  &  0.61  &  \textit{0.43}  &  \textbf{0.72}  &  0.68  &    &  0.67  \\
SO ($6 \rightarrow 5$)  & 
0.68  &  \textbf{0.73}  &  \textit{0.34}  &  \textbf{0.80}  &  0.54  &  \textbf{0.79}  &  \textbf{0.76}  &  0.67  &    \\
\hline
number above 0.7  &  7  &  8  &  1  &  4  &  3  &  7  &  6  &  4  &  4  \\

\enddata
\end{deluxetable}

\clearpage
\end{landscape}

%% file: ncol_corrtable.tex
\clearpage
\begin{deluxetable}{ccccccccc}
\tablecolumns{12}
\tabletypesize{\scriptsize}
\tablecaption{Optically Thin Column Density Correlation Coefficients\label{t:ncol_thin}}
\tablewidth{0pt} 
\tablehead{
\colhead{Molecule}        &
\colhead{H$_2$}    &
\colhead{\hcop}    &
\colhead{\nthp}    &
\colhead{CCH}    &
\colhead{HNC}    &
\colhead{CS}    &
\colhead{SO} }
\startdata

\hcop\  &   \textbf{0.76}  &   &  0.69  &  \textbf{0.84}  &  \textbf{0.91}  &  0.65  &  \textbf{0.77}  \\

\nthp\  &   \textit{0.41}  &  0.69  &    &  \textbf{0.76}  &  \textbf{0.73}  &  \textit{0.45}  &  0.53  \\

CCH  &   \textbf{0.71}  &  \textbf{0.84}  &  \textbf{0.76}  &    &  \textbf{0.88}  &  \textbf{0.75}  &  \textbf{0.70}  \\ 

HNC  &   \textbf{0.71}  &  \textbf{0.91}  &  \textbf{0.73}  &  \textbf{0.88}  &    &  0.67  &  0.68  \\

CS  &   0.62  &  0.65  &  \textit{0.45}  &  \textbf{0.75}  &  0.67  &    &  0.61  \\

SO  &   \textbf{0.78}  &  \textbf{0.77}  &  0.53  &  \textbf{0.70}  &  0.68  &  0.61  &  \\

number above 0.7  &  4  &  3  &  3  &  5  &  3  &  1  &  2  \\

\enddata
\end{deluxetable}


\begin{deluxetable}{cccccccc}
\tablecolumns{12}
\tabletypesize{\scriptsize}
\tablecaption{Optically Thick Column Density Correlation Coefficients\label{t:ncol_thick}}
\tablewidth{0pt} 
\tablehead{
\colhead{Molecule}        &
\colhead{H$_2$}    &
\colhead{\hcop}    &
\colhead{\nthp}    &
\colhead{CCH}    &
\colhead{HNC}    &
\colhead{CS}    &
\colhead{SO} }
\startdata

\hcop\  &  0.66  &    &  0.67  &  \textbf{0.77}  &  0.54  &  \textit{0.35}  &  \textit{0.48}  \\ 

\nthp\  &  \textit{0.41}  &  0.67  &    &  \textbf{0.76}  &  \textit{0.39}  &  \textit{0.20}  &  \textit{0.29}  \\ 

CCH  &  \textbf{0.71}  &  \textbf{0.77}  &  \textbf{0.76}  &    &  0.58  &  0.54  &  0.60  \\ 

HNC  &  0.54  &  0.54  &  \textit{0.39}  &  0.58  &    &  0.54  &  0.51  \\ 

CS  &  0.56  &  \textit{0.35}  &  \textit{0.20}  &  0.54  &  0.54  &    &  0.50  \\ 

SO  &  \textbf{0.73}  &  \textit{0.48}  &  \textit{0.29}  &  0.60  &  0.51  &  0.50  	&  \\ 

number above 0.7  &  2  &  1  &  1  &  2  &  0  &  0  &  0  \\

\enddata
\end{deluxetable}


%% file: n2hp_obs_table.tex
\begin{deluxetable}{lcccc}
\tablecolumns{5}
\footnotesize
\tablecaption{Observed Source Properties for \nthp\ $J=1\rightarrow 0$ \label{t:nthp_properties}}
\tablewidth{0pt} 
\tablehead{
\colhead{Source}  &
\colhead{$I(T_{mb})$}          &
\colhead{$T_{mb}$}       & 
\colhead{$\Delta v$}               &
\colhead{$\mean{R}$}           \\
\colhead{} &
\colhead{(K km/s)} &
\colhead{(K)} & 
\colhead{(km/s)} & 
\colhead{(pc)}
}
\startdata
121.30+0.66  &  49.84(6.03)  &  9.94(1.22)  &  1.83(0.30)  &  0.11(0.004)  \\
123.07-6.31  &  44.36(5.40)  &  7.30(0.93)  &  2.57(0.30)  &  0.26(0.004)  \\
W3(OH)  &  57.50(6.95)  &  9.96(1.22)  &  2.36(0.30)  &  M\tablenotemark{a}  \\
S231  &  76.92(9.26)  &  12.72(1.56)  &  2.59(0.30)  &  0.25(0.004)  \\
S252A  &  19.76(2.45)  &  4.86(0.79)  &  1.36(0.30)  &  0.16(0.010)  \\
RCW142  &  119.74(14.40)  &  13.46(1.64)  &  3.20(0.30)  &  0.36(0.004)  \\
W28A2  &  119.96(14.42)  &  14.98(1.82)  &  3.56(0.30)  &  0.28(0.004)  \\
M8E  &  36.76(4.46)  &  7.98(1.00)  &  1.40(0.30)  &  0.17(0.004)  \\
8.67-0.36  &  126.38(15.20)  &  11.08(1.36)  &  3.80(0.30)  &  M\tablenotemark{a}  \\ 
10.60-0.49  &  84.12(10.15)  &  11.70(1.45)  &  2.95(0.30)  &  0.83(0.010)  \\
12.89+0.49  &  82.46(9.92)  &  11.14(1.36)  &  3.57(0.30)  &  0.33(0.005)  \\
W33A  &  106.52(12.82)  &  14.16(1.71)  &  3.17(0.30)  &  0.73(0.005)  \\
24.49-0.04  &  110.34(13.27)  &  13.50(1.64)  &  3.93(0.30)  &  0.29(0.002)  \\
W43S  &  36.42(4.44)  &  6.22(0.76)  &  2.31(0.30)  &  M\tablenotemark{a}  \\
31.41+0.31  &  47.40(5.73)  &  5.74(0.73)  &  3.64(0.30)  &  0.94(0.025)  \\
W44  &  141.78(17.04)  &  18.06(2.18)  &  3.54(0.30)  &  0.67(0.006)  \\ 
40.50+2.54  &  165.40(19.86)  &  24.46(2.94)  &  2.77(0.30)  &  0.23(0.001)  \\
35.20-0.74  &  77.84(9.39)  &  9.90(1.26)  &  2.77(0.30)  &  0.55(0.008)  \\
59.78+0.06  &  65.42(7.90)  &  14.24(1.75)  &  1.12(0.30)  &  0.26(0.004)  \\
ON1  &  81.62(9.82)  &  10.28(1.26)  &  3.09(0.30)  &  0.80(0.007)  \\
ON2S  &  73.52(8.86)  &  8.94(1.09)  &  3.80(0.30)  &  M\tablenotemark{a}  \\ 
W75N  &  79.58(9.59)  &  12.70(1.55)  &  2.89(0.30)  &  0.48(0.005)  \\ 
W75OH  &  155.40(18.67)  &  22.94(2.77)  &  2.67(0.30)  &  0.56(0.003)  \\ 
S140  &  74.06(8.92)  &  17.00(2.05)  &  1.48(0.30)  &  0.19(0.003)  \\ 
CepA  &  60.90(7.36)  &  9.60(1.20)  &  1.96(0.30)  &  0.14(0.002)  \\ 
NGC7538-IRS9  &  38.96(4.74)  &  6.80(0.88)  &  2.13(0.30)  &  M\tablenotemark{a}  \\ 
S157  &  26.14(3.23)  &  4.66(0.64)  &  2.74(0.30)  &  0.24(0.012)  \\ 
\enddata
\tablenotetext{a}{M = multiple peaks}
\end{deluxetable}

%% file: hcop32_obs_table.tex
\begin{deluxetable}{lcccc}
\tablecolumns{5}
\footnotesize
\tablecaption{Observed Source Properties for \hcop\ $J=3\rightarrow 2$ \label{t:hcop32_properties}}
\tablewidth{0pt} 
\tablehead{
\colhead{Source}  &
\colhead{$I(T_{mb})$}          &
\colhead{$T_{mb}$}       & 
\colhead{$\Delta v$}               &
\colhead{$\mean{R}$}           \\
\colhead{} &
\colhead{(K km/s)} &
\colhead{(K)} & 
\colhead{(km/s)} & 
\colhead{(pc)}
}
\startdata
121.30+0.66  &  46.12(4.73)  &  11.23(1.15)  &  5.3(0.22)  &  0.08(0.008)  \\
123.07-6.31  &  43.60(4.47)  &  7.28(0.78)  &  5.8(0.23)  &  0.13(0.019)  \\ 
W3(OH)  &  80.68(8.16)  &  12.92(1.33)  &  7.0(0.22)  &  0.23(0.017)  \\
S231  &  49.97(5.18)  &  10.83(1.15)  &  4.4(0.24)  &  0.25(0.022)  \\ 
S252A  &  31.55(3.38)  &  7.88(0.89)  &  3.8(0.23)  &  0.11(0.020)  \\ 
RCW142  &  129.33(13.01)  &  20.37(2.08)  &  5.9(0.21)  &  0.12(0.011)  \\ 
W28A2  &  187.17(18.79)  &  16.73(1.70)  &  8.9(0.26)  &  0.16(0.004)  \\ 
M8E  &  36.48(3.76)  &  8.40(0.90)  &  4.0(0.22)  &  0.15(0.021)  \\ 
8.67-0.36  &  44.40(4.67)  &  7.18(0.89)  &  9.0(0.54)  &  0.80(0.086)  \\ 
10.60-0.49  &  195.88(19.67)  &  19.68(2.00)  &  10.9(0.29)  &  0.37(0.011)  \\ 
12.89+0.49  &  18.33(2.06)  &  4.32(0.48)  &  7.4(0.28)  &  0.27(0.136)  \\ 
W33A  &  61.48(6.31)  &  6.83(0.81)  &  8.2(0.32)  &  0.38(0.022)  \\ 
24.49-0.04  &  31.78(3.37)  &  3.33(0.41)  &  6.1(0.27)  &  0.20(0.036)  \\ 
W43S  &  65.20(6.62)  &  10.55(1.10)  &  5.6(0.22)  &  0.39(0.046)  \\ 
31.41+0.31  &  16.47(1.92)  &  2.10(0.44)  &  11.4(0.66)  &  0.63(0.152)  \\ 
W44  &  111.62(11.22)  &  19.53(1.98)  &  10.0(1.51)  &  0.41(0.014)  \\ 
40.50+2.54  &  65.18(6.57)  &  11.77(1.21)  &  4.5(0.21)  &  0.23(0.014)  \\ 
35.20-0.74  &  72.30(7.35)  &  10.48(1.13)  &  11.5(0.26)  &  0.32(0.025)  \\ 
59.78+0.06  &  34.23(3.53)  &  7.48(0.83)  &  4.3(0.23)  &  0.24(0.032)  \\ 
ON1  &  59.42(6.08)  &  9.48(1.00)  &  4.4(0.27)  &  0.50(0.037)  \\ 
ON2S  &  78.05(7.88)  &  10.33(1.11)  &  6.5(0.22)  &  0.57(0.024)  \\ 
W75N  &  155.98(15.64)  &  19.87(2.01)  &  7.5(0.22)  &  0.22(0.006)  \\ 
W75OH  &  120.95(12.16)  &  22.90(2.31)  &  8.0(1.02)  &  0.26(0.012)  \\ 
S140  &  120.92(12.16)  &  27.17(2.77)  &  3.8(0.22)  &  0.10(0.003)  \\ 
CepA  &  121.38(12.22)  &  14.50(1.47)  &  9.4(0.22)  &  0.08(0.002)  \\ 
NGC7538-IRS9  &  36.83(3.86)  &  5.58(0.65)  &  6.5(0.25)  &  0.33(0.041)  \\ 
S157  &  26.90(2.84)  &  6.62(0.90)  &  4.4(0.25)  &  0.26(0.041)  \\ 
\enddata
\end{deluxetable}

%% file: n2hp32_obs_table.tex
\begin{deluxetable}{lcccc}
\tablecolumns{5}
\footnotesize
\tablecaption{Observed Source Properties for \nthp\ $J=3\rightarrow 2$ \label{t:n2hp32_properties}}
\tablewidth{0pt} 
\tablehead{
\colhead{Source}  &
\colhead{$I(T_{mb})$}          &
\colhead{$T_{mb}$}       & 
\colhead{$\Delta v$}               &
\colhead{$\mean{R}$}           \\
\colhead{} &
\colhead{(K km/s)} &
\colhead{(K)} & 
\colhead{(km/s)} & 
\colhead{(pc)}
}
\startdata
121.30+0.66  &  15.74(1.42)  &  4.68(0.45)  &  2.61(0.20)  &  0.06(0.014)  \\ 
123.07-6.31  &  22.00(1.92)  &  5.36(0.54)  &  4.03(0.23)  &  0.09(0.030)  \\ 
W3(OH)  &  20.29(1.74)  &  4.51(0.47)  &  4.86(0.83)  &  0.25(0.044)  \\ 
S231  &  24.26(2.05)  &  4.71(0.54)  &  3.35(0.10)  &  0.14(0.023)  \\ 
S252A  &  12.41(1.27)  &  2.91(0.39)  &  2.87(0.22)  &  0.08(0.034)  \\ 
RCW142  &  33.76(2.81)  &  4.67(0.58)  &  5.08(0.11)  &  0.29(0.017)  \\ 
W28A2  &  40.39(3.28)  &  6.76(0.67)  &  4.45(0.05)  &  0.18(0.018)  \\ 
M8E  &  11.01(0.97)  &  3.51(0.35)  &  2.47(0.27)  &  0.13(0.032)  \\ 
8.67-0.36  &  37.62(4.50)  &  6.43(1.00)  &  5.87(0.23)  &  M\tablenotemark{a}  \\ 
10.60-0.49  &  31.93(2.63)  &  3.55(0.43)  &  7.06(0.52)  &  0.51(0.059)  \\ 
12.89+0.49  &  15.11(1.36)  &  3.45(0.48)  &  4.52(0.19)  &  0.26(0.061)  \\ 
W33A  &  24.95(2.05)  &  4.21(0.46)  &  4.58(0.13)  &  0.58(0.034)  \\ 
24.49-0.04  &  26.27(3.18)  &  4.75(0.72)  &  4.93(0.21)  &  0.24(0.034)  \\ 
W43S  &  10.11(0.94)  &  1.43(0.22)  &  3.01(0.10)  &  M\tablenotemark{a}  \\ 
31.41+0.31  &  11.86(1.48)  &  1.84(0.37)  &  5.88(0.43)  &  0.31(13.19)  \\ 
W44  &  46.53(5.50)  &  7.18(0.94)  &  5.70(0.33)  &  0.67(0.037)  \\ 
40.50+2.54  &  53.55(4.31)  &  10.63(1.00)  &  4.14(0.05)  &  0.17(0.011)  \\ 
35.20-0.74  &  30.03(3.61)  &  5.30(0.72)  &  5.08(0.15)  &  0.42(0.035)  \\ 
59.78+0.06  &  26.68(3.22)  &  7.18(0.97)  &  3.30(0.15)  &  0.14(0.024)  \\ 
ON1  &  22.50(1.85)  &  3.89(0.39)  &  4.39(0.11)  &  0.67(0.042)  \\ 
ON2S  &  35.03(2.87)  &  5.74(0.57)  &  4.14(0.12)  &  0.61(0.028)  \\ 
W75N  &  36.97(4.37)  &  8.43(1.03)  &  3.60(0.07)  &  0.30(0.020)  \\ 
W75OH  &  65.92(7.76)  &  11.37(1.46)  & 	4.99(0.30)  &  0.45(0.024)  \\ 
S140  &  32.28(3.87)  &  8.07(1.15)  &  3.07(0.18)  &  M\tablenotemark{a}  \\ 
CepA  &  28.59(2.33)  &  7.39(0.66)  &  3.97(0.17)  &  0.09(0.007)  \\ 
NGC7538-IRS9  &  16.75(2.09)  &  3.55(0.61)  &  3.89(0.31)  &  0.27(0.042)  \\ 
S157  &  9.61(0.86)  &  1.87(0.26)  &  3.14(0.21)  &  0.14(0.002)  \\ 
\enddata
\tablenotetext{a}{M = multiple peaks}
\end{deluxetable}

%% file: cch32_obs_table.tex
\begin{deluxetable}{lcccc}
\tablecolumns{5}
\footnotesize
\tablecaption{Observed Source Properties for CCH $J=3\rightarrow 2$ \label{t:cch32_properties}}
\tablewidth{0pt} 
\tablehead{
\colhead{Source}  &
\colhead{$I(T_{mb})$}          &
\colhead{$T_{mb}$}       & 
\colhead{$\Delta v$}               &
\colhead{$\mean{R}$}           \\
\colhead{} &
\colhead{(K km/s)} &
\colhead{(K)} & 
\colhead{(km/s)} & 
\colhead{(pc)}
}
\startdata
121.30+0.66  &  12.06(0.73)  &  1.41(0.08)  &  2.5(0.10)  &  0.05(0.009)  \\ 
123.07-6.31  &  13.92(0.81)  &  1.53(0.08)  &  3.1(0.14)  &  0.13(0.026)  \\ 
W3(OH)  &  24.43(1.21)  &  2.30(0.10)  &  4.6(0.13)  &  0.28(0.025)  \\ 
S231  &  15.49(0.87)  &  1.55(0.08)  &  3.6(0.13)  &  0.16(0.020)  \\ 
S252A  &  12.78(1.06)  &  1.32(0.10)  &  2.8(0.21)  &  0.10(0.023)  \\ 
RCW142  &  33.00(1.65)  &  2.58(0.15)  &  6.0(0.19)  &  0.16(0.015)  \\ 
W28A2  &  76.09(3.43)  &  6.66(0.30)  &  4.1(0.13)  &  0.17(0.009)  \\ 
M8E  &  12.73(2.74)  &  2.71(0.30)  &  2.8(0.29)  &  0.15(0.046)  \\ 
8.67-0.36  &  29.43(1.49)  &  2.63(0.14)  &  4.4(0.24)  &  0.56(0.069)  \\ 
10.60-0.49  &  62.10(3.07)  &  4.21(0.22)  &  7.0(0.03)  &  0.43(0.026)  \\ 
12.89+0.49  &  11.22(1.03)  &  1.54(0.11)  &  3.8(0.27)  &  0.23(0.072)  \\ 
W33A  &  16.64(1.27)  &  1.46(0.11)  &  4.1(0.11)  &  0.35(0.057)  \\ 
24.49-0.04  &  11.48(0.86)  &  0.97(0.08)  &  5.4(0.38)  &  0.24(0.044)  \\ 
W43S  &  22.40(1.48)  &  2.28(0.14)  &  4.2(0.19)  &  0.64(0.093)  \\ 
31.41+0.31  &  15.48(0.84)  &  1.19(0.08)  &  5.9(0.32)  &  0.44(0.057)  \\
W44  &  49.29(2.07)  &  3.96(0.17)  &  6.0(0.13)  &  0.44(0.012)  \\ 
40.50+2.54  &  32.19(1.44)  &  3.43(0.15)  &  3.0(0.07)  &  0.25(0.012)  \\ 
35.20-0.74  &  23.07(1.04)  &  1.94(0.09)  &  4.6(0.16)  &  0.31(0.016)  \\ 
59.78+0.06  &  12.87(1.62)  &  1.28(0.22)  &  2.3(0.36)  &  0.13(0.027)  \\ 
ON1  &  18.70(1.01)  &  1.90(0.10)  &  4.1(0.25)  &  0.36(0.044)  \\ 
ON2S  &  23.06(1.16)  &  2.25(0.11)  &  4.2(\nodata)  &  0.59(0.043)  \\ 
W75N  &  27.85(1.22)  &  2.60(0.12)  &  4.6(0.06)  &  0.33(0.015)  \\ 
W75OH  &  46.57(2.18)  &  4.09(0.19)  &  4.6(0.12)  &  E\tablenotemark{a} \\ 
S140  &  25.21(1.26)  &  3.04(0.14)  &  2.9(3.81)  &  0.08(0.012)  \\ 
CepA  &  31.17(1.35)  &  3.47(0.14)  &  3.5(0.08)  &  0.10(0.004)  \\ 
NGC7538-IRS9  &  10.17(0.67)  &  1.06(0.07)  &  3.7(0.24)  &  0.26(0.035)  \\ 
S157  &  14.12(1.37)  &  1.33(0.13)  &  4.1(0.44)  &  0.21(0.037)  \\ 
\enddata
\tablenotetext{a}{E = extended emission, no well-defined source boundary}
\end{deluxetable}

%% file: hnc32_obs_table.tex
\begin{deluxetable}{lcccc}
\tablecolumns{5}
\footnotesize
\tablecaption{Observed Source Properties for HNC $J=3\rightarrow 2$ \label{t:cch32_properties}}
\tablewidth{0pt} 
\tablehead{
\colhead{Source}  &
\colhead{$I(T_{mb})$}          &
\colhead{$T_{mb}$}       & 
\colhead{$\Delta v$}               &
\colhead{$\mean{R}$}           \\
\colhead{} &
\colhead{(K km/s)} &
\colhead{(K)} & 
\colhead{(km/s)} & 
\colhead{(pc)}
}
\startdata
121.30+0.66  &  11.63(0.47)  &  3.19(0.15)  &  3.5(0.06)  &  0.08(0.006)  \\ 
123.07-6.31  &  10.33(0.39)  &  2.53(0.12)  &  3.8(0.08)  &  0.15(0.014)  \\
W3(OH)  &  18.53(0.68)  &  3.81(0.19)  &  4.8(0.07)  &  0.24(0.017)  \\ 
S231  &  14.03(0.48)  &  3.53(0.16)  &  3.5(0.07)  &  0.18(0.012)  \\ 
S252A  &  7.04(0.37)  &  2.27(0.13)  &  3.1(0.12)  &  0.12(0.023)  \\ 
RCW142  &  36.61(1.21)  &  5.76(0.27)  &  5.9(0.12)  &  0.25(0.007)  \\ 
W28A2  &  94.85(2.99)  &  12.25(0.56)  &  6.1(0.12)  &  0.17(0.005)  \\ 
M8E  &  13.20(0.96)  &  3.25(0.32)  &  3.3(0.27)  &  0.12(0.022)  \\ 
8.67-0.36  &  22.44(0.87)  &  4.25(0.22)  &  5.0(0.24)  &  0.68(0.048)  \\ 
10.60-0.49  &  77.16(2.42)  &  8.40(0.42)  &  8.7(0.09)  &  0.44(0.016)  \\ 
12.89+0.49  &  14.50(0.68)  &  2.63(0.19)  &  5.3(0.22)  &  0.21(0.056)  \\ 
W33A  &  27.41(0.88)  &  4.50(0.23)  &  5.3(0.16)  &  0.40(0.021)  \\ 
24.49-0.04  &  16.37(0.54)  &  2.62(0.14)  &  5.1(0.15)  &  0.20(0.018)  \\ 
W43S  &  31.29(1.02)  &  5.31(0.27)  &  5.2(0.09)  &  0.44(0.040)  \\ 
31.41+0.31  &  8.44(0.34)  &  1.18(0.08)  &  9.1(0.46)  &  0.63(0.054)  \\ 
W44  &  51.45(1.58)  &  7.93(0.32)  &  6.2(0.06)  &  0.41(0.008)  \\ 
40.50+2.54  &  26.97(0.87)  &  6.38(0.27)  &  3.9(0.04)  &  0.24(0.008)  \\ 
35.20-0.74  &  23.52(0.74)  &  3.67(0.15)  &  7.0(0.09)  &  0.37(0.012)  \\ 
59.78+0.06  &  11.79(0.42)  &  3.42(0.15)  &  3.1(0.19)  &  E\tablenotemark{a}  \\ 
ON1  &  19.03(0.63)  &  3.42(0.15)  &  5.2(0.13)  &  0.49(0.031)  \\ 
ON2S  &  20.48(0.66)  &  3.53(0.16)  &  4.6(0.07)  &  E\tablenotemark{a}  \\ 
W75N  &  36.55(1.11)  &  6.34(0.27)  &  5.1(0.04)  &  0.26(0.006)  \\ 
W75OH  &  35.64(1.18)  &  7.26(0.33)  &  6.0(0.92)  &  0.40(0.018)  \\ 
S140  &  29.29(0.91)  &  8.23(0.33)  &  3.0(0.03)  &  0.12(0.003)  \\ 
CepA  &  31.50(1.00)  &  7.21(0.30)  &  4.7(0.06)  &  0.09(0.000)  \\ 
NGC7538-IRS9  &  11.34(0.41)  &  2.71(0.12)  &  3.9(0.11)  &  0.34(0.027)  \\ 
S157  &  6.15(0.48)  &  1.12(0.12)  &  3.5(0.57)  &  0.19(0.059)  \\ 
\enddata
\tablenotetext{a}{E = extended emission, no well-defined source boundary}
\end{deluxetable}

%% file: so65_obs_table.tex
\begin{deluxetable}{lcccc}
\tablecolumns{5}
\footnotesize
\tablecaption{Observed Source Properties for SO $N_J=6_7\rightarrow 5_6$ \label{t:so65_properties}}
\tablewidth{0pt} 
\tablehead{
\colhead{Source}  &
\colhead{$I(T_{mb})$}          &
\colhead{$T_{mb}$}       & 
\colhead{$\Delta v$}               &
\colhead{$\mean{R}$}           \\
\colhead{} &
\colhead{(K km/s)} &
\colhead{(K)} & 
\colhead{(km/s)} & 
\colhead{(pc)}
}
\startdata
121.30+0.66  &  4.67(0.32)  &  1.10(0.07)  &  2.8(0.17)  &  0.05(0.014)  \\ 
123.07-6.31  &  6.63(0.38)  &  1.12(0.07)  &  5.3(0.20)  &  0.12(0.018)  \\ 
W3(OH)  &  27.20(1.12)  &  4.03(0.17)  &  5.8(0.06)  &  0.15(0.005)  \\ 
S231  &  6.95(0.49)  &  1.23(0.09)  &  4.6(0.23)  &  0.21(0.024)  \\ 
S252A  &  5.06(0.54)  &  0.99(0.10)  &  2.9(0.19)  &  0.09(0.028)  \\ 
RCW142  &  12.17(1.76)  &  2.27(0.31)  &  5.6(0.79)  &  0.10(0.030)  \\ 
W28A2  &  48.97(2.53)  &  5.29(0.37)  &  8.5(0.47)  &  0.10(0.007)  \\ 
M8E  &  9.94(0.87)  &  2.65(0.22)  &  2.6(0.35)  &  0.07(0.025)  \\ 
8.67-0.36  &  5.65(0.53)  &  1.06(0.11)  &  5.7(0.50)  &  0.52(0.175)  \\ 
10.60-0.49  &  30.09(1.49)  &  4.08(0.22)  &  7.5(0.19)  &  0.27(0.023)  \\ 
12.89+0.49  &  4.41(0.48)  &  0.79(0.10)  &  3.8(0.35)  &  0.21(0.090)  \\ 
W33A  &  8.35(0.54)  &  1.00(0.09)  &  8.3(0.94)  &  0.20(0.047)  \\ 
24.49-0.04  &  2.11(0.64)  &  0.59(0.12)  &  7.6(1.23)  &  0.08(0.140)  \\ 
W43S  &  8.35(1.26)  &  1.51(0.20)  &  5.2(0.56)  &  0.30(0.210)  \\ 
31.41+0.31  &  7.16(0.40)  &  0.96(0.07)  &  7.0(0.29)  &  0.18(0.091)  \\ 
W44  &  26.16(1.05)  &  3.84(0.16)  &  6.5(0.08)  &  0.21(0.007)  \\ 
40.50+2.54  &  9.49(0.48)  &  2.21(0.11)  &  3.7(0.14)  &  0.13(0.016)  \\ 
35.20-0.74  &  6.80(0.33)  &  1.09(0.07)  &  6.0(0.20)  &  0.23(0.020)  \\ 
59.78+0.06  &  7.18(1.31)  &  1.60(0.25)  &  2.3(0.52)  &  U\tablenotemark{a}  \\ 
ON1  &  5.08(0.49)  &  0.87(0.10)  &  5.2(0.41)  &  0.34(0.081)  \\ 
ON2S  &  6.84(0.67)  &  1.44(0.11)  &  4.9(0.36)  &  0.50(0.067)  \\ 
W75N  &  30.41(1.26)  &  5.24(0.22)  &  5.2(0.09)  &  0.17(0.005)  \\ 
W75OH  &  30.85(2.06)  &  4.43(0.36)  &  6.4(0.30)  &  0.17(0.018)  \\ 
S140  &  21.14(0.88)  &  6.93(0.28)  &  3.3(0.08)  &  0.09(0.002)  \\ 
CepA  &  15.12(0.71)  &  2.52(0.12)  &  5.4(0.15)  &  0.05(0.004)  \\ 
NGC7538-IRS9  &  5.57(0.36)  &  0.74(0.07)  &  4.8(0.30)  &  0.24(0.029)  \\ 
S157  &  4.98(0.46)  &  1.05(0.10)  &  3.5(0.39)  &  0.13(0.073)  \\ 
\enddata
\tablenotetext{a}{U = unresolved}
\end{deluxetable}

%% file: chemfigs1.tex
\begin{figure*}
\figurenum{1}
\epsscale{0.9}
\includegraphics[angle=0, scale=1]{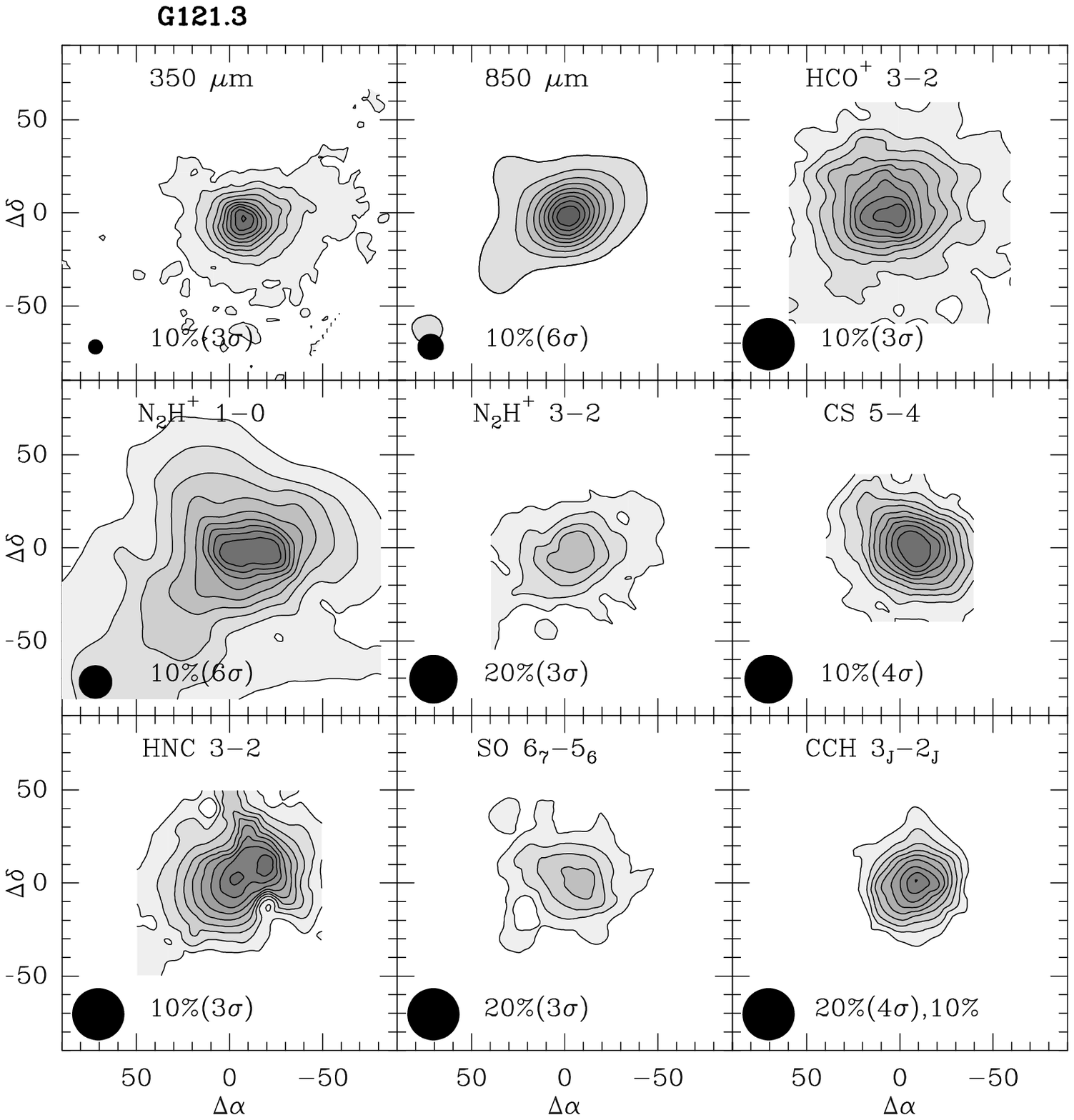}\label{f:g121.3}
\end{figure*}

\begin{figure*}
\figurenum{2}
\epsscale{0.9}
\includegraphics[angle=0, scale=1]{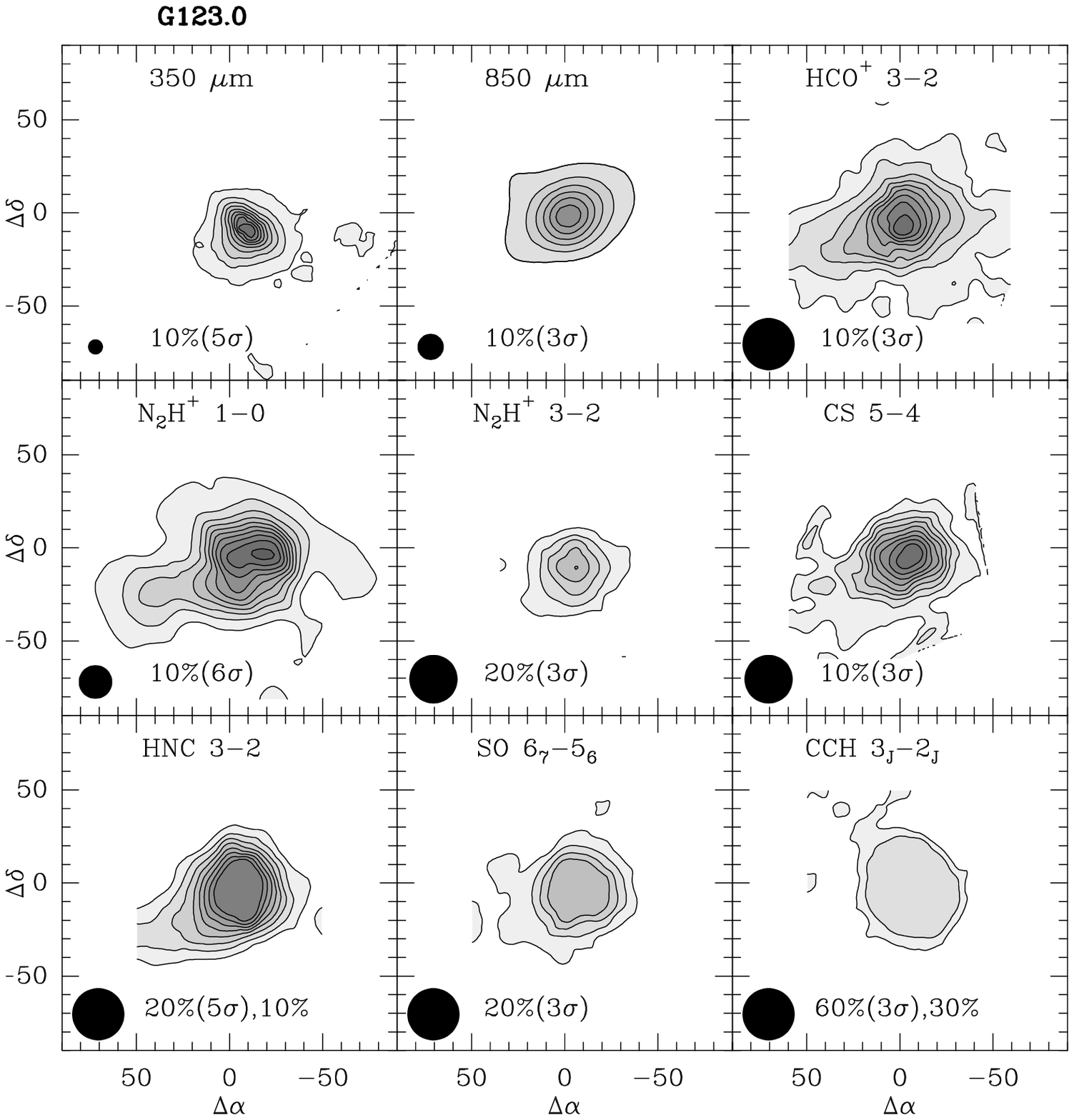}\label{f:g123.0}
\end{figure*}

\begin{figure*}
\figurenum{3}
\epsscale{0.9}
\includegraphics[angle=0, scale=1]{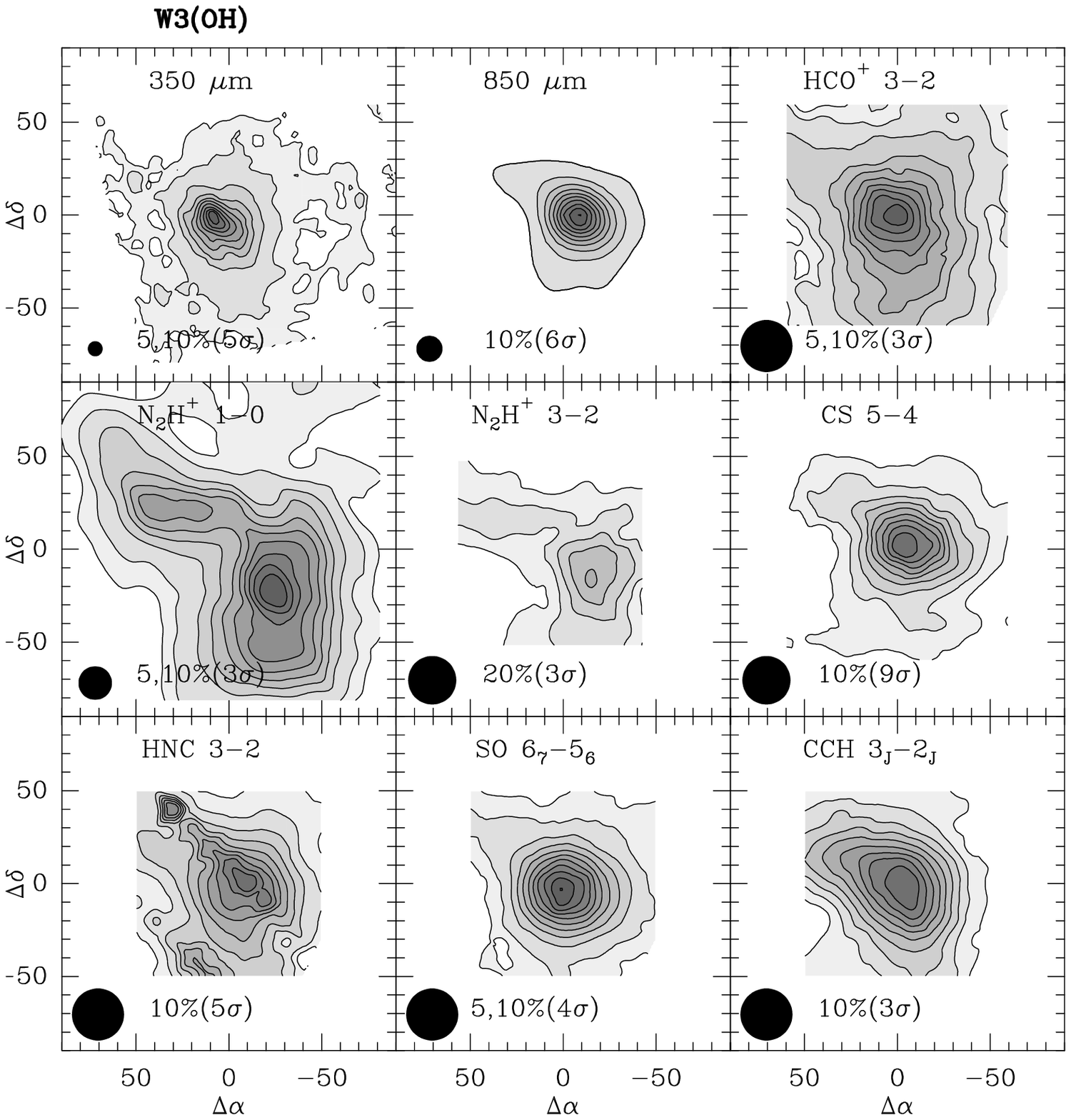}\label{f:w3oh}
\end{figure*}

\begin{figure*}
\figurenum{4}
\epsscale{0.9}
\includegraphics[angle=0, scale=1]{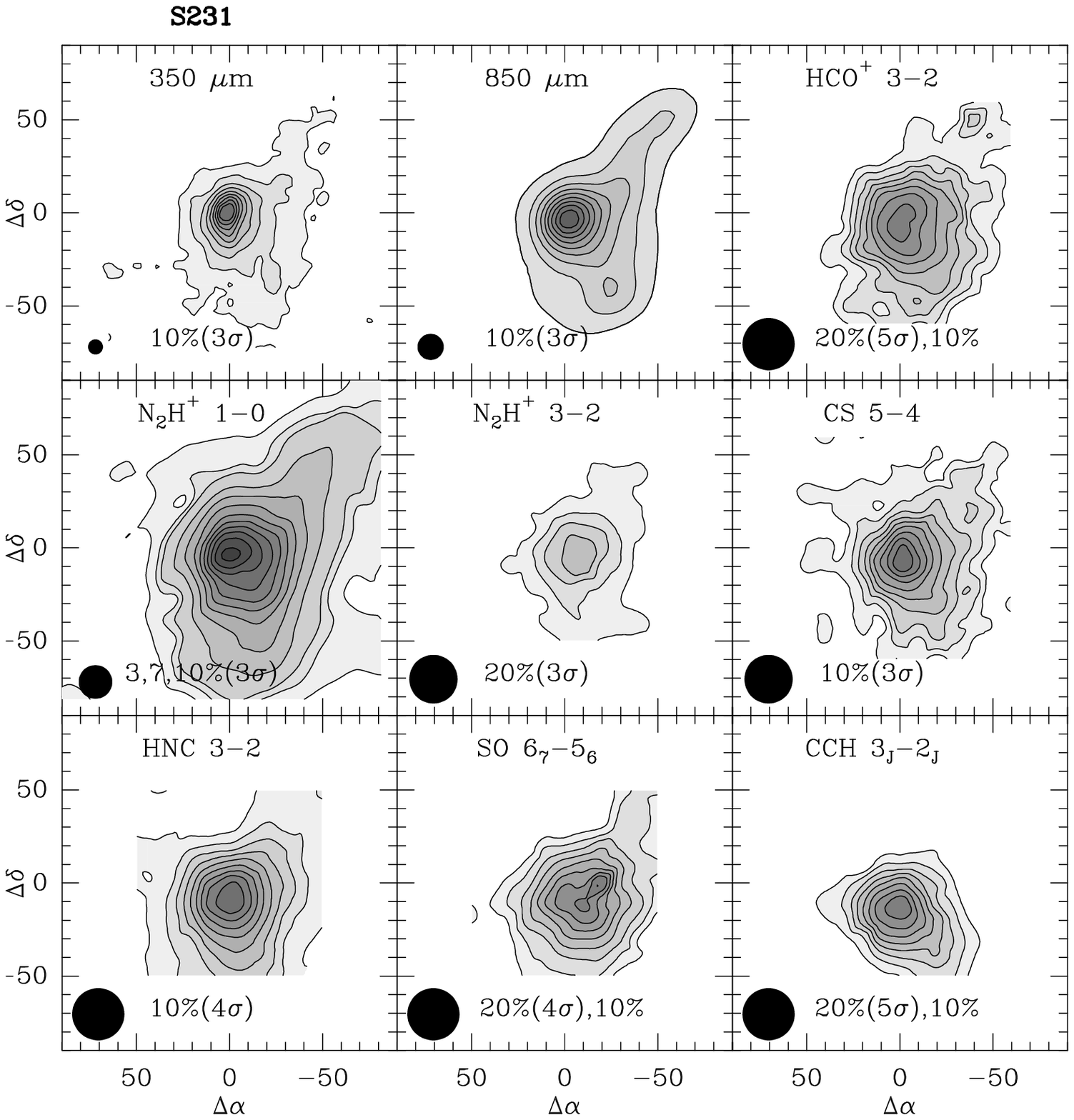}\label{f:s231}
\end{figure*}

\begin{figure*}
\figurenum{5}
\epsscale{0.9}
\includegraphics[angle=0, scale=1]{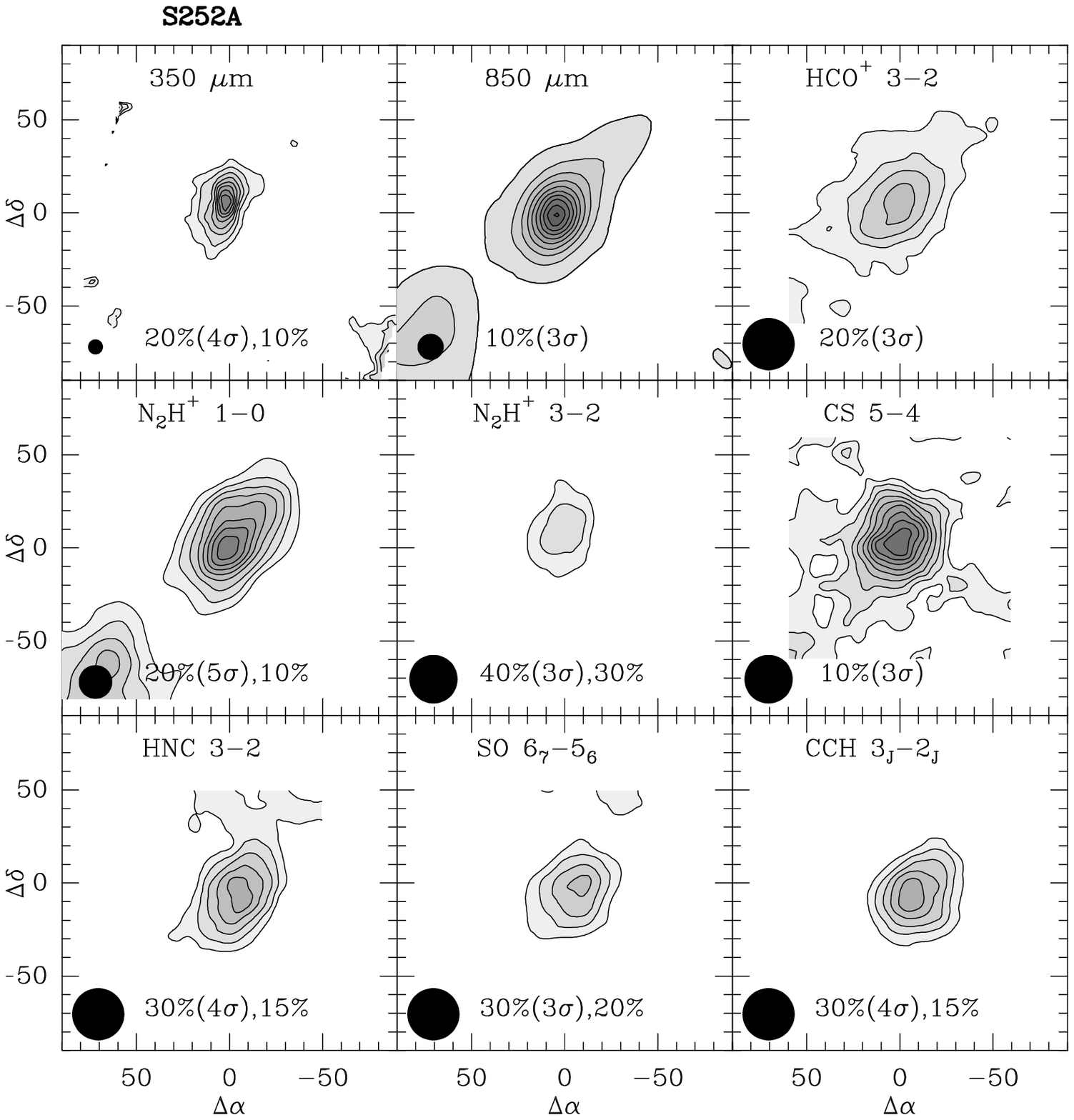}\label{f:s252a}
\end{figure*}

\begin{figure*}
\figurenum{6}
\epsscale{0.9}
\includegraphics[angle=0, scale=1]{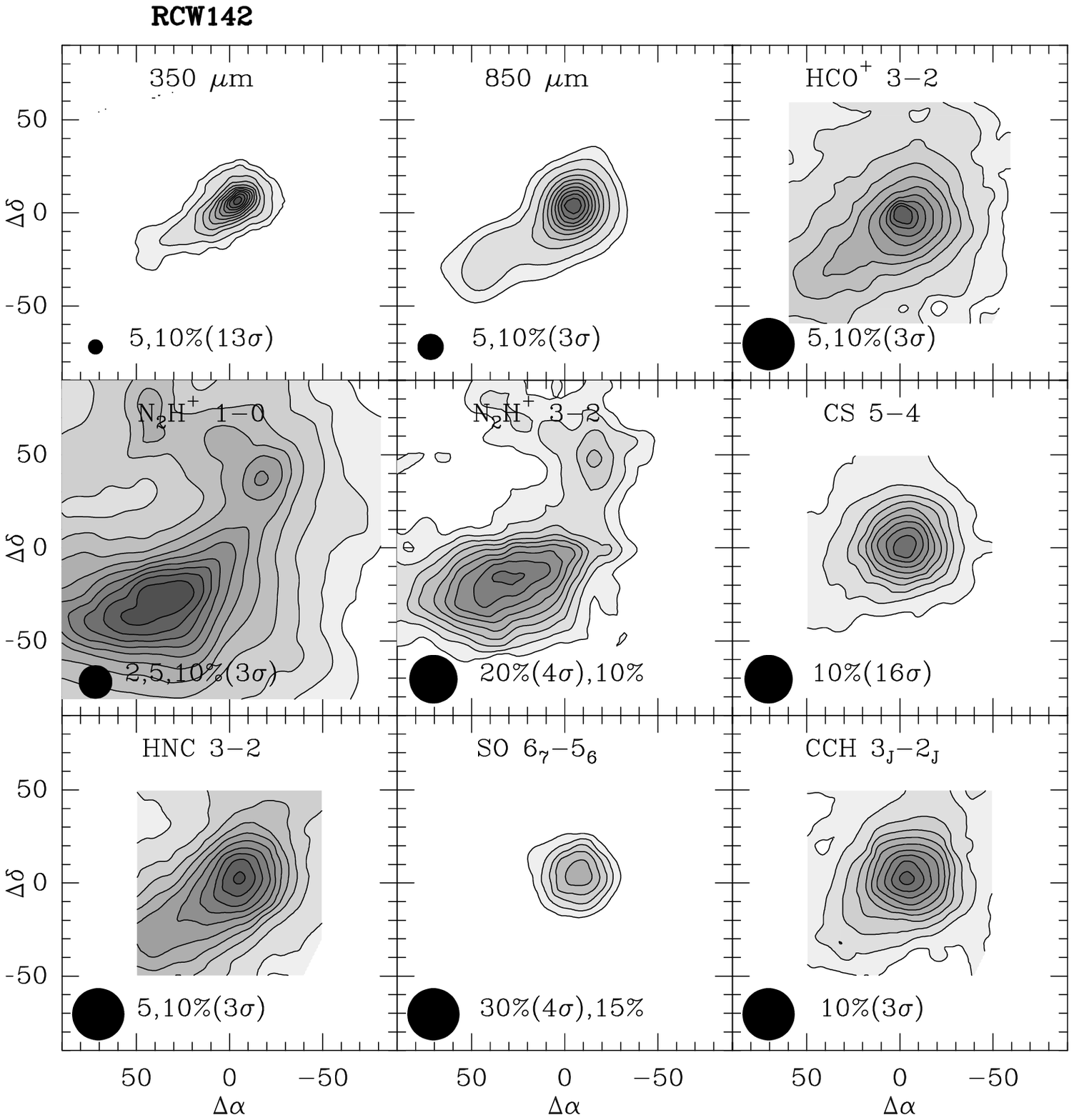}\label{f:rcw142}
\end{figure*}

\begin{figure*}
\figurenum{7}
\epsscale{0.9}
\includegraphics[angle=0, scale=1]{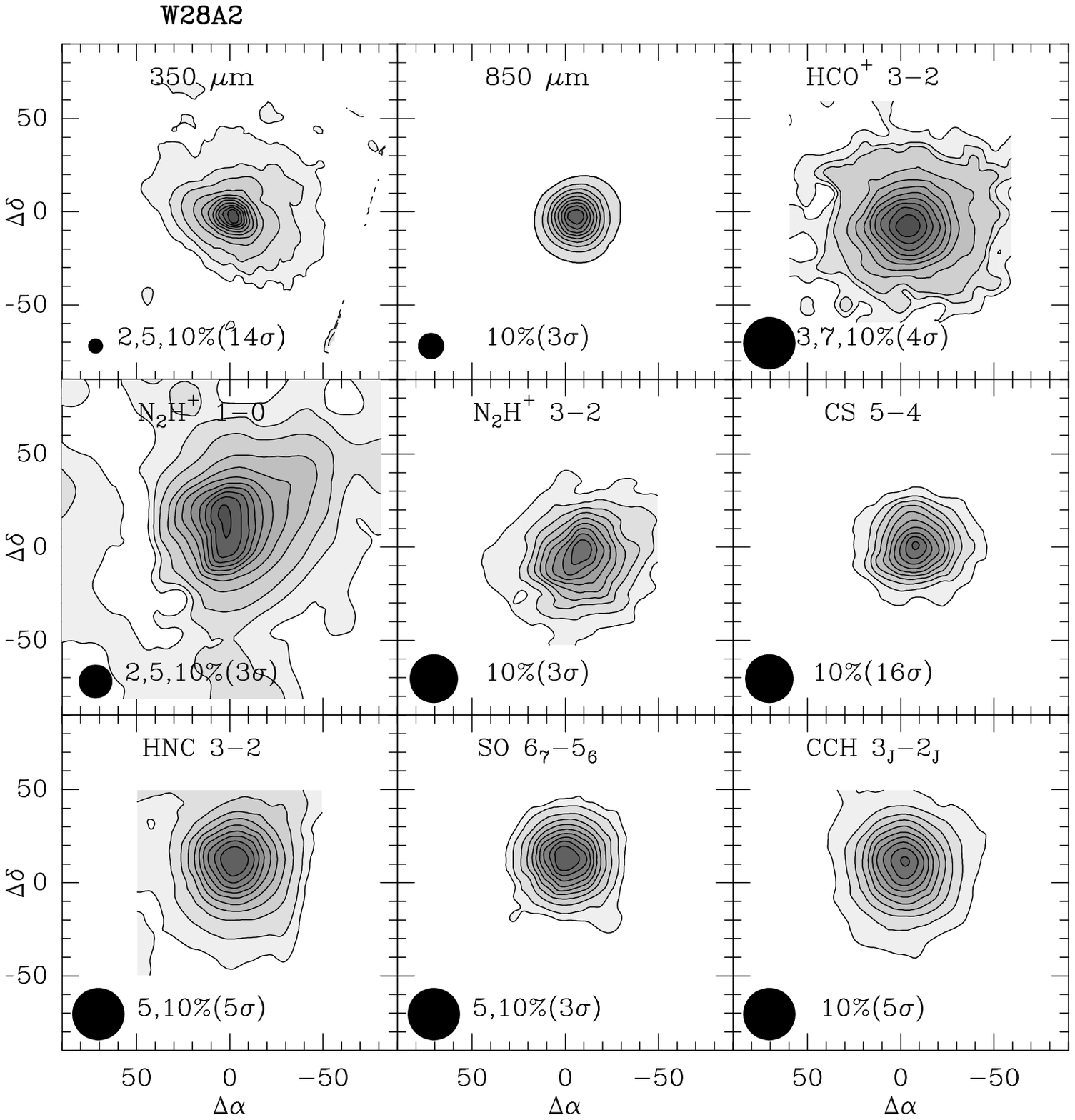}\label{f:w28a2}
\end{figure*}

\begin{figure*}
\figurenum{8}
\epsscale{0.9}
\includegraphics[angle=0, scale=1]{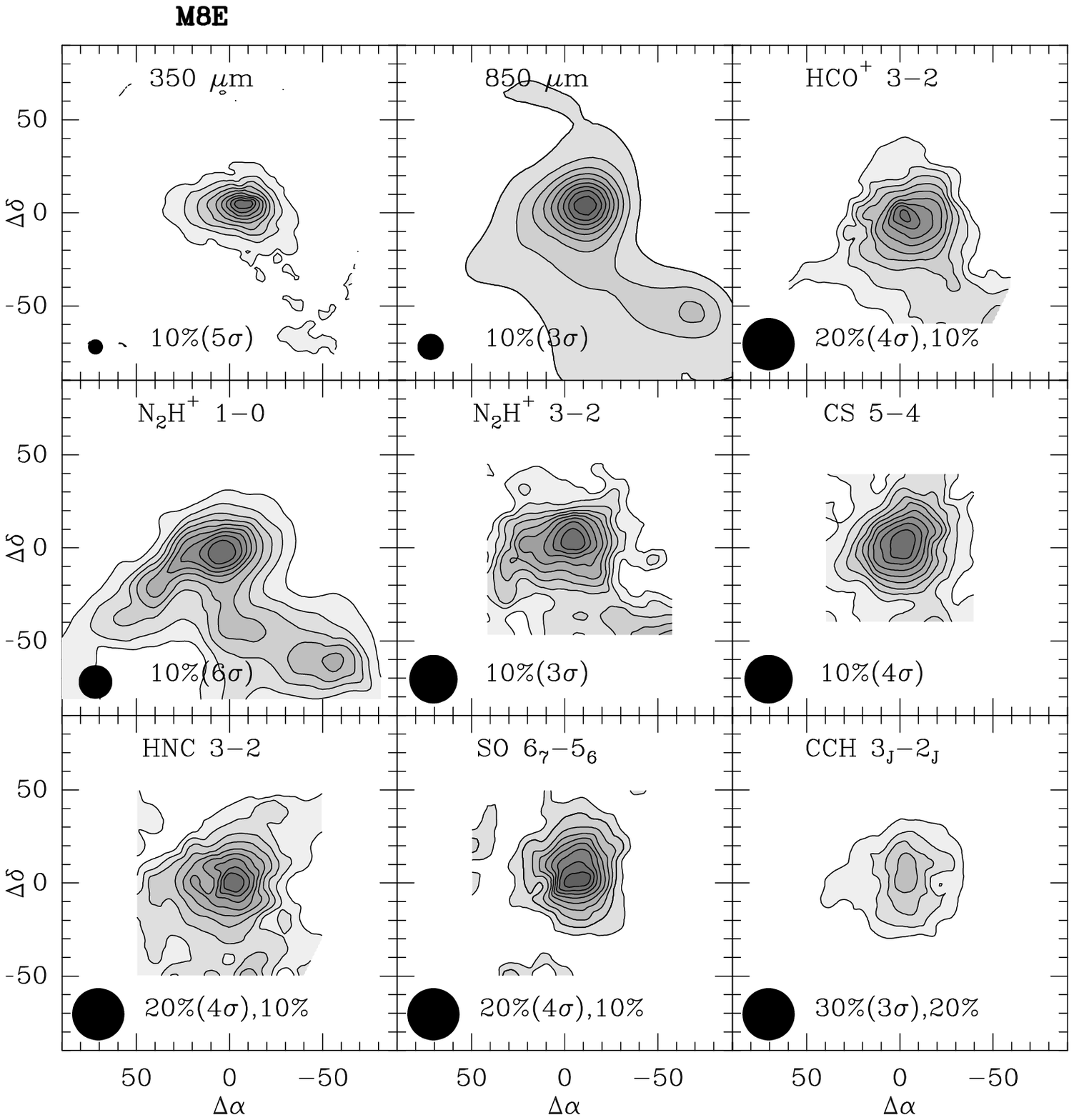}\label{f:m8e}
\end{figure*}

\begin{figure*}
\figurenum{9}
\epsscale{0.9}
\includegraphics[angle=0, scale=1]{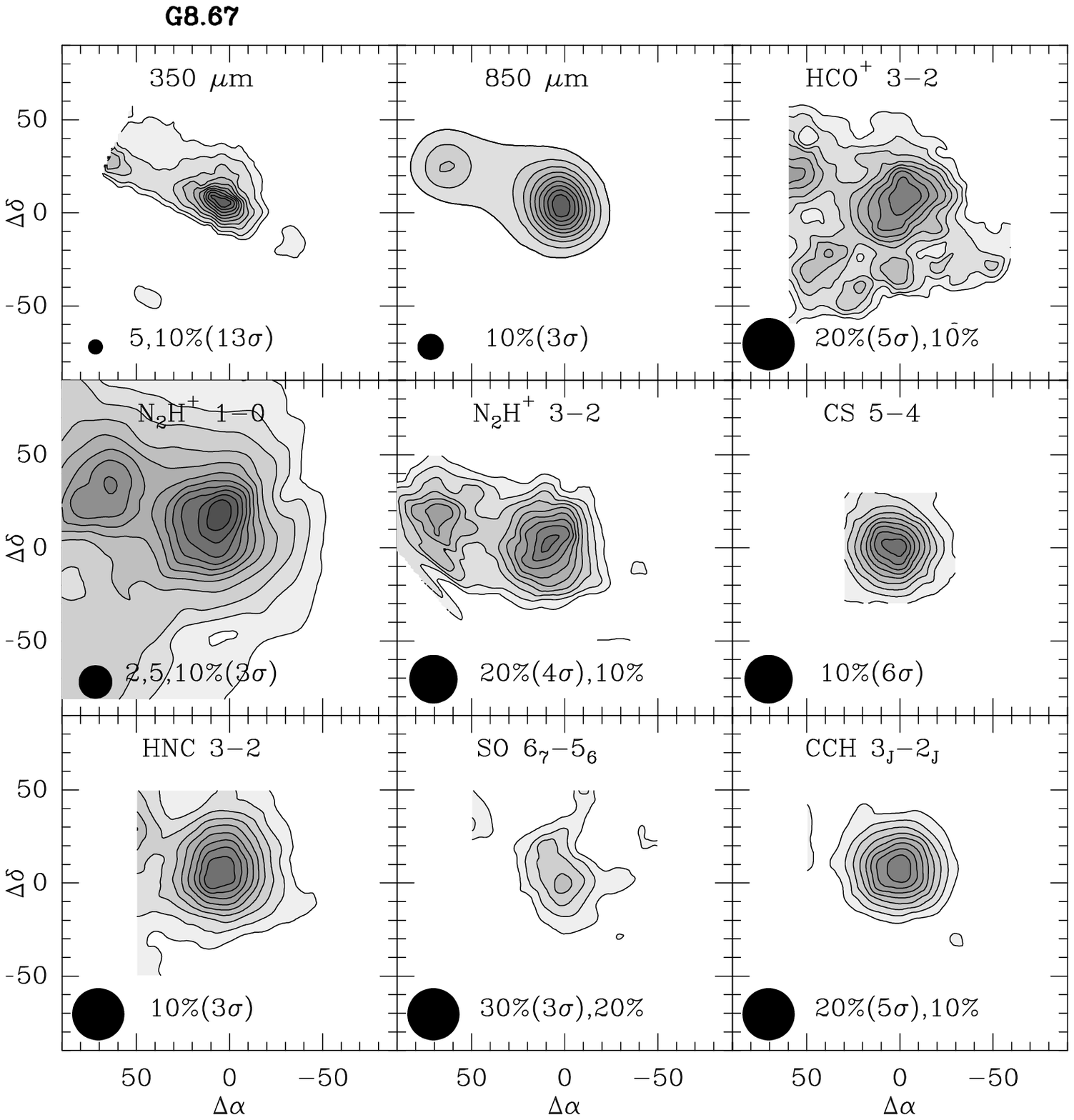}\label{f:g8.67}
\end{figure*}

\begin{figure*}
\figurenum{10}
\epsscale{0.9}
\includegraphics[angle=0, scale=1]{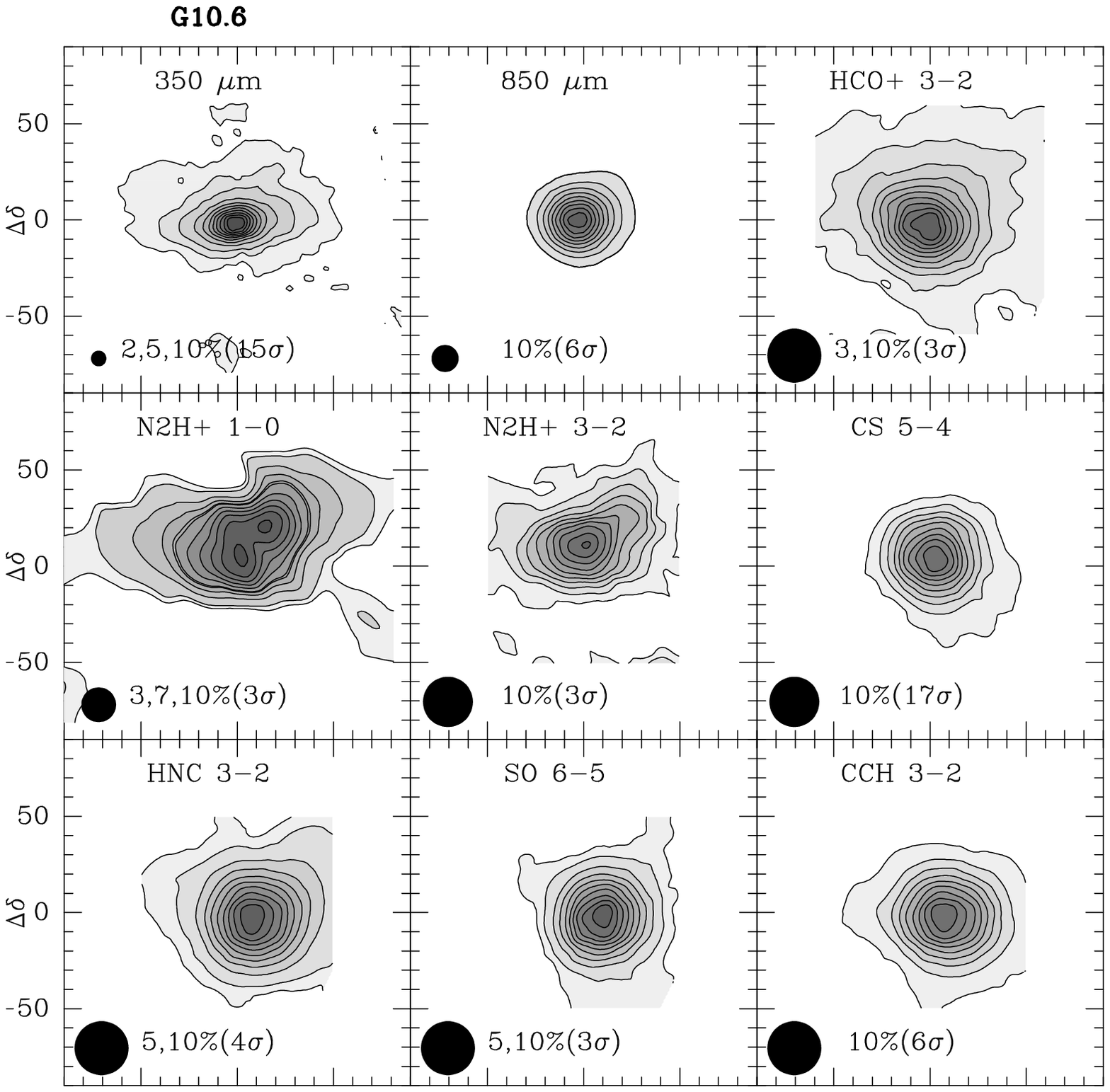}\label{f:g10.6}
\end{figure*}

%% file: chemfigs2.tex
\begin{figure*}
\figurenum{11}
\epsscale{0.9}
\includegraphics[angle=0, scale=1]{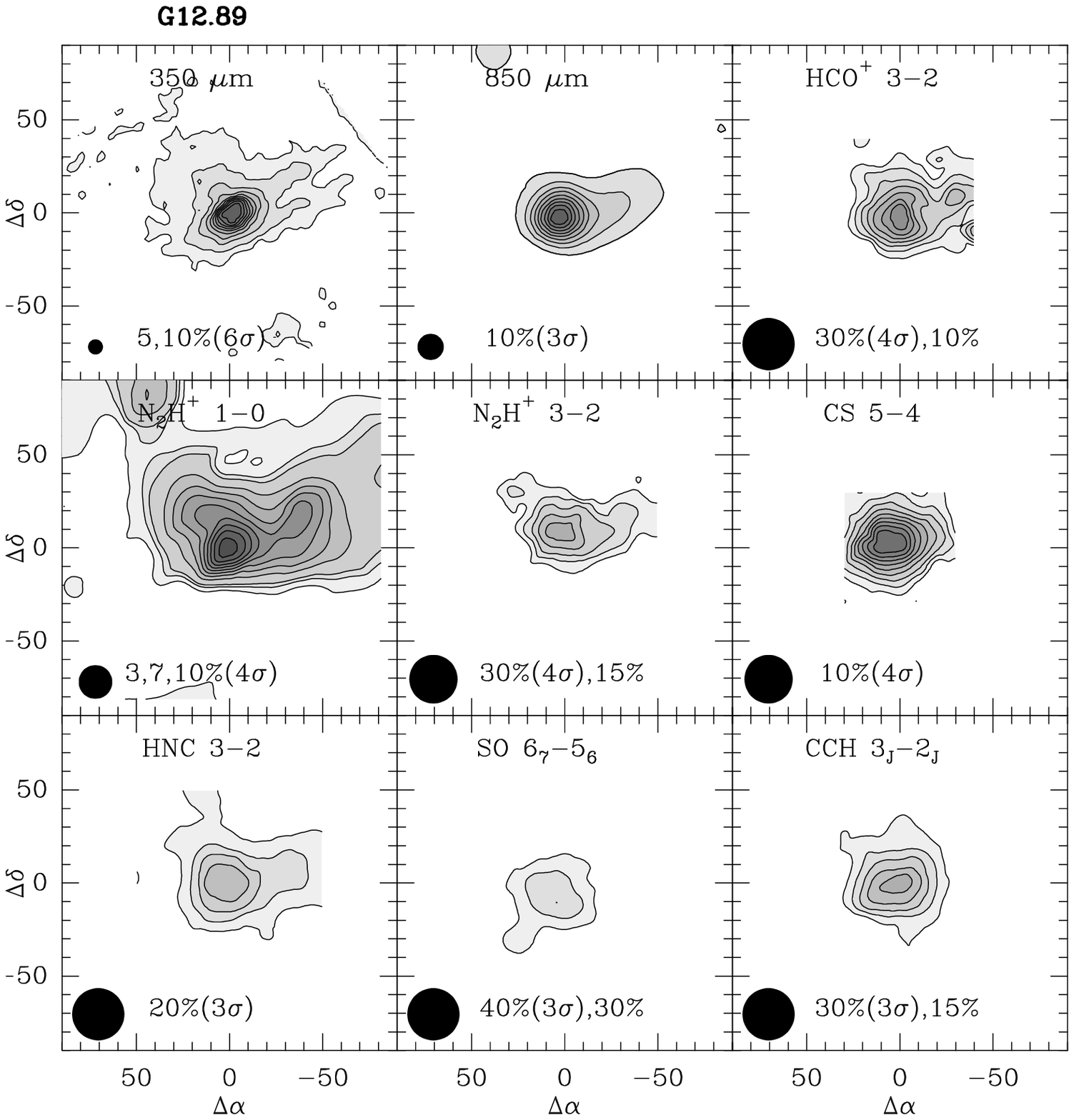}\label{f:g12.89}
\end{figure*}

\begin{figure*}
\figurenum{12}
\epsscale{0.9}
\includegraphics[angle=0, scale=1]{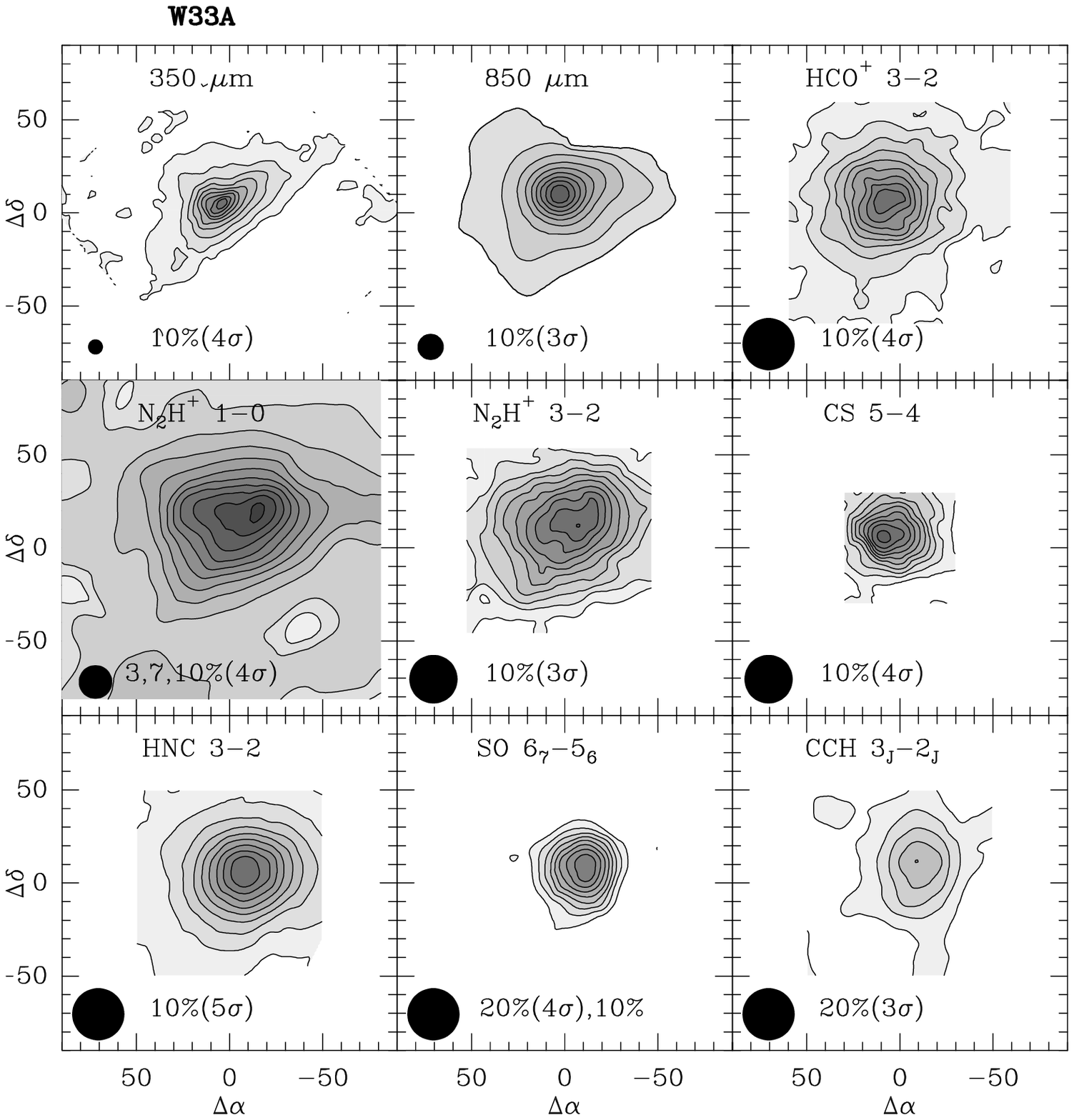}\label{f:w33a}
\end{figure*}

\begin{figure*}
\figurenum{13}
\epsscale{0.9}
\includegraphics[angle=0, scale=1]{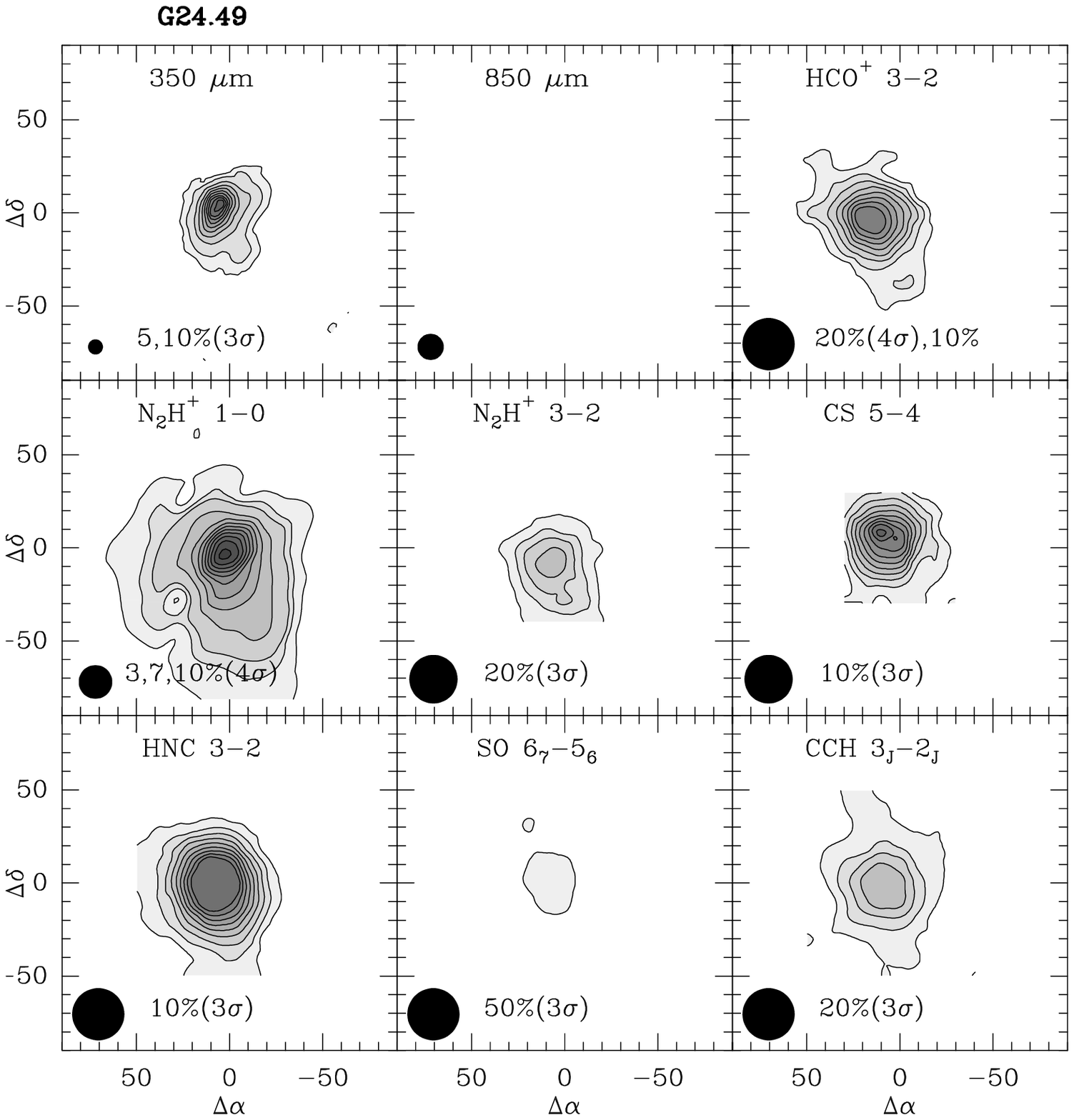}\label{f:g24.49}
\end{figure*}

\begin{figure*}
\figurenum{14}
\epsscale{0.9}
\includegraphics[angle=0, scale=1]{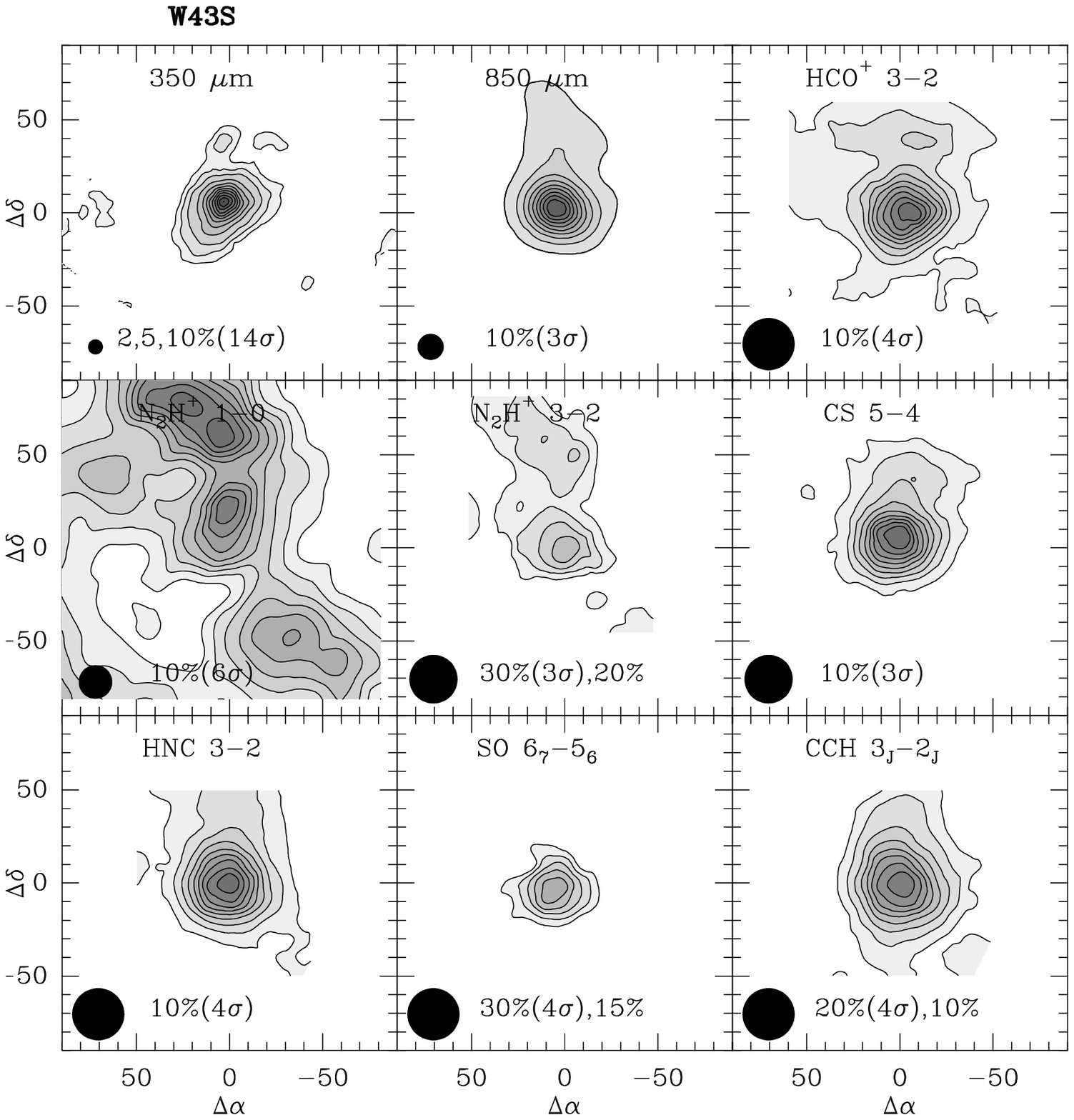}\label{f:w43s}
\end{figure*}

\begin{figure*}
\figurenum{15}
\epsscale{0.9}
\includegraphics[angle=0, scale=1]{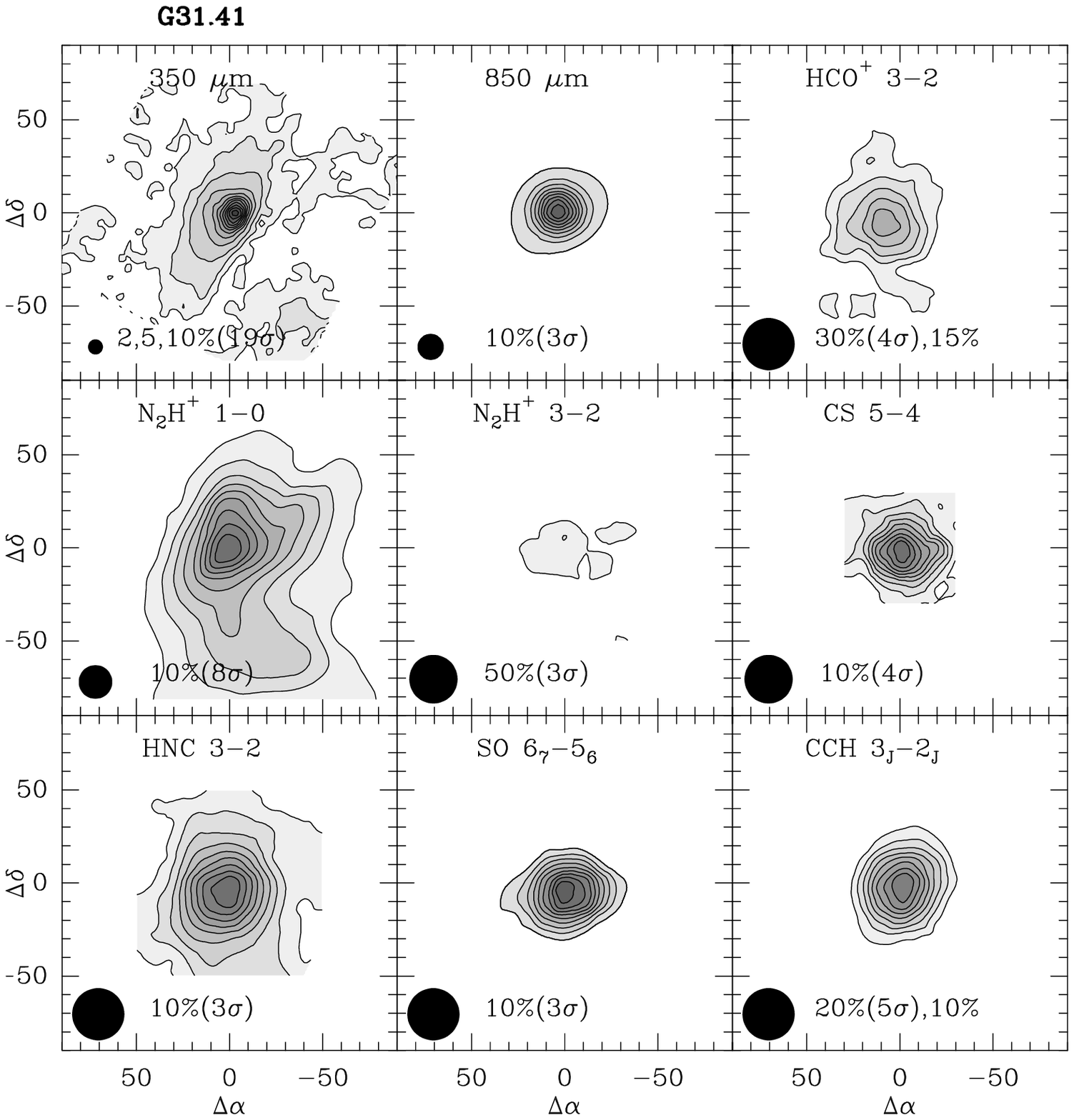}\label{f:g31.41}
\end{figure*}

\begin{figure*}
\figurenum{16}
\epsscale{0.9}
\includegraphics[angle=0, scale=1]{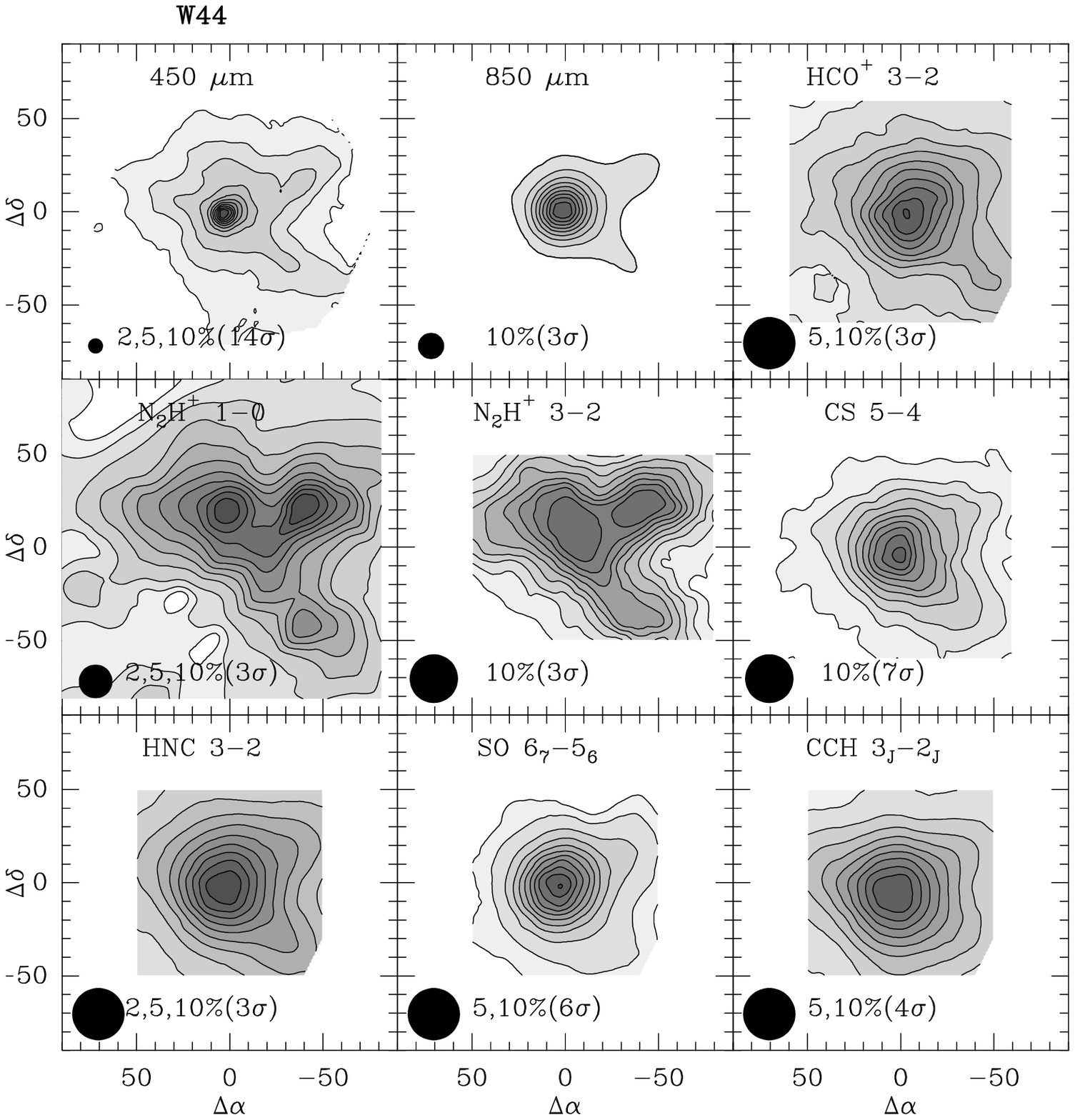}\label{f:w44}
\end{figure*}

\begin{figure*}
\figurenum{17}
\epsscale{0.9}
\includegraphics[angle=0, scale=1]{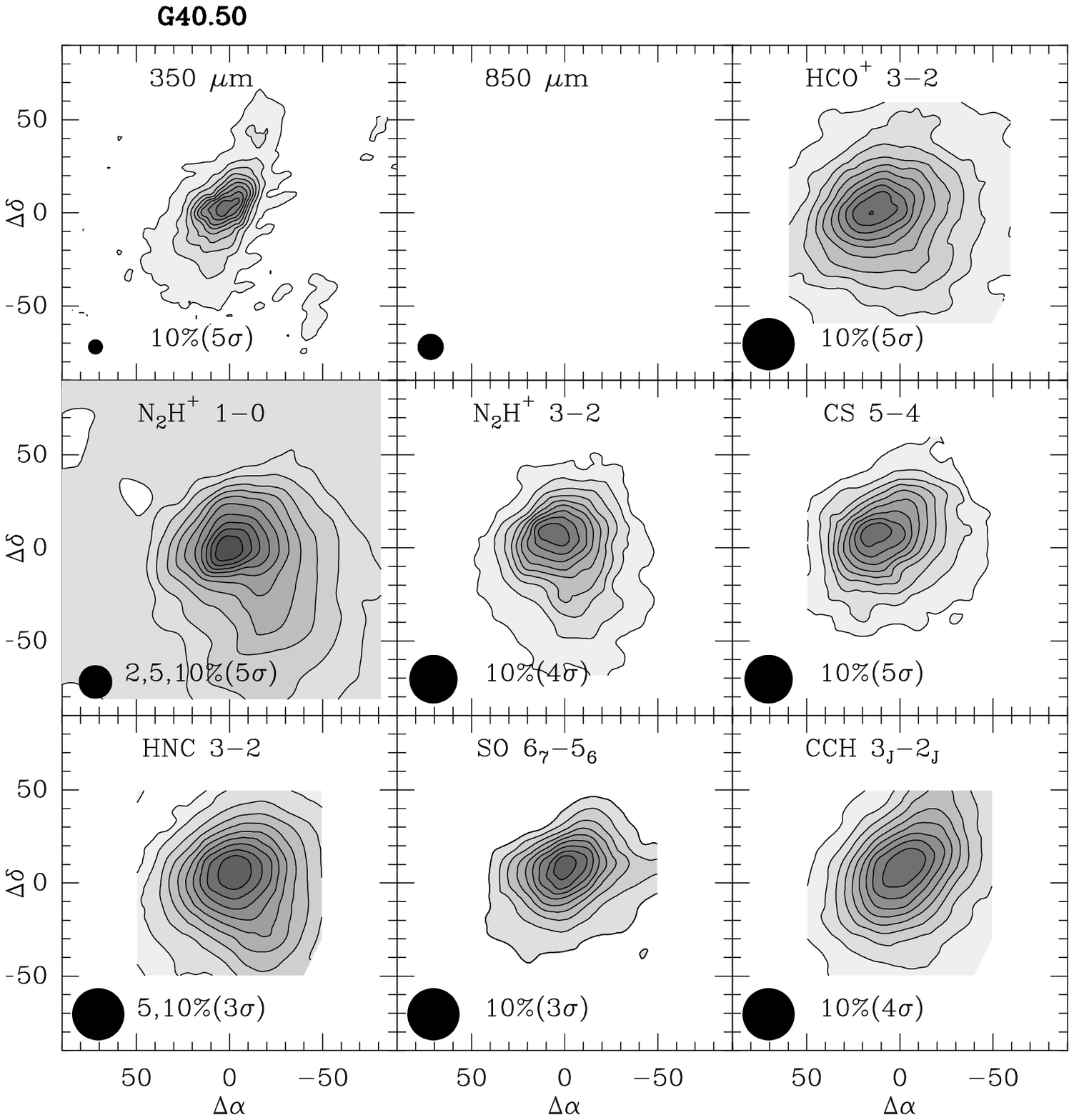}\label{f:g40.50}
\end{figure*}

\begin{figure*}
\figurenum{18}
\epsscale{0.9}
\includegraphics[angle=0, scale=1]{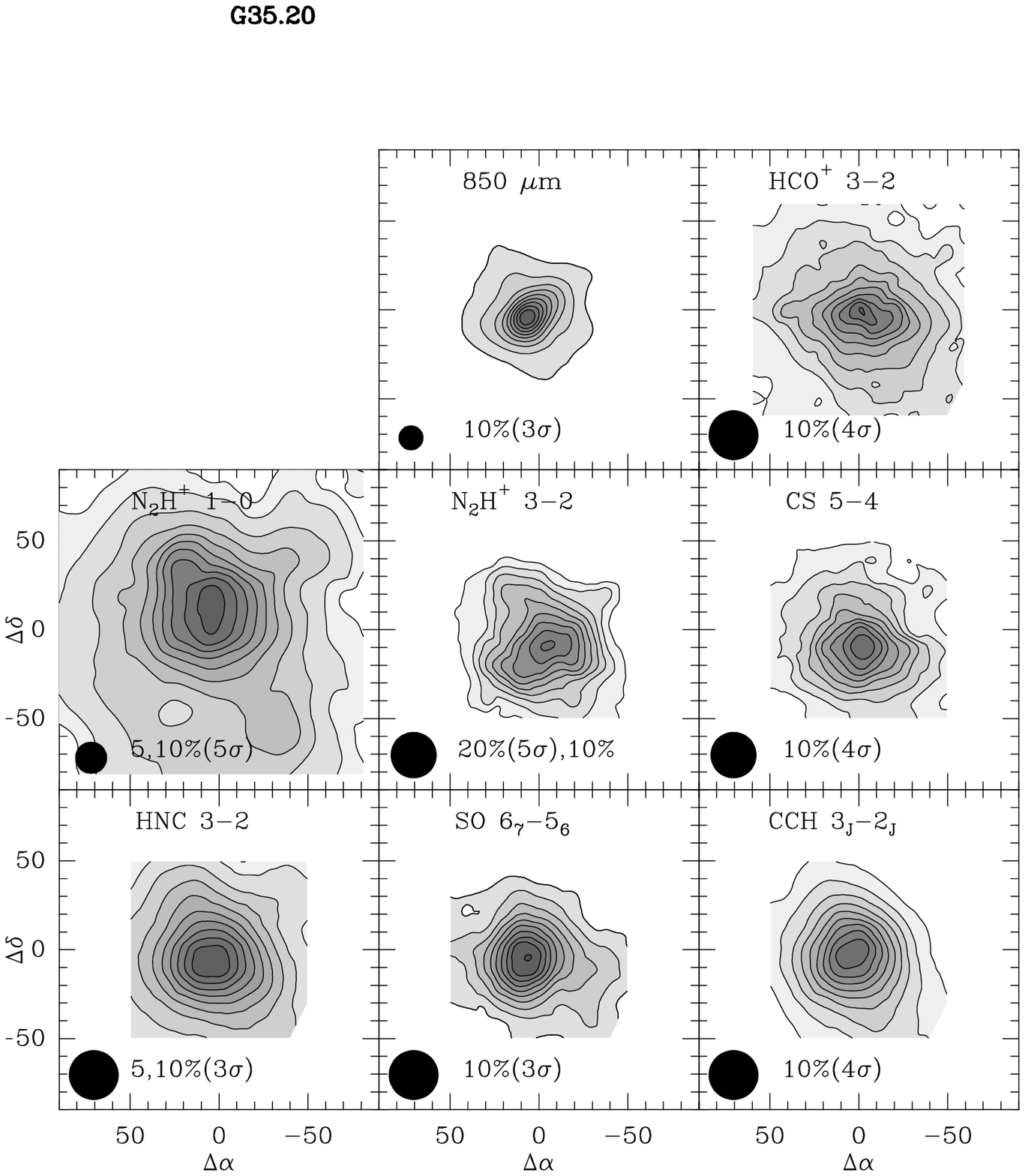}\label{f:g35.20}
\end{figure*}

\begin{figure*}
\figurenum{19}
\epsscale{0.9}
\includegraphics[angle=0, scale=1]{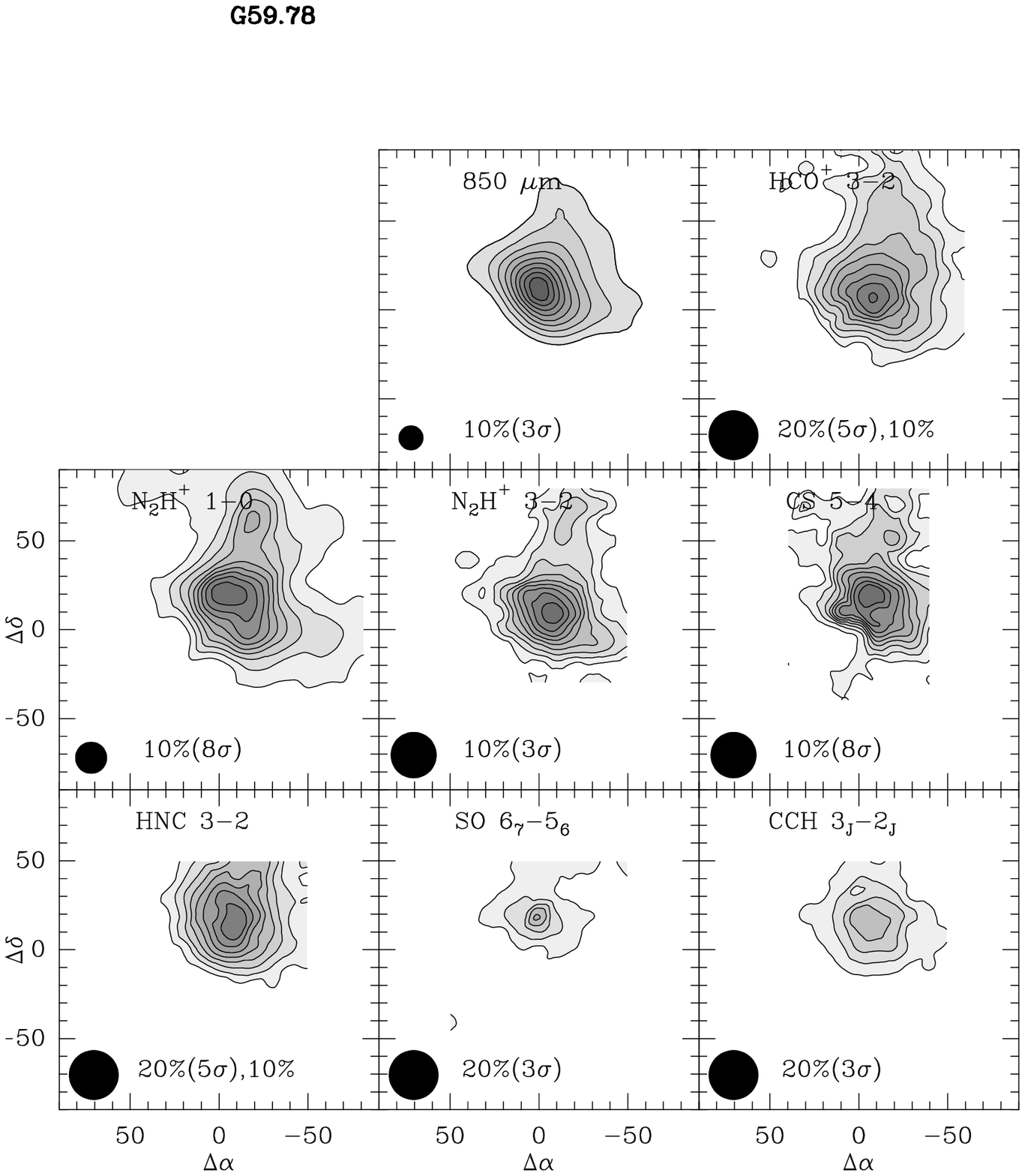}\label{f:g59.78}
\end{figure*}

\begin{figure*}
\figurenum{20}
\epsscale{0.9}
\includegraphics[angle=0, scale=1]{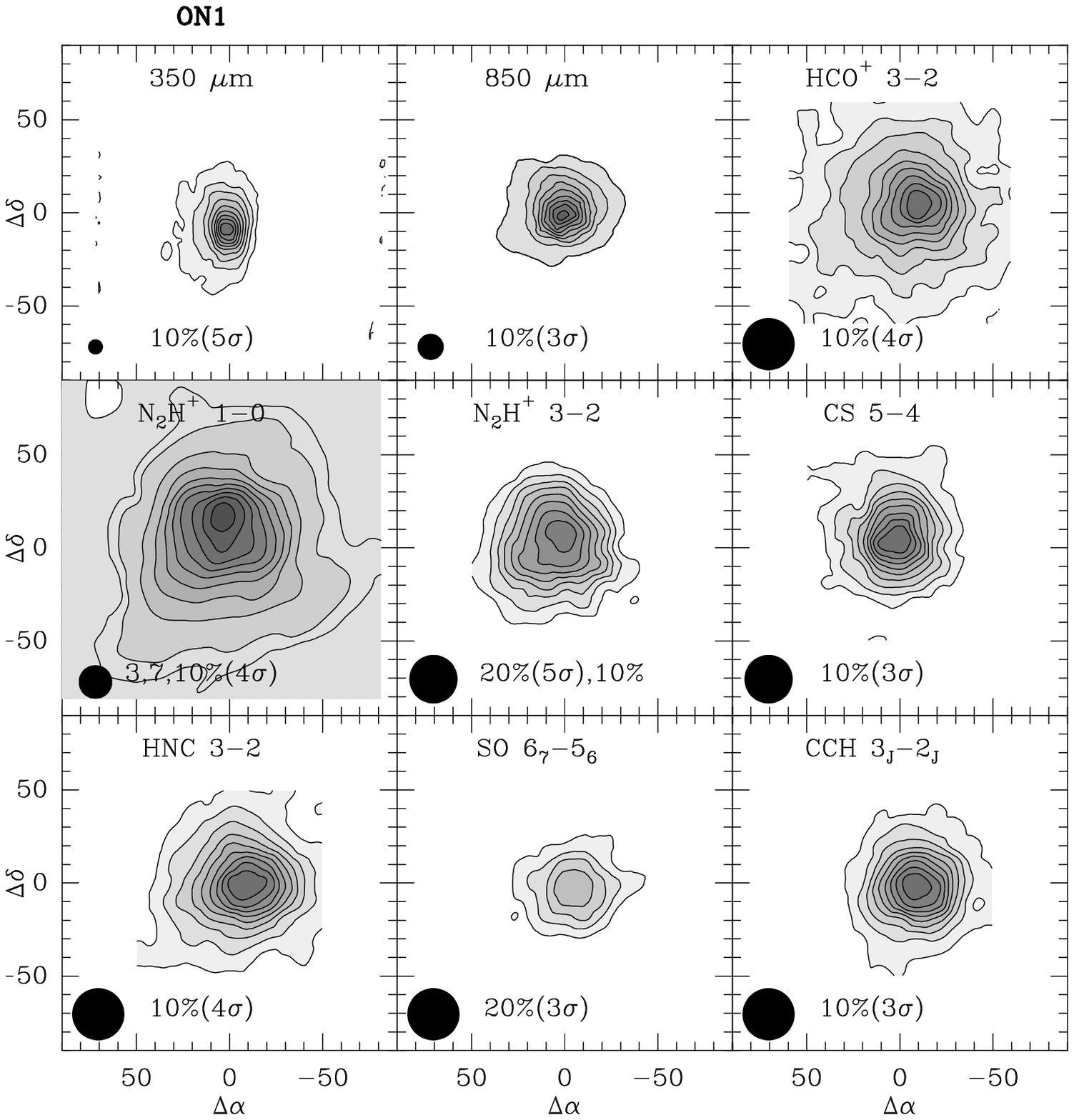}\label{f:on1}
\end{figure*}

%% file: chemfigs3.tex
\begin{figure*}
\figurenum{21}
\epsscale{0.9}
\includegraphics[angle=0, scale=1]{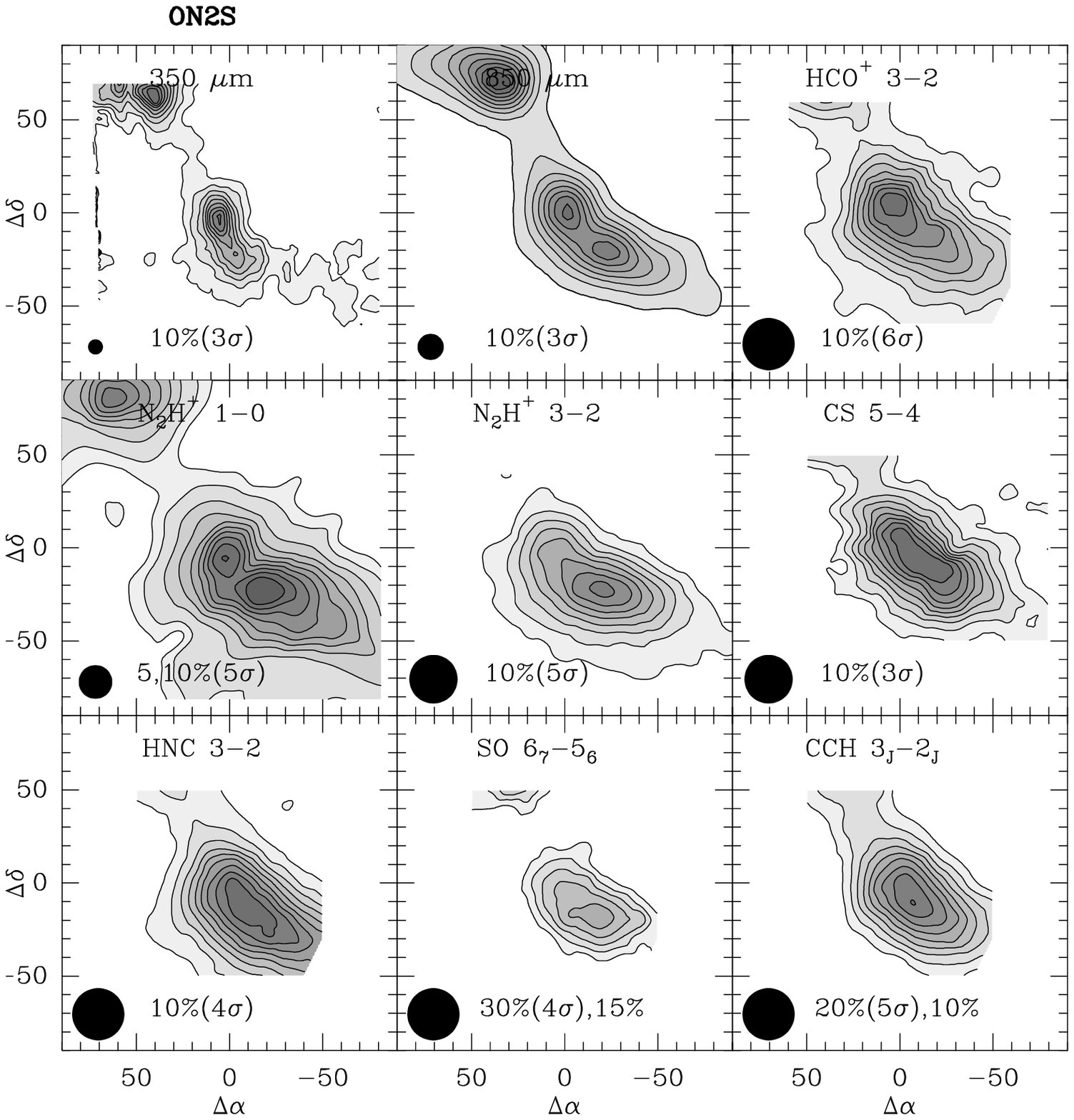}\label{f:on2s}
\end{figure*}

\begin{figure*}
\figurenum{22}
\epsscale{0.9}
\includegraphics[angle=0, scale=1]{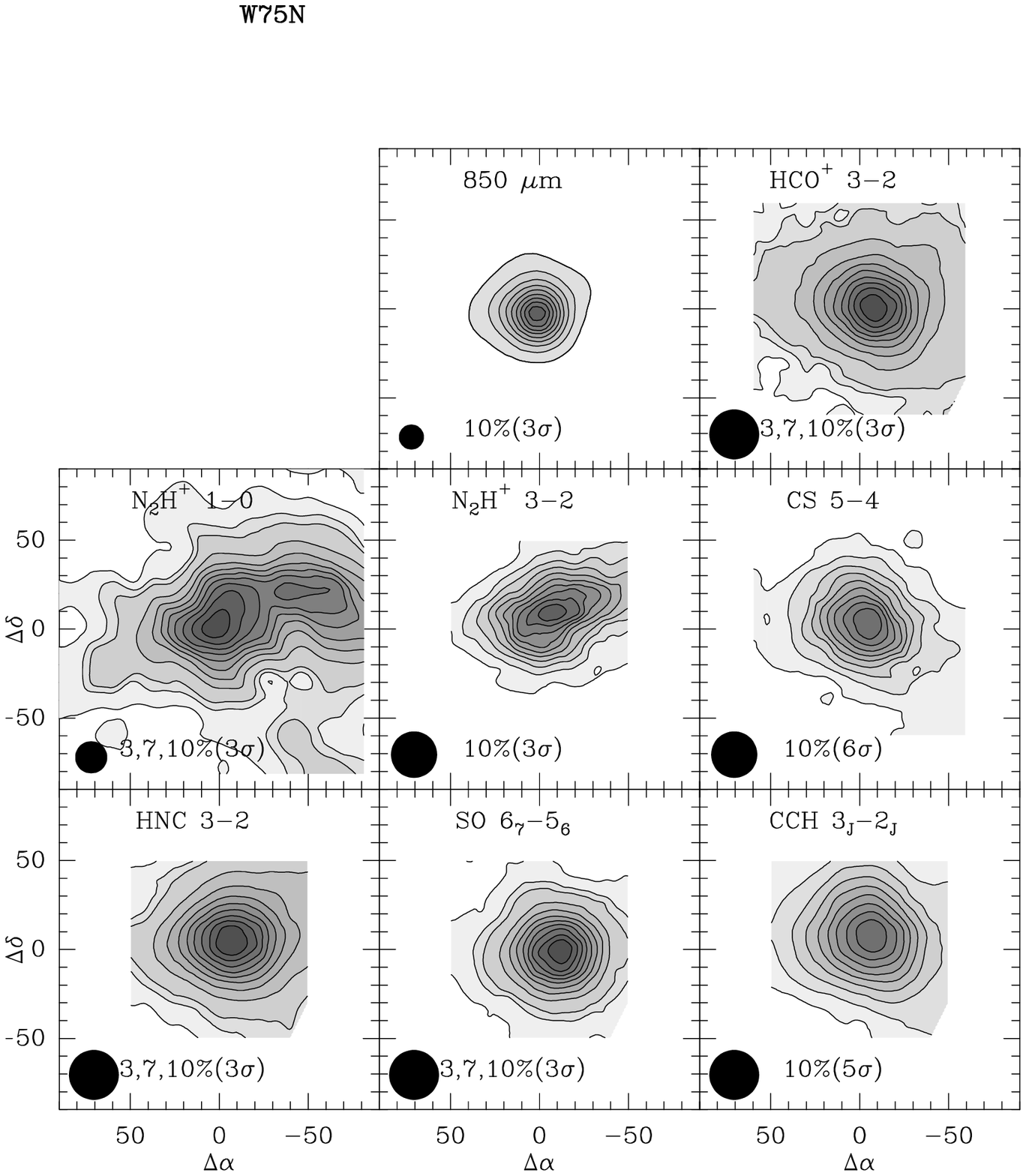}\label{f:w75n}
\end{figure*}

\begin{figure*}
\figurenum{23}
\epsscale{0.9}
\includegraphics[angle=0, scale=1]{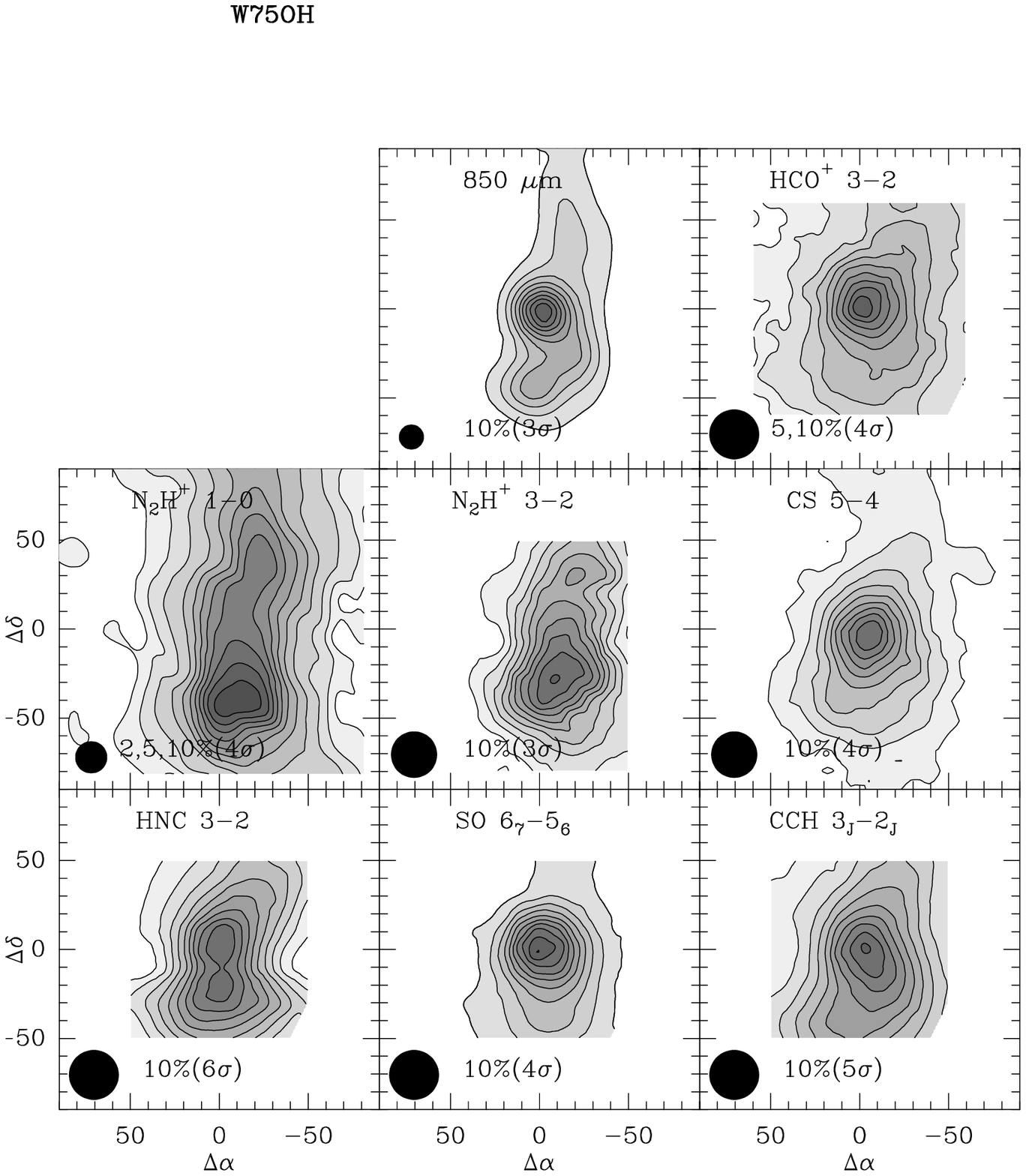}\label{f:w75oh}
\end{figure*}

\begin{figure*}
\figurenum{24}
\epsscale{0.9}
\includegraphics[angle=0, scale=1]{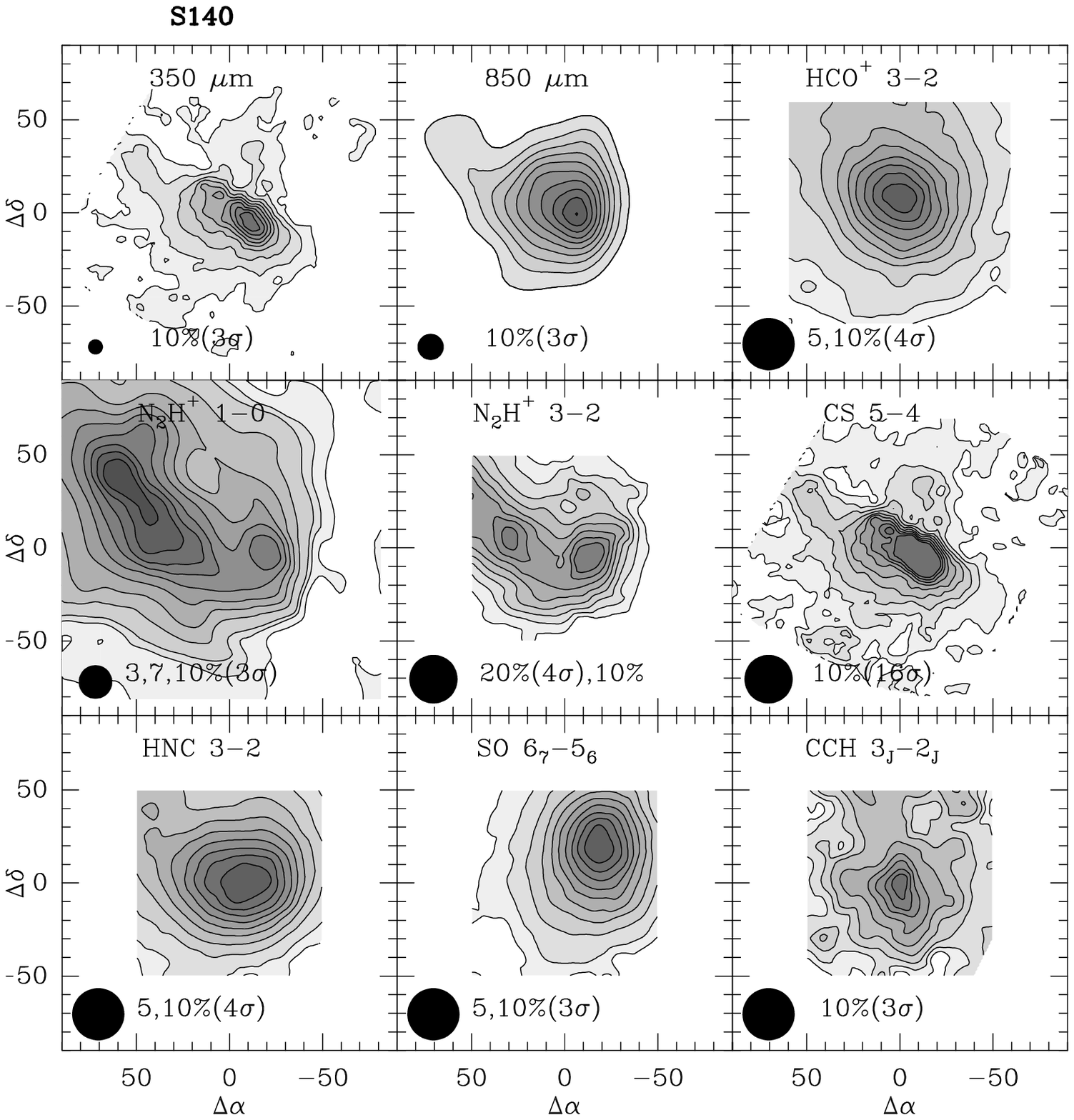}\label{f:s140}
\end{figure*}

\begin{figure*}
\figurenum{25}
\epsscale{0.9}
\includegraphics[angle=0, scale=1]{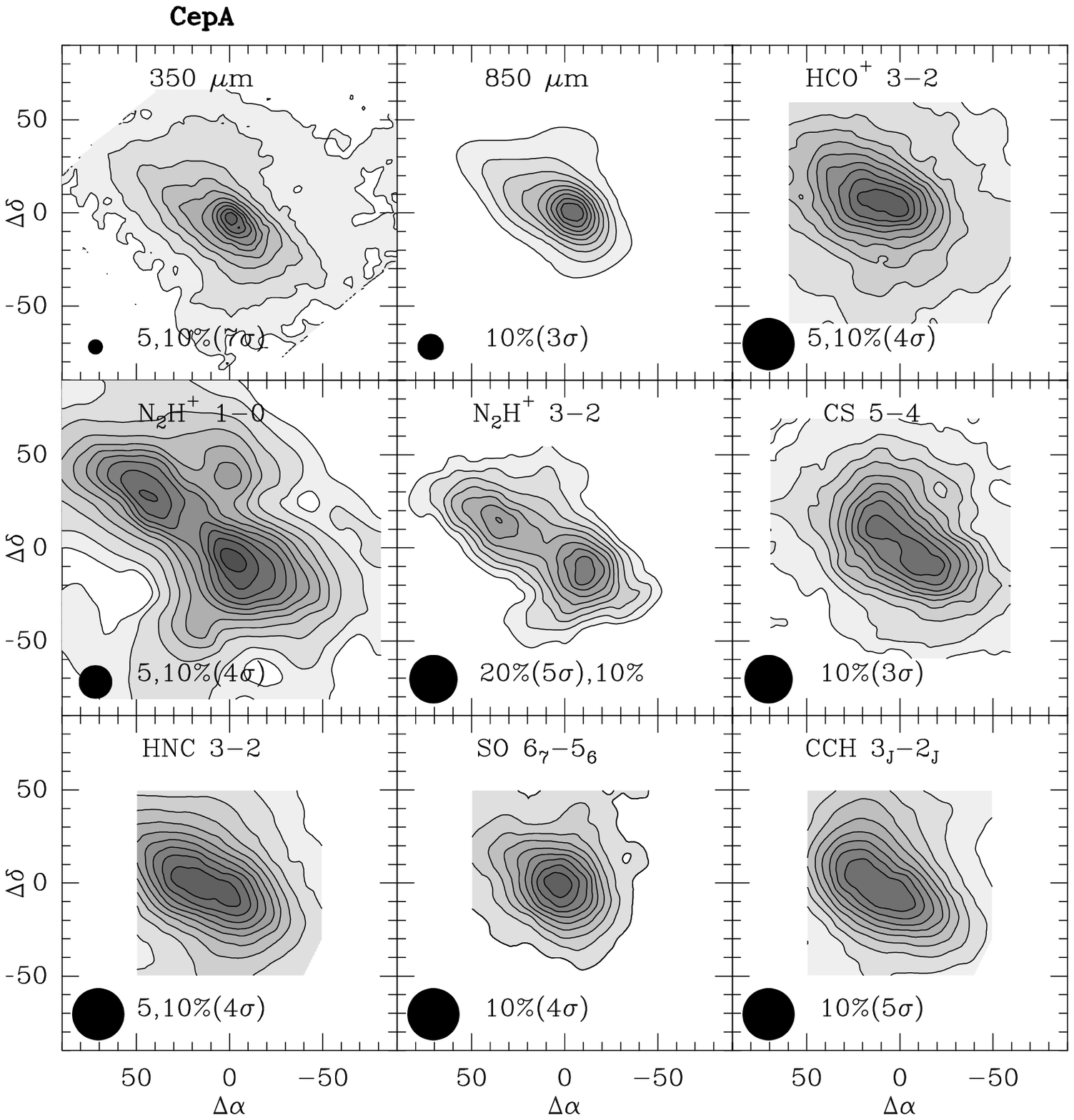}\label{f:cepa}
\end{figure*}

\begin{figure*}
\figurenum{26}
\epsscale{0.9}
\includegraphics[angle=0, scale=1]{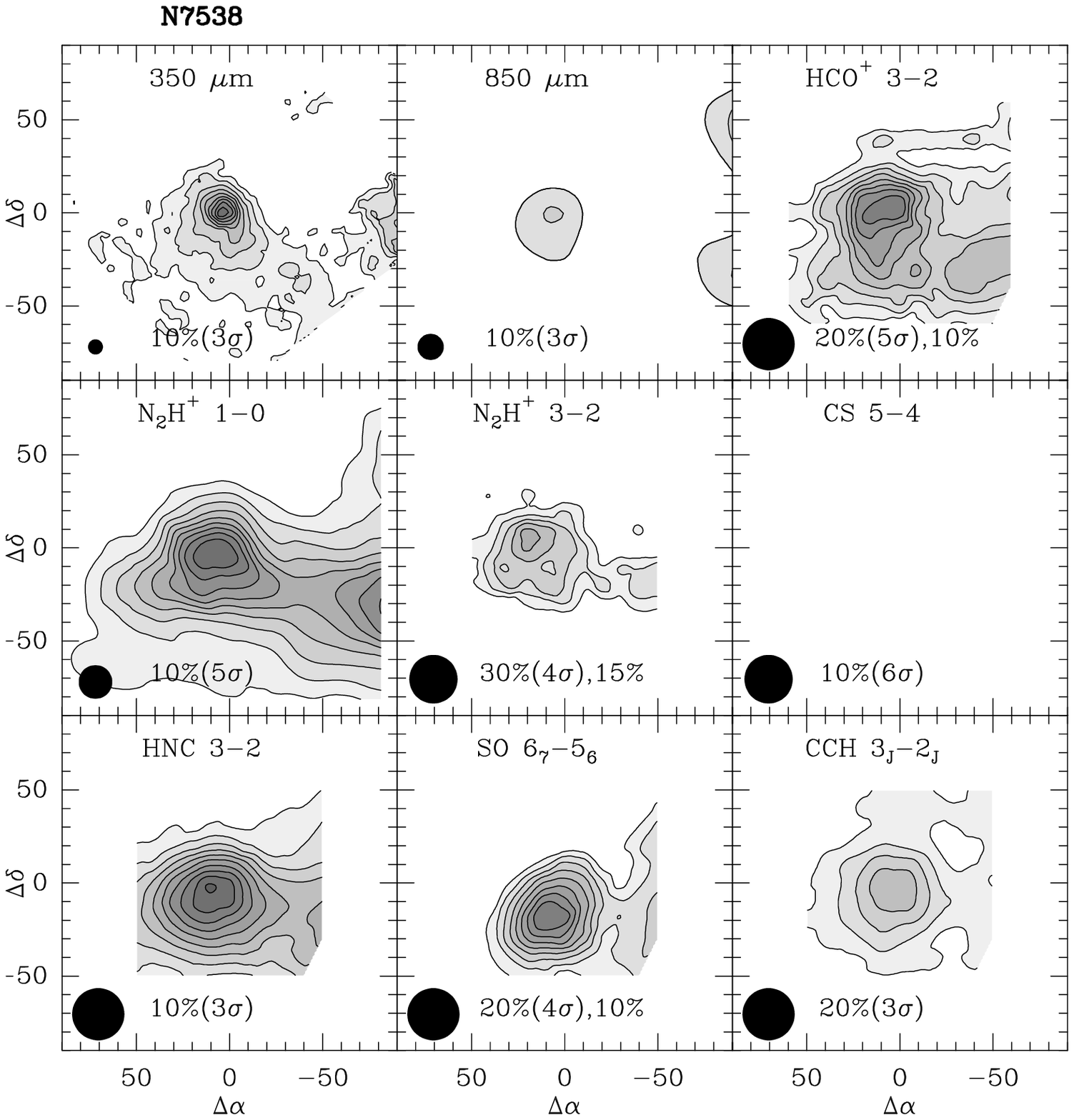}\label{f:n7538}
\end{figure*}

\begin{figure*}
\figurenum{27}
\epsscale{0.9}
\includegraphics[angle=0, scale=1]{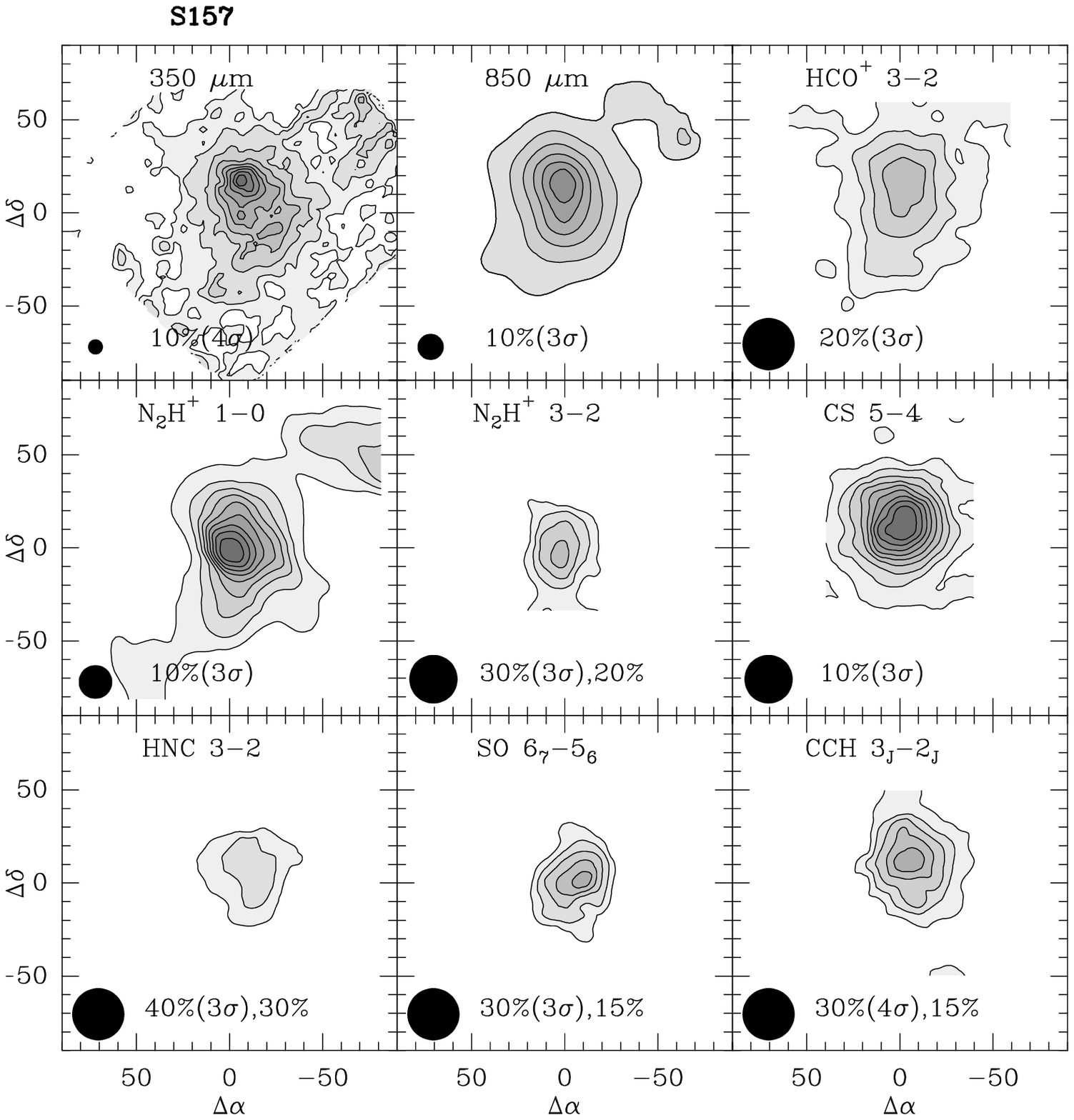}\label{f:s157}
\end{figure*}